%% file: OptTrt_Revision_2.tex
\documentclass[11pt]{article}
\usepackage[margin=1in]{geometry}
\usepackage{amsmath}
\usepackage{graphicx}
\usepackage{enumerate}
\usepackage{enumitem}
\usepackage{natbib}
\usepackage{url} % not crucial - just used below for the URL 

\usepackage{algorithmic} %For computer algorithm code in LATEX
\usepackage{algorithm} %For algorithm box around the algorithmic
\usepackage{setspace} %For setting spacing (e.g., double spacing, use \doublespacing)
\usepackage{rotating} %For rotating tables (provides sidewaysfigure and sidewaystable)

\usepackage{amsfonts, amssymb} %Either provides Real number and other font'related stuff.
\usepackage{amsthm}

\usepackage{epstopdf} %for adding eps in pdfLatex	
\usepackage{hyperref} %For adding hyperlinks in documents
\usepackage{bm} %Bolds everything when uses \bm as a command
\usepackage{bbm}
\usepackage{multirow} %Have multiple rows in a table
\usepackage{latexsym} %Latex-based math symbols. Generally load this in conjunction with amssymb
\usepackage{rotating} %arbitrary rotate any object
\usepackage{titlesec} %centering section headings
\usepackage{caption} %captioning for tables
\usepackage{mathtools} %ceiling/floor commands
\usepackage{color} %Provides color for text.
\usepackage{diagbox}
\usepackage{pict2e}
\usepackage{etoolbox}
\usepackage[english]{babel}
\usepackage{empheq}

\usepackage{cases}
\usepackage{tikz}
\usetikzlibrary{arrows, chains, positioning, quotes, shapes.geometric, arrows.meta}

\usepackage{lato}

\newtheorem{theorem}{Theorem}[section]

\newtheorem{lemma}{Lemma}[section]

\theoremstyle{definition}

\theoremstyle{remark}

\newcommand{\NC}{N}
\newcommand{\NI}{n}
\newcommand{\UNI}{M}

\newcommand{\bO}{\bm{O}}
\newcommand{\bo}{\bm{o}}
\newcommand{\bY}{\bm{Y}}

\newcommand{\bA}{\bm{A}}
\newcommand{\ba}{\bm{a}}
\newcommand{\bX}{\bm{X}}
\newcommand{\bx}{\bm{x}}

\newcommand{\mbO}{\overline{\bm{O}}}
\newcommand{\mY}{\overline{Y}}
\newcommand{\mA}{\overline{A}}
\newcommand{\ma}{\overline{a}}
\newcommand{\mbX}{\overline{\bm{X}}}
\newcommand{\mbx}{\overline{\bm{x}}}
\newcommand{\tbX}{{\bm{X}}}
\newcommand{\tbx}{{\bm{x}}}

\newcommand{\Beta}{\bm{\eta}}

\newcommand{\Y}{\mathcal{Y}}

\newcommand{\mOR}{\mu}
\newcommand{\PS}{e}

\newcommand{\thr}{\mathcal{T}}
\newcommand{\F}{\Theta}
\newcommand{\RKHS}{\mathcal{H}}
\newcommand{\X}{\mathcal{X}}

\newcommand{\pot}[2]{#1^{(#2)}}
\newcommand{\zoset}{\mathcal{A}}
\newcommand{\eij}{{i(-j)}}

\newcommand{\indirect}{\text{IND}}
\newcommand{\LB}{\theta}

\newcommand{\tLB}{\theta}
\newcommand{\loss}{L}
\newcommand{\losszo}{\nu}
\newcommand{\risk}{R}
\newcommand{\trisk}{R}
\newcommand{\kernel}{\mathcal{K}}
\newcommand{\kvec}{\bm{k}}
\newcommand{\kmat}{K}
\newcommand{\winsor}{\mathcal{W}}

\newcommand{\R}{ \mathbb{R} }
\newcommand{\T}{ ^{\intercal} }
\newcommand{\ind}{\mathbbm{1}}
\newcommand{\indep}{\rotatebox[origin=c]{90}{$\models$} \, }
\newcommand{\con}{\, ; \,}
\newcommand{\cond}{\, \big| \,}
\newcommand{\EXP}{{\rm E}}
\newcommand{\VAR}{{\rm Var}}

\newcommand{\SO}{\text{\rm SO}}
\newcommand{\OV}{\text{\rm OV}}

\newcommand{\OTR}{\text{OTR}}
\newcommand{\OAP}{\text{OAP}}
\newcommand{\OURR}{\text{MRTP}}

\DeclareMathOperator*{\argmax}{arg\,max}
\DeclareMathOperator*{\argmin}{arg\,min}
\DeclareMathOperator*{\median}{median}

\hypersetup{
colorlinks=true,
linkcolor=blue,
filecolor=blue,
urlcolor=blue,
citecolor=blue
}

\graphicspath{ {plot/} }% plot path

\definecolor{red1}{RGB}{255,204,204}
\definecolor{blue1}{RGB}{204,204,255}

\numberwithin{table}{section}
\numberwithin{figure}{section}

\usepackage{rotating}
\usepackage{caption}

\definecolor{red1}{RGB}{255,204,204}
\definecolor{blue1}{RGB}{204,204,255}
\definecolor{light-gray}{gray}{0.7}

\defcitealias{SEN_DHS}{ANSD and ICF, 2020}

\begin{document}

\setlength{\abovedisplayskip}{8pt}
\setlength{\belowdisplayskip}{8pt}
\setlength{\abovedisplayshortskip}{8pt}
\setlength{\belowdisplayshortskip}{8pt}

\title{\vspace*{-1cm}Minimum Resource Threshold Policy Under Partial Interference}
 \author{
  Chan Park$^{a}$, 
  Guanhua Chen$^{b}$, Menggang Yu$^{b}$, and Hyunseung Kang$^{c}$
  \\[0.2cm]
  {\small $^{a}$Department of Statistics and Data Science, The Wharton School, University of Pennsylvania} \\
  {\small $^{b}$Department of Biostatistics and Medical Informatics, University of Wisconsin--Madison}\\
  {\small $^{c}$Department of Statistics, University of Wisconsin--Madison} \\
    }
 \date{}
  \maketitle
\begin{abstract}
When developing policies for prevention of infectious diseases, policymakers often set specific, outcome-oriented targets to achieve. For example, when developing a vaccine allocation policy, policymakers may want to distribute them so that at least a certain fraction of individuals in a census block are disease-free and spillover effects due to interference within blocks are accounted for. The paper proposes methods to estimate a block-level treatment policy that achieves a pre-defined, outcome-oriented target while accounting for spillover effects due to interference. Our policy, the minimum resource threshold policy (MRTP), suggests the minimum fraction of treated units required within a block to meet or exceed the target level of the outcome. We estimate the MRTP from empirical risk minimization using a novel, nonparametric, doubly robust loss function. We then characterize statistical properties of the estimated MRTP in terms of the excess risk bound. We apply our methodology to design a water, sanitation, and hygiene allocation policy for Senegal with the goal of increasing the proportion of households with no children experiencing diarrhea to a level exceeding a specified threshold. Our policy outperforms competing policies and offers new approaches to design allocation policies, especially in international development for communicable diseases. 
\end{abstract}
\noindent%

\input{Main.tex}

\newpage

\input{Supp.tex}

\newpage

\bibliographystyle{apa}
\bibliography{OptTrt.bib}

\end{document}

%% file: Main.tex
\section{Introduction} \label{sec:intro} 
A classic problem in individualized treatment regimes is learning an optimal static, single-time policy from data \citep{KM2015, Tsiatis2019}. Briefly, given $N$ units indexed by $i=1,\ldots,N$, the problem is characterized as finding a function $\theta_{\OTR}^*$, which takes in some characteristics of unit $i$, denoted by $\bX_{i} \in \mathcal{X}$, and outputs the level of treatment that maximizes unit $i$'s expected potential outcome; a bit more formally: 
\begin{align}	\label{eq-OTR}
\theta_{\OTR}^*( \bX_i ) = \argmax_{\theta \in \F} \Y \big( \theta(\bX_i) \big) \text{ for all } i=1,\ldots,N \ .
\end{align}
Here, $\F$ is a collection of all policy functions and $\Y(\theta(\bX_i))$ is the expected potential outcome under policy $\theta$ for unit $i$ with characteristic $\bX_i$. For example, consider our application where we design Senegal's water, sanitation, and hygiene (WASH) policy and the investigator defines a unit $i$ to be the $i$th census block of Senegal. The function $\theta(\bX_i)$ is a WASH policy that suggests what proportion of households in the $i$th census block should get WASH facilities based on characteristics of the $i$th census block, $\Y(\theta(\bX_i))$ is the expected diarrhea-free incidence rate within the $i$th census block under a WASH policy $\theta$, and $\theta_{\OTR}^*(\bX_i)$ is the optimal WASH policy for the $i$th census block that maximizes the expected diarrhea-free incidence rate; see Sections \ref{sec:notation} and \ref{sec:SenegalData} for formal definitions about these variables.  Unfortunately, because WASH facilities are well-known to prevent communicable diseases \citep{Diarrhea_New5, Diarrhea_New6,Diarrhea_New3,WASH_Effective2015,McMichael2019}, $\theta_{\OTR}^*$ may suggest all households get WASH facilities. More generally, a direct application of standard methods for learning a static, single-time policy to settings where the treatment/intervention/resource is known to benefit almost all members of the population (e.g., vaccines, mosquito nets, literacy programs) may lead to over-allocation of treatment; see Section \ref{sec:Beneficial Examples} of the Supplementary Materials for some examples.

A common solution in these settings is to incorporate treatment costs with an optimal allocation policy \citep{Richter1999, Laber2018, Ren2022}. Specifically, an optimal allocation policy, denoted as $\theta_{\OAP}^*$, maximizes the total expected outcome across $N$ units under some (often linear) cost constraint  $\mathcal{C}$: 
\begin{align}	
& 
\theta_{\OAP}^* = \argmax_{\theta \in \F} \sum_{i=1}^{N} \Y \big( \theta(\bX_i) \big) \quad{} \text{ subject to } \quad{}
\big\{ \theta(\bX_1),\ldots,\theta(\bX_N) \big\} \in \mathcal{C} \ .
\label{eq-OTA}
\end{align}
Due to the cost constraint, $\theta_{\OAP}^*$ will usually suggest less treatment than $\theta_{\OTR}^*$ for each policy target, and $\theta_{\OAP}^*$ can actually under-allocate treatment if the costs are significant. For example, for WASH policies, $\theta_{\OAP}^*$ may allocate fewer WASH facilities to small, rural census blocks where the cost of installing WASH facilities is high compared to urban, well-developed census blocks where the installation cost is small due to existing infrastructure. In other words, $\theta_{\OAP}^*$ may maximize the ``greatest good for the greatest number'' at the expense of certain, costly census blocks and may lead to an inequitable, unfair policy. However, estimating $\theta_{\OAP}^*$ is often computationally demanding, requiring mixed integer programming or Monte Carlo, Bayesian optimization. 

The main goal of the paper is to propose an alternative solution based on outcome targets, which, like cost constraints, are widely available; see Section \ref{sec:Outcome Examples} of the Supplementary Material and \citet{Chen2022} for examples. For instance, when developing a WASH policy for diarrhea prevention (i.e., the outcome), the World Health Organization's Integrated Global Action Plan for Pneumonia and Diarrhea \citep{Diarrhea_New8} recommends reducing the incidence of severe diarrhea to at least 75\% below country-specific levels in 2010. Let $\thr$ denote such targets for the outcome, specifically the smallest value of the outcome that is deemed to be acceptable. Given $\thr$, we propose a policy $\theta_\OURR^*(\bX_i)$, which suggests the smallest amount of treatment that is necessary to meet or exceed the target outcome level for the $i$th policy target, i.e.,
\begin{align}	\label{eq-Ours} 
\theta_\OURR^*(\bX_i) = \argmin_{\theta \in \F} \theta (\bX_i)  \ \text{ subject to }  \ \Y (\theta(\bX_i)) \geq \thr \ \text{ for all } \ i=1,\ldots,N \ .
\end{align}
We refer to $\theta_\OURR^*(\bX)$ as the \textit{minimum resource threshold policy} ({\OURR}). In the WASH example, if $\thr = 0.75$, $\theta_{\OURR}^*$ suggests the minimum proportion of households that need WASH facilities in order to achieve a diarrhea-free incidence rate of at least 75\%. A census block that is economically well-developed may not need any WASH facilities (i.e., $\theta_{\OURR}^* = 0$) to ensure that at least 75\% of its households are diarrhea-free. But, an under-developed block may need $90$\% of its households to have WASH facilities  (i.e., $\theta_{\OURR}^* = 0.9$) in order to achieve the same, disease-free incidence rate. In contrast, an optimal treatment rule $\theta_{\OTR}^*$ may suggest providing WASH facilities to all households (i.e., $\theta_{\OTR}^* = 1$). Also, an optimal allocation policy $\theta_{\OAP}^*$ may suggest that less than 90\% of households in under-developed blocks receive WASH facilities if the installation cost in these blocks is high. Or $\theta_{\OAP}^*$ may suggest that a non-zero fraction of households in well-developed blocks receive WASH facilities, even though these blocks do not need WASH facilities to achieve the target incidence rate. Computationally, estimating $\theta_{\OURR}^*$ is much simpler than estimating $\theta_{\OAP}^*$, as the proposed estimator of $\theta_{\OURR}^*$ is an instance of a computationally feasible empirical risk minimization problem.

The paper also addresses another related problem when developing the MRTP for prevention of communicable diseases, the presence of within-block spillover effects through partial interference. Briefly, interference is a phenomenon where the outcome of a study unit is affected by others' treatment \citep{Cox1958, Rubin1986} and partial interference \citep{Sobel2006} is a type of interference where interference occurs only within non-overlapping blocks of study units. In the WASH example, having a sufficient number of households with WASH facilities is believed to have a multiplier effect for the entire census block \citep{Diarrhea2}. Unfortunately, some existing approaches  (e.g., Section 2.3 of \citet{Imbens2009} and \citet{Kilpatrick2021}) to account for partial interference through a block-level analysis (i.e., analysis using only block-level summaries of the data) can lead to misleading policies; see Section \ref{sec:aggregation} of the Supplementary Material for details. Recent works by \citet{Laber2018},  \citet{Su2019}, \citet{Viviano2021}, and \citet{Jiang2022} proposed identification and estimation of $\theta_{\OTR}^*$, $\theta_{\OAP}^*$, or their dynamic counterparts under interference. \citet{Su2019} and \citet{Jiang2022} focused on $\theta_{\OTR}^*$ at a single time point and at multiple time points, respectively, and their estimators are instances of semiparametric Q- and A-learning. \citet{Laber2018} focused on $\theta_{\OAP}^*$ at multiple time points and the estimator is based on a Bayesian, Monte Carlo framework. \citet{Viviano2021} estimated $\theta_{\OTR}^*$ and $\theta_{\OAP}^*$ using cross-fitting \citep{Victor2018} and mixed-integer programming. To the best of our knowledge, none of these existing works in interference addressed identifying or estimating the policy $\theta_{\OURR}^*$. 

Our work builds on an earlier work by \citet{Chen2022} on estimating a range of effective doses. Specifically, we extend \citet{Chen2022} to identify, estimate, and theoretically characterize the smallest effective dose of treatment when partial interference is present and when the nuisance functions are nonparametrically estimated. In particular, the presence of partial interference necessitates a new estimand (see Sections \ref{sec:notation}-\ref{sec:2.3}), new estimators (see Sections \ref{sec:Risk}-\ref{sec:ObservationalStudy}), and a new theoretical analysis (see Section \ref{sec:theory}); in contrast, \citet{Chen2022} assumed no interference, which enabled using well-known estimators and theoretical analysis.

The rest of the paper is divided as follows. Section \ref{sec:setup} discusses identifying $\theta_{\OURR}^*$ in the presence of partial interference. Section \ref{sec:Method} discusses estimating $\theta_{\OURR}^*$ based on a nonparametric, empirical risk minimization problem with a novel loss function.  Notably, the loss function has a double-robustness property, which results in a better convergence rate compared to those using either the outcome or the propensity score model alone. % see Theorem \ref{thm-ER-3} for details. 
Section 	\ref{sec:Simulation} applies the proposed method to simulated data and Section \ref{sec:SenegalData} uses the proposed method to develop a WASH policy for Senegal.

\section{Setup}		\label{sec:setup}

\subsection{Review: Notation, Interference, and Overall Outcomes} \label{sec:notation}

Let $\NC$ be the number of blocks and each block is indexed by $i=1,\ldots,\NC$. For each block, let $\NI_i$ be the number of study units in block $i$ and study units in a block are indexed by $j=1,\ldots, \NI_i$. We remark that unlike Section \ref{sec:intro} where the review of works on optimal treatment regimes focused on the case without interference, we now use $\NC$ to denote the total number of blocks, not the total number of units. We also introduce $\NI_i$ to denote the total number of units in block $i$. Without interference, defining a block is not necessary and $\NC$ would usually refer to the total number of units; see Section \ref{sec:intro} and the included references. We assume $\NI_i$ is bounded above by a constant $\UNI$ for all $\NI_i$. This is a common assumption in the partial interference literature to enable ``large $\NC$, small $\NI_i$'' asymptotics; see Section \ref{sec:theory}. Also, the bounded $\NI_i$ assumption can be satisfied by some sampling designs, including the one used in our application; see Section \ref{sec:BOAssessment} of the Supplementary Materials. 

For each study unit $j$ in block $i$, we observe $\bO_{ij} = (Y_{ij}, A_{ij}, \bX_{ij})$ where $Y_{ij}$ is the unit-level outcome that takes a value over compact support, $A_{ij}$ is the unit-level treatment indicator with $A_{ij} = 1$ denoting that the unit was treated and $A_{ij} = 0$ denoting that unit was not treated, and $\bX_{ij}$ is the unit- and cluster-level pre-treatment covariates where the dimension of the covariates is finite. Without loss of generality, we assume a larger $Y_{ij}$ is preferred and $Y_{ij}$ is bounded in the range $[0,1]$. Also, let $\bO_i = (\bY_i  , \bA_i  , \bX_i )$ be the collection of all study units' observed data in block $i$ where $\bY_i = (Y_{i1},\ldots,Y_{i \NI_i} )\T$, $\bA_i = (A_{i1}, \ldots, A_{i \NI_i})\T$, and $\bX_i = (\bX_{i1}\T, \ldots, \bX_{i \NI_i}\T)\T$. Let $\bA_{\eij} = (A_{i1},\ldots,A_{i,j-1},A_{i,j+1},\ldots,A_{i\NI_i})\T$ be the vector of treatment indicators of all study units in block $i$ that excludes study unit $j$, $\bX_{\eij} =  (\bX_{i1}\T,\ldots,\bX_{i,j-1}\T,\bX_{i,j+1}\T,\ldots,\bX_{i\NI_i}\T)\T$ be the vector of covariates of all study units in block $i$ that excludes study unit $j$, and $\bO_{\eij} =  (\bO_{i1},\ldots,\bO_{i,j-1},\bO_{i,j+1},\ldots,\bO_{i\NI_i})$ be the vector of observed data of all study units in block $i$ that excludes study unit $j$. Let $S_i = \sum_{j=1}^{\NI_i} A_{ij}$ be the sum of treatment indicators in block $i$ and let $S_\eij = \sum_{\ell \neq j} A_{i\ell}$ be the sum of treatment indicators in block $i$ that excludes study unit $j$. Lastly, let $\mA_i = S_i / \NI_i$ and $\mA_{\eij} = S_{\eij} / (\NI_i-1)$ be the proportion of treated units in block $i$ and the proportion of treated peers of unit $j$ in block $i$, respectively.

In the WASH example, $Y_{ij}$ is a binary indicator where $Y_{ij}=1$ indicates that no child has diarrhea in household $j$ from census block $i$ and $Y_{ij}=0$ indicates the opposite. Also, $A_{ij}=1$ indicates that the household has a WASH facility whereas $A_{ij}=0$ indicates that the household does not have a WASH facility. Section \ref{sec:DataDesc} lists the covariates in $\bX_{ij}$ and $\bO_i$ is all the observed data from $n_j$ households in census block $i$.

To define causal parameters under partial interference, we adopt the potential outcomes framework. Let $\zoset(t)$ be a collection of $t$-dimensional binary vectors (e.g., $\zoset(2) = \{(0,0), (0,1), (1,0), (1,1) \}$). Let $\pot{ Y_{ij} }{ a_{ij}, \ba_{\eij} } $ be the potential outcome of study unit $j$ in block $i$ which one would have observed if, possibly contrary to the fact, study unit $j$ in block $i$ had been assigned to treatment level $a_{ij} \in \{ 0,1\}$ and the other study units in the same block had been assigned to treatment level $\ba_{\eij} \in \zoset(\NI_i-1)$. In other words, treating a study unit influences not only its own outcome but also the outcomes of other study units in the same block. But, the treatment of a study unit cannot influence the outcome of study units in a different block and thus, partial interference may not be appropriate if there are significant interactions among study units in different blocks. Also, for $\ba_i \in \zoset(\NI_i)$, let $\pot{ \bY_{i} }{ \ba_i } = (\pot{Y_{i1}}{a_{i1}, \ba_{i(-1)}}, \ldots, \pot{Y_{i \NI_i }}{a_{i \NI_i }, \ba_{i(- \NI_i )}} ) \T$ be the collection of all potential outcomes of study units in block $i$. The paper assumes a super-population setup where each $\pot{ \bY_{i} }{ \ba_i}$ and $\bO_i$ are an independent sample from an infinite population of blocks of bounded size.

Following \citet{HH2008} and \citet{TTV2012}, we define the expected overall potential outcome of a policy $\alpha \in [0,1]$: 
\begin{align}
\Y_{\OV}(\alpha)
= 
\EXP \bigg\{ \frac{1}{\NI_i} \sum_{j=1}^{\NI_i} \sum_{\ba_i \in \zoset(\NI_i)} \pot{Y_{ij}}{a_{ij}, \ba_{\eij} }  \pi(\ba_i \con \alpha) \bigg\}
\ , \
\pi(\ba_i \con \alpha) 
=
\prod_{j=1}^{\NI_i} \alpha^{a_{ij}} (1-\alpha)^{1-a_{ij}}
\ . \label{eq:overall_pot}
\end{align} 
In words, $\Y_{\OV}(\alpha)$ measures the expected potential outcome of a policy that independently assigns treatment to each unit in a block with probability $\alpha$. In the WASH example, $\Y_{\OV}(\alpha)$ measures the expected number of diarrhea-free households in a block if roughly $\alpha$ fraction of households in the block have WASH facilities. \citet{VWTT2011} and \citet{Halloran2019} showed that $\Y_{\OV}(\alpha)$ measures the entire impact of a policy and \citet{Halloran2019} recommended using the overall potential outcome to evaluate policies under interference. Also,  $\pi(\ba_i \con \alpha)$ in $\Y_{\OV}(\alpha)$ can be replaced with a stochastic, dependent version of it \citep{Barkley2020} and in principle, the estimator based on the indirect approach in Section \ref{sec-Indirect} can be constructed. However, for the estimator based on the direct approach in Section \ref{sec:Risk}, a stochastic, dependent $\pi$ requires a different loss function and potentially more modeling assumptions on the propensity score compared to the $\pi$ in equation \eqref{eq:overall_pot} that allows for nonparametric models of the propensity score; see Section \ref{sec:ObservationalStudy}.

Also, we define the following secondary outcome of interest, which we call the ``spillover outcome under policy $\alpha$,''  and is denoted as $\Y_{\SO}(\alpha)$:  
\begin{align}
\Y_{\SO}(\alpha)  
=
\EXP \bigg\{ \frac{1}{\NI_i} \sum_{j=1}^{\NI_i} \sum_{\ba_\eij \in \zoset(\NI_i-1)}  \pot{Y_{ij}}{0, \ba_\eij}  \pi(\ba_\eij \con \alpha) \bigg\} \ .
\label{eq:spilloverall_pot}
\end{align}
In words, $\Y_{\SO}(\alpha)$ is the expected potential outcome of a block when the study unit is left untreated, but their peers are treated with probability $\alpha$; see Sections 3 and 4 of \citet{VWTT2011} for additional discussions about this potential outcome. In the WASH example, $\Y_{\SO}(\alpha)$ is the expected proportion of diarrhea-free households in a block if the household does not have WASH facilities, but roughly $\alpha$ fraction of their peers have WASH facilities. While not the primary outcome of interest in this paper, comparing the policies under $\Y_{\SO}(\alpha)$ and $\Y_{\OV}(\alpha)$ can help investigators explain the mechanism behind the policy under $\Y_{\OV}(\alpha)$; see Section \ref{sec:results}. Also, deriving the policy under $\Y_{\SO}(\alpha)$ allows us to highlight our methods' broad applicability to a wide range of interference-specific potential outcomes and we hope the derivation is useful for future researchers.

Finally, we use the following notations for sums, norms, functions, and limits. Let $\sum_{j}$, $\sum_{j,s}$, and $\sum_{j,a,s}$ be shorthands for  $\sum_{j=1}^{\NI_i}$, $ \sum_{j=1}^{\NI_i}
\sum_{s=0}^{\NI_i-1}$, and $ \sum_{j=1}^{\NI_i}
\sum_{a=0}^{1}
\sum_{s=0}^{\NI_i-1}$, respectively. For any $q \geq 1$, let $\| \bm{v} \|_q$ be the  $q$-norm of a vector $\bm{v}$ and let $\| f \|_{P,q} = \big\{ \int \| f(\bo_i) \|_2^q \, dP(\bo_i) \big\}^{1/q}$ be the $L_q(P)$-norm of a function $f$. For a set $\mathcal{D} \subset \{1,\ldots,\NC\}$, let $\mathcal{D}^c$ be its complement.  Let $O(\cdot)$, $O_P(\cdot)$, $o(\cdot)$, and $o_P(\cdot)$ be the usual big-$O$ and small-$O$ notations.

\subsection{Assumptions} 	\label{sec:Assumption2}
For causal identification, we make the following assumptions common in the partial interference literature \citep{Sobel2006, TTV2012, PH2014, Liu2014, Liu2016, Liu2019, Barkley2020, Kilpatrick2021, Park2022, Qu2022}.
\begin{itemize}[noitemsep,topsep=2pt]
\item[\hypertarget{(A1)}{(A1)}] \textit{Consistency}: $\bY_i =  \pot{ \bY_{i} }{\bA_i} $ almost surely.
\item[\hypertarget{(A2)}{(A2)}] \textit{Conditional Ignorability/Unconfoundedness}: $\pot{\bY_i}{\ba_i} \indep \bA_i \cond \bX_i$ for all $\ba_i \in \zoset(\NI_i)$.
\item[\hypertarget{(A3)}{(A3)}] \textit{Overlap}: For some $c> 0$, we have $c < \Pr \big( \bA_i = \ba_i \cond \bX_i  \big)$ for any $\ba_i \in \zoset(\NI_i)$ almost surely.
\end{itemize}
\noindent
Assumptions \hyperlink{(A1)}{(A1)}-\hyperlink{(A3)}{(A3)} are natural extensions of consistency, conditional ignorability/unconfoundedness, and overlap to observational studies under partial interference; see \citet{HR2020} and \citet{TTV2012} for more discussions. Under Assumptions \hyperlink{(A1)}{(A1)}-\hyperlink{(A3)}{(A3)}, we can identify the potential outcomes in \eqref{eq:overall_pot} and \eqref{eq:spilloverall_pot} from the observed data as $\Y_{\OV}(\alpha)  = \EXP \big\{  \NI_i^{-1}
\sum_{\ba_i \in \zoset(\NI_i)}
\EXP \big(  Y_{ij} \cond \bA_i = \ba_i , \bX_i	\big)
\pi(\ba_i \con \alpha) \big\}$ and \\ $\Y_{\SO}(\alpha)
=
\EXP \big\{  \NI_i^{-1}
\sum_{\ba_\eij \in \zoset(\NI_i-1)}
\EXP \big(  Y_{ij} \cond A_{ij} = 0,  \bA_\eij = \ba_{\eij} , \bX_i	\big)
\pi(\ba_{\eij} \con \alpha) \big\}$. We refer to the conditional expectation $\EXP \big(  Y_{ij} \cond \bA_i, \bX_i	\big)$ as the outcome regression, despite using nonparametric methods to estimate it; see Section \ref{sec:ObservationalStudy}.

Speaking of estimation, while Assumptions \hyperlink{(A1)}{(A1)}-\hyperlink{(A3)}{(A3)} are sufficient for causal identification, estimation, specifically the outcome regression $\EXP \big(  Y_{ij} \cond \bA_i, \bX_i	\big)$, can be challenging due to the curse of dimensionality. To this end, we make two additional assumptions. The first assumption is a type of ``exposure mapping'' in \citet{Aronow2017} where the outcome of a household in block $i$ is a function of $A_{ij}$, $\bX_{i}$, and $\mA_\eij$, a lower-dimensional summary of the treatment vector $\bA_{\eij}$:
\begin{itemize}[noitemsep,topsep=2pt]
\item[\hypertarget{(A4)}{(A4)}] \textit{Conditional Stratified Interference}: There exists a function $\mu^*$ that satisfies $\EXP\big( Y_{ij}  \cond \bA_i, \bX_i  \big) = \mu^* \big( A_{ij}, \mA_\eij, \bX_{ij}, \bX_{\eij} \big)$ for any $\bA_i$ and $\bX_i$.
\end{itemize}	
Assumption \hyperlink{(A4)}{(A4)} is closely related to the stratified interference assumption of \citet{HH2008} where after fixing the study unit's treatment $A_{ij}$ and the proportion of their peers' treatment $\mA_\eij$, the outcome of the study unit $Y_{ij}$ is invariant to who actually received the treatment in the block. Assumption  \hyperlink{(A4)}{(A4)} still allows for nonparametric models of the outcome. However, if the study unit's outcome depends on the treatment status of a few, focal peers, Assumption \hyperlink{(A4)}{(A4)} can be violated and the proposed estimators of the outcome regression in subsequent sections may be mis-specified for some units. The downstream effect of this mis-specification on the final, estimated policy will be most noticeable for the policy estimated using the indirect approach in Section \ref{sec-Indirect}. For the policy estimated using the direct approach in Section \ref{sec:Risk}, some of this downstream effect may be mitigated if the propensity score is correctly specified. In practice, assessing Assumption \hyperlink{(A4)}{(A4)} requires knowledge about the connections between study units, which many real-world datasets (including ours) do not have, and in the absence of such knowledge, Assumption \hyperlink{(A4)}{(A4)} can be considered as a first-order approximation of $\EXP \big( Y_{ij} \cond \bA_i, \bX_i \big)$. Theoretically, variants of Assumption  \hyperlink{(A4)}{(A4)} have been used in prior works on partial interference to achieve consistency, non-degenerate limiting distributions, and/or consistent variance estimation  \citep{Liu2014, vvLaan2014, Sofrygin2016, Forastiere2020}, including the recent work on optimal treatment regimes under interference \citep{Viviano2021}. 

Lastly, we make the following monotonicity assumption:
\begin{itemize}[noitemsep,topsep=2pt]
\item[\hypertarget{(A5)}{(A5)}] \textit{Monotonic Response}: $ \sum_{j} \mu^*(a, a', \bX_{ij}, \bX_{\eij})/\NI_i$ is non-decreasing in $(a,a')$ for any $\bX_i$.\hspace*{-2cm}
\end{itemize}
Assumption \hyperlink{(A5)}{(A5)} implies that the average of the expected outcomes of all study units in a block is a non-decreasing function of ego's and peers' treatment status. A sufficient condition for Assumption \hyperlink{(A5)}{(A5)} is that $\mu^*(a,a',\bX_{ij}, \bX_{\eij})$ itself (and not its average) is non-decreasing in $(a,a')$. In practice, Assumption \hyperlink{(A5)}{(A5)} is plausible if the treatment is harmless, as is the case for WASH facilities. But, the assumption may be violated if the treatment had a negative impact on the outcome of some study units. Also, Assumption \hyperlink{(A5)}{(A5)} can be assessed by computing the empirical derivative of the estimated nonparametric regression model for $\mu^*$ with respect to $(a,a')$; see Section \ref{sec:BOAssessment} of the Supplementary Materials for details. Of note, the direct approach relies on Assumption \hyperlink{(A5)}{(A5)} to make the {\OURR} as the global minima of the proposed risk function; see Lemma \ref{lem-minimizer2} below for details. In contrast, the indirect approach remains applicable for estimating {\OURR}, even when Assumption \hyperlink{(A5)}{(A5)} may seem implausible, as it does not employ the risk function.

\subsection{Identification of the {\OURR}}						\label{sec:2.3}

Let $\F = \big\{ \LB \cond \LB(\bx_i) \in [0,1] \}$ be a collection of functions from the support of $\bX_i$ to $[0,1]$. 
A function $\LB \in \F$ is one possible policy that outputs the proportion of study units in block $i$ that should be treated given both block- and unit-level characteristics $\bx_i$. Let $\Y_{\OV}(\LB(\bx_i) )$ be the expected overall outcome under $\LB \in \F$ given $\bx_i$, i.e., $\Y_{\OV}( \theta(\bx_i))  = \NI_i^{-1}
\sum_{\ba_i \in \zoset(\NI_i)}
\EXP \big(  Y_{ij} \cond \bA_i = \ba_i , \bX_i = \bx_i \big)
\pi(\ba_i \con \theta(\bx_i) )$, and let $\thr \in [0,1]$ be the smallest value of the outcome that is deemed acceptable by the investigator. % In the context of the Senegal DHS, the target can be set as the proportion of diarrhea-free children should be at least 70\%. 
Given $\thr$, $\bx_i$, and Assumptions \hyperlink{(A1)}{(A1)}-\hyperlink{(A4)}{(A4)}, the {\OURR} associated with the expected overall outcome, denoted as $\LB_{\OV}^*$ where {\OURR} in the subscript is suppressed for notational simplicity, is identified as follows: 
\begin{align} \nonumber
\hspace*{-0.3cm}
\LB_{\OV}^* (\bx_i) 
\hspace*{-0.05cm}
&= \hspace*{-0.05cm} \argmin_{\theta \in \F} \theta (\bx_i)  \ \text{ subject to }  \ \Y_{\OV} (\theta(\bx_i)) \geq \thr \\
& = 
\hspace*{-0.1cm} \inf_{\alpha \in [0,1]}
\hspace*{-0.1cm}
\bigg\{ \alpha
\, \bigg| \, 
\frac{1}{\NI_i} \sum_{j,a,s}
{\NI_i - 1 \choose s}
\mu^* \Big( a, \frac{s}{\NI_i-1}, \bx_{ij}, \bx_{\eij} \Big)
\alpha^{a+s} (1-\alpha)^{\NI_i - a - s}
\geq \thr
\bigg\} .
\hspace*{-0.1cm}
\label{eq-def:lowerB}
\end{align}
%In words, $\LB_{\OV}^*$ is the minimum proportion of treated study units in block $i$ that achieves the target outcome level $\thr$. 
In the WASH example, $\LB_{\OV}^*(\bx_i)$ is the minimum proportion of households with WASH facilities needed in a census block with characteristic $\bx_i$ in order to meet or exceed the diarrhea-free incidence level $\thr$.  If the set in \eqref{eq-def:lowerB} is empty,  we define $\LB_{\OV}^* (\bx_i)=1$ to maximize the expected average outcome. 

Similarly, the {\OURR} associated with the spillover outcome $\Y_{\SO}(\alpha)$ is identified as follows:
\begin{align}
\label{eq-def:lowerB SO}
    \LB_{\SO}^*(\bx_i)
\hspace*{-0.1cm}
= 
\hspace*{-0.1cm}
\inf_{\alpha \in [0,1]}
\bigg\{ \alpha
\, \bigg| \,
\frac{1}{\NI_i} \sum_{j,s}
{\NI_i-1 \choose s}
\mu^* \bigg( 0, \frac{s}{\NI_i-1}, \bx_{ij}, \bx_{\eij} \bigg)
\alpha^{s} (1-\alpha)^{\NI_i - 1 - s}
\geq \thr
\bigg\} \ , 
\end{align}
where, again, {\OURR} in the subscript is suppressed. In words, $\LB_{\SO}^*(\bx_i)$ is the minimum proportion of treated peers in a block with characteristic $\bx_i$ that is needed in order to achieve the target $\thr$ when the ego is not treated. In the WASH example, $\LB_{\SO}^*(\bx_i)$ is the smallest proportion of neighboring households with WASH facilities needed in a census block with characteristic $\bx_i$ in order to achieve the target diarrhea-free incidence level $\thr$ when the ego's household does not have WASH facilities. Note that under the monotonicity assumption \hyperlink{(A5)}{(A5)}, $\LB_{\OV}^*(\bx_i)$ cannot be smaller than $\LB_{\SO}^*(\bx_i)$.

\section{Estimation}		\label{sec:Method}

\subsection{An Indirect Approach Via Outcome Modeling} \label{sec-Indirect}
We present a naive estimator of the {\OURR} based on indirectly using the outcome regression model, sometimes referred to as the ``indirect'' approach \citep{Moodie2013}; see \citet{Deliu2022} for a recent review. Formally, let $\widehat{\mu}$ be an estimator of $\mu^*$; see Section \ref{sec:ObservationalStudy} on estimating $\mu^*$ under partial interference. Then, under Assumptions \hyperlink{(A1)}{(A1)}-\hyperlink{(A4)}{(A4)}, an estimator of $\LB^*$ for a given $\bx_i$, denoted as $\widehat{\LB}_{\indirect}(\bx_i)$, is obtained by replacing $\mu$ with $\widehat{\mu}$ in \eqref{eq-def:lowerB} and \eqref{eq-def:lowerB SO} and doing a grid search on the unit interval:
\begin{align}										\label{eq-IndirectRule}
\hspace*{-0.4cm}
\widehat{\LB}_{\OV, \indirect} ( \bx_i )
& =
\hspace*{-0.2cm}
\inf_{\alpha \in [0,1]}
\bigg\{ \alpha
\, \bigg| \, 
\frac{1}{\NI_i} \sum_{j,a,s}
{\NI_i - 1 \choose s}
\widehat{\mu} \bigg( a, \frac{s}{\NI_i-1}, \bx_{ij}, \bx_{\eij} \bigg)
\alpha^{a+s} (1-\alpha)^{\NI_i - a - s}
\geq \thr
\bigg\} \ , \nonumber \\
\hspace*{-0.4cm}		
\widehat{\LB}_{\SO,\indirect} ( \bx_i )
& =
\hspace*{-0.2cm}
\inf_{\alpha \in [0,1]}
\bigg\{ \alpha
\, \bigg| \, 
\frac{1}{\NI_i} \sum_{j,s}
{\NI_i - 1 \choose s}
\widehat{\mu} \bigg( 0, \frac{s}{\NI_i-1}, \bx_{ij}, \bx_{\eij} \bigg)
\alpha^{s} (1-\alpha)^{\NI_i - 1 - s}
\geq \thr
\bigg\} .
\end{align}
We again remark that Assumption \hyperlink{(A5)}{(A5)} is not necessary for the indirect approach. However, one notable limitation of the indirect approach is that it relies on the correct specification of the outcome model $\mu^*$. Also, similar to \citet{Zhao2012}'s observation about the indirect approach under no interference, our indirect approach may have poor finite-sample properties compared to an approach that directly estimates $\LB^*$. In Sections \ref{sec:Simulation} and \ref{sec:SenegalData}, we reconfirm this observation under partial interference and, consequently, we recommend the direct approach for estimating the {\OURR}.

\subsection{A Direct Approach Via Risk Minimization} \label{sec:Risk}
In this section, we propose our preferred estimation approach based on directly estimating the {\OURR}. We do this by recasting $\LB_{\OV}^*$ and $\LB_{\SO}^*$ as solutions to risk minimization problems with specialized loss functions $L_{\OV}(\LB,\bO_i)$ and $L_{\SO} (\LB, \bO_i)$, respectively. Notably, by reframing the original problem as a risk minimization problem, we can leverage a wealth of methods in empirical risk minimization to directly obtain estimators of $\LB_{\OV}^*$ and $\LB_{\SO}^*$.  

\begin{figure}[!htb]
\centering
\includegraphics[width=0.45\textwidth]{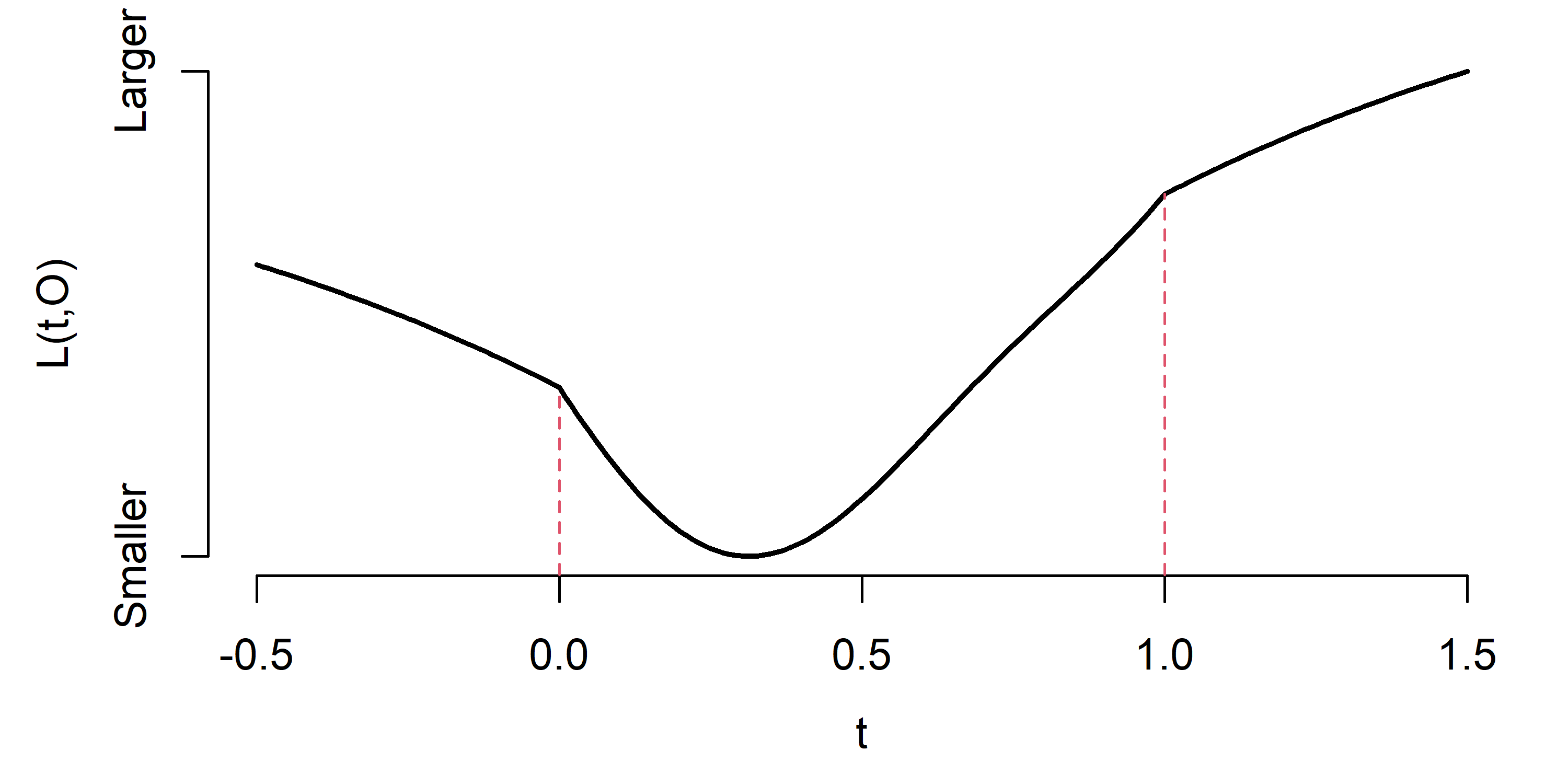}
\vspace*{-0.25cm}
\caption{\footnotesize A Graphical Illustration of the Loss Function $L(t,\bO_i)$}
\label{Fig-Loss}
\vspace*{-0.25cm}
\end{figure}

Formally, for $a=0,1$ and $s=0,1,\ldots,\NI_i-1$, let $e^* ( a, s \cond \bX_i ) = \Pr \big( A_{ij} = a, S_\eij = s \cond \bX_i)$ be the propensity score associated with the ego's treatment status (i.e., $A_{ij}$) and the number of treated peers (i.e., $S_{\eij}$). For a given $t \in \R$ and data $\bO_i$, consider the following loss function:
{\small
\begin{subequations}								\label{eq-L}
\begin{equation}
\loss_{z} (t, \bO_i)
=
\left\{
\begin{array}{lll}
\losszo_{z} (0,\bO_i) + \delta - \delta e^t & \text{ if } -\infty < t < 0		\\
\losszo_{z} (t, \bO_i) & \text{ if } 0 \leq t \leq 1  
&
\ , \
z \in \{ \OV, \SO \}
\\
\losszo_{z} (1,\bO_i) + \delta - \delta e^{-t + 1} & \text{ if } 1 < t < \infty
\end{array}
\right.
\label{eq-L01}
\end{equation}
\vspace*{-1.3cm}
\begin{align}
&
\losszo_z (t, \bO_i)
\label{eq-L02}
\\
&
=
\left\{
\begin{array}{ll}
\\[-1cm]
\frac{1}{\NI_i}
\sum_{j,a,s} {\NI_i - 1 \choose s} \psi_\text{DR} (a, s, \bO_{ij}, \bO_\eij) 
\sum_{\ell=0}^{\NI_i - a - s} {\NI_i - a - s \choose \ell} \frac{ (-1)^{\ell}  t^{\ell + a + s + 1} }{\ell + a + s + 1} - \thr t + C_0
&
\text{if $z=\OV$}
\\
\frac{1}{\NI_i}
\sum_{j,s} {\NI_i - 1 \choose s} \psi_\text{DR} (0, s, \bO_{ij}, \bO_\eij) \sum_{\ell=0}^{\NI_i - 1 - s} {\NI_i - 1 - s \choose \ell} \frac{ (-1)^{\ell}  t^{\ell + s + 1} }{\ell + s + 1} - \thr t + C_0 
& 
\text{if $z=\SO$}
\end{array}
\right.
\nonumber	
\end{align}
\vspace*{-1.3cm}
\begin{align}
\label{eq-L03}
&
\psi_{\text{DR}} (a , s, \bO_{ij}, \bO_\eij )
\\
&
=
\frac{ \big\{ Y_{ij} - \mu^*(a, \frac{s}{\NI_i-1},\bX_{ij}, \bX_\eij) \big\} \ind \big\{ A_i = a , S_\eij = s \big\} }{\PS^*(a , s \cond \bX_i)} + \mu^* \bigg( a ,  \frac{s}{\NI_i-1}, \bX_{ij}, \bX_\eij \bigg).
\nonumber
\end{align}
\end{subequations}}%
Here, $\delta > 0$ is any positive constant, and $C_0$ is any large constant guaranteeing $\losszo(t, \bO_i) \geq 0$; see Figure \ref{Fig-Loss} for a graphical illustration. 

The loss function $L_z(t,\bO_i)$ is strictly decreasing over $t \in (-\infty,0)$ and increasing over $t \in (1,\infty)$. Critically, the loss function is equal to $\nu_z(t,\bO_i)$ in \eqref{eq-L02} when $t \in [0,1]$, and it consists of two key parts: the function $\psi_{\text{DR}}$ and the binomial coefficients with the linear term $\thr t$. The function  $\psi_{\text{DR}}$ replaces the outcome regression $\mu^*$ in \eqref{eq-def:lowerB} and is based on the influence function of the doubly robust estimator of the overall causal effect under partial interference \citep{Liu2019,Park2022,Qu2022}. If desired, other $\psi$ functions based on the influence functions for the inverse probability-weighted (IPW) estimator or the outcome regression estimator are possible. The term $\sum_{\ell=0}^{\NI_i - 1 - s} {\NI_i - 1 - s \choose \ell} { (-1)^{\ell} t^{\ell + s + 1} }/({\ell + s + 1})$ with $\thr t$ is the integral of the term $\alpha^{a+s} (1-\alpha)^{\NI_i - a - s}$ with threshold $\thr$ in equation \eqref{eq-def:lowerB}; see Section \ref{sec:otherloss} of the Supplementary Material for details. More importantly, the expected loss function (i.e., risk function) achieves its minimum when $t$ in the loss function $L_z(t,\bO_i)$ is equal to the {\OURR}. The following lemma formalizes the connection between the loss function and the {\OURR} as follows.
\begin{lemma}[Risk Minimization Property of the {\OURR}]		\label{lem-minimizer2}
Suppose Assumptions \hyperlink{(A1)}{(A1)}-\hyperlink{(A5)}{(A5)} hold. Then, for $z \in \{\OV,\SO\}$, the {\OURR} $\LB_z^*$ defined in \eqref{eq-def:lowerB} and \eqref{eq-def:lowerB SO} is the minimizer of the risk $\risk_z(\LB) :=\EXP \big\{ \loss_z \big( \LB(\bX_i), \bO_i \big) \big\}$, i.e., $\risk_z (\LB_z^*) \leq \risk_z (\LB)$ for all $\LB \in \F$.
\end{lemma}
\noindent Hereafter, we suppress the subscript $z$ for notational brevity, and the results in subsequent sections apply to the {\OURR}s under both potential outcomes in \eqref{eq:overall_pot} and \eqref{eq:spilloverall_pot}. 

Methodologically, Lemma \ref{lem-minimizer2} provides a direct approach to estimate $\LB^*$ by minimizing the risk function $\risk(\LB)$ and this connection to risk minimization allows investigators to use a variety of methods in empirical risk minimization with the proposed loss function $\loss$.  As a concrete example, consider a support vector machine (SVM) estimator of $\LB^*$ using a Gaussian kernel $\kernel(\tbx, \tbx') = \text{exp} \big\{ - \| \tbx - \tbx' \|_2^2 / \gamma_\NC^2 \big\}$ with bandwidth $\gamma_\NC$. Let $\RKHS_\kernel$ denote the corresponding reproducing kernel Hilbert space (RKHS). Of note, a Gaussian kernel implicitly implies that the domain has a unified dimension $d$ for all $i$; see Section \ref{sec:ObservationalStudy} for further discussions. If a Gaussian kernel is not appropriate, other kernel functions that are symmetric, continuous, and positive definite can be used. The representer theorem \citep{KW1970,SHS2001} states that the estimated policy can be written as $\widetilde{\LB} (\bx) 	=
\sum_{j=1}^\NC \widehat{\eta}_j \kernel(\tbx, \tbx_j) 	+	\widehat{b}$ where $\widehat{\eta}_j$ and $\widehat{b}$ solve
\begin{align}									\label{eq-SVM2}
\hspace*{-0.25cm}
\min_{\LB \in \RKHS_\kernel} 
\bigg\{
\frac{1}{\NC} \sum_{i=1}^\NC \loss \big( \LB(\bX_i) , \bO_i \big) + \frac{\lambda_\NC}{2}  \big\| \LB \big\|_{\RKHS_\kernel}^2
\bigg\}
\hspace*{-0.1cm}
=
\hspace*{-0.1cm}
\min_{\bm{\eta}, b}
\bigg\{
\frac{1}{\NC} \sum_{i=1}^\NC \loss \big(  \kvec_i\T \Beta + b , \bO_i \big) + \frac{\lambda_\NC}{2} \Beta\T \kmat \Beta \bigg\} .
\hspace*{-0.25cm}
\end{align}
Here, $\| \LB \|_{\RKHS_\kernel}$ is a seminorm of $\LB$ in $\RKHS_\kernel$, $\lambda_\NC$ is a regularization parameter, $\Beta = ( \eta_1,\ldots,\eta_\NC )\T \in \R^{\NC}$, $\kvec_i = \big[ \kernel(\tbx_i,\tbx_1) , \ldots, \kernel(\tbx_i,\tbx_\NC) \big] \T \in \R^{\NC}$ for $i=1,\ldots,\NC$, and $\kmat = \big[ \kvec_1,\ldots,\kvec_\NC \big] \in \R^{\NC \times \NC}$. Note that the optimization problem in \eqref{eq-SVM2} is nonconvex, but this particular nonconvex function can be efficiently solved; see Section \ref{sec:supp-SVMsolve} of the Supplementary Material for details. Finally, we winsorize the estimated policy $\widetilde{\LB}$ from \eqref{eq-SVM2} to satisfy the support restriction on $\alpha$, i.e., $
\widehat{\LB} 
= \winsor ( \widetilde{\LB} )$ where $\winsor(\widetilde{\LB} ) = \widetilde{\LB} \ind\{ \widetilde{\LB}  \in [0,1] \} + \ind ( \widetilde{\LB}  > 1 )$. Investigators can then use the winsorized policy $\widehat{\LB}$ as an estimator of the {\OURR}.

We conclude with some remarks about the direct approach. First, while there are theoretically optimal values of the tuning parameters $\gamma$ and $\lambda$ for SVMs (see Theorem \ref{thm-ER-3} below), in practice, cross-validation is used to choose them where one starts with a grid of values for $(\gamma, \lambda)$ and finds the value that minimizes the cross-validated empirical risk \citep{SVM, ESL}. Our empirical analysis also uses cross-validation to choose $(\gamma, \lambda)$ and Section \ref{sec:supp-cv} of the Supplementary Material lays out the computational details. In general, the tuning parameter $\gamma$ controls the bandwidth of the SVM kernel whereas the tuning parameter $\lambda$ regularizes the SVM coefficients $\Beta$ in \eqref{eq-SVM2}. Second, investigators can choose function spaces other than the RKHS. For instance, if an investigator prefers a simple, linear {\OURR}, the investigator can replace the RKHS with a space of linear functions, i.e., $\big\{ \bm{\beta}\T \bx \cond \bm{\beta} \in \R^{\text{dim}(\bx)} \big\}$, and minimize with respect to the new function space. While the estimated policy may be more interpretable under a linear space, it may have a higher excess risk bound than the policy estimated under RKHS if the true policy is not well-approximated by the space of linear functions; in fact, in this case, the excess risk bound will be bounded away from $0$ even though the number of blocks $N$ grows to infinity. Third, the difference of convex functions (DC) algorithm \citep{DCalgorithm} can be used to efficiently solve \eqref{eq-SVM2} and the computational details are in Section \ref{sec:supp-SVMsolve} of the Supplementary Materials. Fourth, unlike \citet{Chen2022}, which used a surrogate loss to obtain an approximation of the target loss function, our approach directly uses the target loss function in \eqref{eq-L02} and does not require a surrogate loss function. 

%In Theorem %\ref{thm-ER-1} and \ref{thm-ER-3}, the leading order of the excess risk bound is minimized by choosing $\gamma_\NC \propto \NC^{ - 1/(2\beta + d) }$ and $\lambda_\NC \propto \NC^{ - (\beta+d)/(2\beta + d) }$. Despite the theoretical guarantee, in practice, investigators often use data-driven methods to choose the tuning parameters, usually via cross-validation. For our problem, we consider a set of candidate values for $(\gamma_\ell,\lambda_\ell)$. Without loss of generality, let the estimation data fold be $\mathcal{D}_1=\mathcal{D}_2^c$ and, as a consequence, observations in $\mathcal{D}_2$ is used to evaluate the estimated loss function $\widehat{\loss}_{(-1)}(t,\bO_i)$ for $i \in \mathcal{D}_2$. We further split $\mathcal{D}_2$ into training and tuning sets based on the number of cross-validation folds. For each candidate parameter $(\gamma_\ell,\lambda_\ell)$, we estimate the {\OURR}\ $\widehat{\LB}_\text{train}^W$ by only using the training set and obtain the empirical risk using the tuning set. The optimal choice of $(\gamma, \lambda)$ is the minimizer of the average of the empirical risks across the tuning sets $\sum_{i \in \text{tuning}} \widehat{\loss}(\widehat{\LB}_\text{train}^W(\mbX_i), \bO_i)/\NC_\text{tuning}$. 

\subsection{Estimating the Nuisance Functions}						\label{sec:ObservationalStudy}

Under partial interference, some care must be exercised when estimating the two nuisance functions, the outcome regression and the propensity score. %unlike indirect methods under no interference, 
For example, for the outcome regression $\mu^*$, naively using some popular nonparametric (or parametric) regression methods where $Y_{ij}$ is the dependent variable and $(A_{ij}, \mA_\eij, \bX_{ij}, \bX_\eij)$ are the independent variables will often be infeasible because the dimension of $\bX_\eij$ varies for each study unit $j$ and block $i$. Thankfully, this issue is not new in modern machine learning and we briefly review some common solutions. Also, following a recent trend in causal inference using machine learning methods (e.g., \citet{Victor2018}), our theoretical results are agnostic to specific estimators of the nuisance functions so long as they are estimated consistently; see Section \ref{sec:theory}.

We start with estimating the outcome regression $\mu^*$. In data-rich environments where $\NC$ is large and $\mu^*$ is estimated using overparametrized neural networks, a common trick is to pad the vector $\bX_{\eij}$ with zeros so that all the independent variables have identical length across blocks \citep[Chapter 10]{DLbook2016}. Alternatively, in data-scarce environments where  $\NC$ is small to moderate or the investigator prefers simpler, nonparametric methods, it is common to modify the original $\bX_\eij$ to have a common dimension $d$ and then use ``classic'' nonparametric regression estimators, such as the Nadaraya-Watson kernel regression estimator \citep{Nadaraya1964, Watson1964}. Some common ways of modifying $\bX_\eij$ in the literature include (a) taking a random subsample of a fixed number of households from each block, (b) using covariates from the ``average'' or the  ``extreme'' households, such as 
$\bX_\eij = ( \bX_\eij^{(\max)}, \bX_\eij^{(\min)} )$ where $\bX_\eij^{(\max)}$ and $\bX_\eij^{(\min)}$ are the collection of maximum and minimum values, respectively, or (c) generating a $d$-dimensional, low-rank summary of $\bX_\eij$ such as the cluster-level average $\mbX_\eij = \sum_{\ell \neq j} \bX_{i\ell}/(\NI_i-1)$. Investigators can also use an ensemble of learners based on different modifications of $\bX_{i(-j)}$ via the super learner algorithm \citep{SL2007, Polley2010}; see Section \ref{sec:supp-ORest} of the Supplementary Material for details. Given the moderate sample size of our data, we use approach (c) to train $\mu^*$.

We remark that estimation of $\mu^*$ is closely tied to correctly modeling the pattern of partial interference within a block. So long as the investigator uses a sufficiently flexible estimator for $\mu^*$ that can correctly model a wide variety of interference patterns in a block, the risk of the estimated policy will converge to that of the true policy; see Theorem \ref{thm-ER-3}. In the off-chance that the investigator has a priori knowledge about the interference pattern, incorporating this knowledge into estimating $\mu^*$ can improve the finite-sample performance of the estimated policy.

For the propensity score $\PS^*(a, s \cond \bx_i)$, we suggest nonparametric and semiparametric approaches. The nonparametric approach is similar to fitting the outcome regression model %under additional assumption
where we use $A_{ij}$ as the dependent variable and $\bX_{ij}$ as the independent variable inside the super learner algorithm to estimate the propensity score $\Pr (A_{ij} = a_{ij} \cond \bX_{ij} )$. Then, we use the estimated propensity score to obtain an estimated $e^*(a , s \cond \bX_{ij}, \bX_{\eij})$ as follows:
\begin{align*}
\widehat{\PS} \big( a ,  s \cond \bX_{ij}, \bX_{\eij} \big)
=
\widehat{\Pr} (A_{ij} = a \cond \bX_{ij}  )
\bigg\{
\sum_{\ba_\eij } \ind \Big(
\sum_{\ell \neq j} a_{i\ell} = s
\Big) \prod_{\ell \neq j}^{\NI_i} \widehat{\Pr} \big( A_{i\ell} = a_{i\ell} \cond \bX_{i\ell} \big)
\bigg\}
\ .
\end{align*}
Implicitly, the estimator $\widehat{e}$ assumes that the propensity score $\Pr(\bA_i = \ba_i \cond \bX_i)$ can be decomposed into a product of individual propensity scores $\Pr (A_{ij} = a_{ij} \cond \bX_{ij})$. If this assumption is implausible or $\widehat{e}$ is numerically unstable due to the product term, an alternative approach is to use a semiparametric model. Specifically, we decompose the propensity score as  $\PS^*(a, s \cond \bX_{ij}, \bX_{\eij})
=
\Pr \big\{ A_{ij} = a \cond \mA_{\eij} = s/(\NI_i-1) , \bX_{ij}, \bX_{\eij} \big\}
\Pr \big\{ \mA_{\eij}=s/(\NI_i-1) \cond  \bX_{ij}, \bX_{\eij} \big\}$ and estimate the first conditional probability using the same nonparametric approach discussed for estimating the outcome regression. For the second conditional probability, we bin $\mA_\eij$ into $(\UNI+1)$ equi-spaced bins and consider the following ordinal regression model for $\mA_\eij$ parametrized by $( {\beta}_{0,0}, \ldots, {\beta}_{0,\UNI}, {\bm{\beta}}_{\text{ego}}, {\bm{\beta}}_{\text{peer}}  )$:
\begin{align*}
\Pr \bigg\{ \mA_\eij \leq \frac{t}{\UNI+1} \, \bigg| \, \bX_{ij} , \bX_{\eij} \bigg\}
=
\text{expit} \big(
\beta_{0,t} +
\bX_{ij}\T \bm{\beta}_{\text{ego}} + 
\tbX_{\eij} \T \bm{\beta}_{\text{peer}}
\big)
\ , \
t=0,\ldots,\UNI \ .
\end{align*}
Here, the intercept parameters satisfy $\beta_{0,t} \leq \beta_{0,t+1}$. We then estimate $( \widehat{\beta}_{0,0}, \ldots, \widehat{\beta}_{0,\UNI}, \widehat{\bm{\beta}}_{\text{ego}}, \widehat{\bm{\beta}}_{\text{peer}}  )$ via the likelihood principle. The final estimator of $\PS^*(a, s \cond \bX_{ij}, \bX_{\eij})$ is simply a plug-in estimator based on the estimated $\beta$, i.e., $\widehat{\PS} (a, s \cond \bX_{ij}, \bX_{\eij} ) = {\rm expit}(\widehat{\beta}_{0,t} + \tbX_{ij} \T \widehat{\bm{\beta}}_{\text{ego}} + \tbX_{\eij}\T \widehat{\bm{\beta}}_{\text{peer}}) -  {\rm expit}(\widehat{\beta}_{0,t'} + \tbX_{ij} \T \widehat{\bm{\beta}}_{\text{ego}} + \tbX_{\eij}\T \widehat{\bm{\beta}}_{\text{peer}})$. For our data example, we used the semiparametric approach as it was more numerically stable than the nonparametric approach, especially after taking measures to prevent overfitting; see the next paragraph. 
 
We also discuss two common techniques that we used to prevent overfitting the nuisance functions. First, we randomly remove observations so that the number of study units in blocks is roughly similar to each other. This adjustment is often referred to as undersampling in multilevel studies \citep[Chapter 5.2]{Fernandez2018} and the adjustment prevents a few large, outlying blocks from having a dominant effect on the estimated nuisance function. Second, for the two nuisance functions in the loss function in \eqref{eq-L}, we use cross-fitting by  \citet{Victor2018}  where we train the two nuisance functions and the optimal policy $\LB^*$ from two different subsamples of the data. Specifically, suppose we split the data into two folds $\mathcal{D}_1$ and $\mathcal{D}_2$ and let $\widehat{\mOR}_{(-\ell)}$, $\widehat{e}_{(-\ell)}$, $\widehat{\loss}_{(-\ell)}, \ell \in \{1,2\}$, be the estimated outcome regression, propensity score, and loss function, respectively, by using $\mathcal{D}_\ell^c$. We estimate the {\OURR}\ by solving the SVM in \eqref{eq-SVM2} where $\widehat{\loss}_{(-\ell)}$ is  evaluated in $\mathcal{D}_\ell$. Again, we take the winsorized policy $\widehat{\LB}_{(-\ell)}$ to make sure the optimal policy is bounded between $0$ and $1$, i.e., $	\widehat{\LB}_{(-\ell)} = \winsor \big( \widetilde{\LB}_{(-\ell)} \big)$. Investigators may use either $\widehat{\LB}_{(-1)}$, $\widehat{\LB}_{(-2)}$, or the winsorized average of the two policies, i.e., $\winsor\big( \{ \widehat{\LB}_{(-1)} + \widehat{\LB}_{(-2)} \} /2 \big) $, as the final estimator of the {\OURR}. Finally, we remark that it is common to repeat undersampling and cross-fitting multiple times and aggregate the final estimators to remove finite-sample effects from random sampling; see Section \ref{sec:supp-cf} of the Supplementary Material for details. 

\subsection{Theoretical Properties}										\label{sec:theory}
In this section, we study the performance of the estimated policy $\widehat{\LB}_{(-\ell)}$ from the previous section. To proceed, for $\ell=1,2$, let $\rho_{(-\ell),\PS,\NC} = \big\|	 \widehat{\PS}_{(-\ell)} \big(A_{ij} , S_{\eij} \cond \bX_{ij}, \bX_\eij \big) - \PS^* \big(A_{ij} , S_{\eij} \cond \bX_{ij}, \bX_\eij \big) \big\|_{P,2}$ and $\rho_{(-\ell),\mOR,\NC} = \big\|\widehat{\mOR}_{(-\ell)} \big(A_{ij} , S_{\eij} / ( \NI_i - 1 ) , \bX_{ij}, \bX_\eij \big) - \mOR^* \big(A_{ij} , S_{\eij} / ( \NI_i - 1 ) , \bX_{ij}, \bX_\eij \big) \big\|_{P,2}$ be the convergence rates in $L_2(P)$-norm of the estimated nuisance functions. We make the following assumptions about the estimated nuisance functions $(\widehat{\mOR}_{(-\ell)}, \widehat{\PS}_{(-\ell)})$ and the corresponding convergence rates.
\begin{itemize}[noitemsep,topsep=2pt]
\item[\hypertarget{(E1)}{(E1)}] \textit{Overlap of $\widehat{\PS}_{(-\ell)}$}: There exists a constant $c' > 0$ such that $c' < \widehat{\PS}_{(-\ell)} \big( a , s \cond \bX_{ij}, \bX_{\eij} \big)  $ for any $(a,s) \in \{0,1\} \otimes \{0,1,\ldots,\NI_i-1\}$ and $(\bX_{ij}, \bX_\eij)$. 
\item[\hypertarget{(E2)}{(E2)}] \textit{Bounded $\widehat{\mOR}_{(-\ell)}$}: $\widehat{\mOR}_{(-\ell)}(a, s/(\NI_i-1), \bX_{ij}, \bX_\eij)$ is bounded for any $(a,s) \in \{0,1\} \otimes \{0,1,\ldots,\NI_i-1\}$ and $(\bX_{ij}, \bX_\eij)$. 
\item[\hypertarget{(E3)}{(E3)}] \textit{Consistency of $\widehat{\PS}_{(-\ell)}$ and $\widehat{\mOR}_{(-\ell)}$}: Let $r_{e,\NC}$, $r_{\mu,\NC}$, and $\Delta_\NC$ be sequences of real numbers, which are $o(1)$ as $N \rightarrow \infty$. Then, for $\ell=1,2$, we have $\rho_{(-\ell),\PS,\NC} \leq r_{e,\NC}$ and $\rho_{(-\ell),\mu,\NC} \leq r_{\mu,\NC}$ with probability greater than $1-\Delta_\NC$.
\end{itemize}
\noindent Assumption \hyperlink{(E1)}{(E1)} implies that the estimated propensity score satisfies overlap and %overlap assumption, similar to the overlap assumption for true propensity score in Assumption \hyperlink{(A3)}{(A3)}. 
Assumption \hyperlink{(E2)}{(E2)} implies that the estimated outcome regression is uniformly bounded. For the WASH example, Assumption \hyperlink{(E2)}{(E2)} is satisfied because the outcome is binary. Assumption \hyperlink{(E3)}{(E3)} states that, with high probability, the estimated nuisance functions $\widehat{\PS}_{(-\ell)}$ and $\widehat{\mOR}_{(-\ell)}$ converge to the true nuisance functions in $L_2(P)$-norm sense with rates $O_P(r_{e,\NC})$ and $O_P(r_{\mu,\NC})$, respectively. We remark that similar conditions are introduced in previous works that use nonparametric methods to estimate the nuisance components; see, for example, \citet{Victor2018} under no interference with independent and identically distributed data and \citet{Sofrygin2016} and \citet{Park2022} under partial interference with independent, but not necessarily identically distributed data.
%In contrast, we only need convergence at some rate because our goal is characterize the estimator's excess risk, a common metric of performance in the optimal treatment regime literature \citep{Zhao2012, Chen2016, Dias2018}.  %Kennedy2016 %For example, $r_{\PS,\NC} r_{\mOR, \NC} = o(\NC^{-1/2})$ is required for consistency of the DR estimator for the average treatment effect, implying that at least one nuisance function must be estimated with a rate faster than $\NC^{-1/4}$-rate.
%In particular, Assumption \hyperlink{(E3)}{(E3)} holds even if both nuisance functions are estimated at rates slower than $\NC^{-1/4}$.

In Theorem \ref{thm-ER-3}, we characterize the excess risk of $\widehat{\LB}_{(-\ell)}$, a common metric of performance in this literature (e.g., \citet{Zhao2012, Dias2018, KitagawaTetenov2018}).
%Before we establish the excess risk of $\widehat{\LB}_{(-\ell)}^W$, we define the risk function and the {\OURR}\ associated with the estimated loss function. Let $\widehat{\risk}_{(-\ell)}(\LB) = \EXP \big\{ \widehat{\loss}_{(-\ell)} \big( \LB(\mbX_i), \bO_i \big) \cond \mathcal{D}_\ell^c \big\} $ be the estimated risk function where the expectation is taken with respect to $\bO_i$ while $\widehat{\loss}_{(-\ell)}$ is considered as a fixed function which is clarified by denoting $\mathcal{D}_\ell^c$ in the conditioning statement. Accordingly, let $\LB_{(-\ell)}^*$ be the approximated {\OURR}\ which is the minimizer of $\widehat{\risk}_{(-\ell)}(\LB)$, i.e., $\widehat{\risk}_{(-\ell)}(\LB_{(-\ell)}^*) \leq \widehat{\risk}_{(-\ell)}(\LB) $ for all $\LB \in \F^{[0,1]}$. Using $\LB_{(-\ell)}^*$ as the intermediate quantities, we can establish the excess risk of $\widehat{\LB}_{(-\ell)}^W$ under Assumption \ref{assumption-1}. Theorem \ref{thm-ER-3} formally states the result.
\begin{theorem}								\label{thm-ER-3}
Suppose that Assumptions \hyperlink{(A1)}{(A1)}-\hyperlink{(A5)}{(A5)} and \hyperlink{(E1)}{(E1)}-\hyperlink{(E3)}{(E3)} hold. In addition, suppose that $\LB^*$ belongs to a Besov space $\mathcal{B}_{1,\infty}^\beta (\R^d) = \{ \tLB \in L_{\infty}(\R^d) \cond \sup_{t>0} t^{-\beta} \{ \omega_{r, L_1(\R^d) } (\tLB,t) \}	< \infty	,  r > \beta \}$ where $\omega_r$ is the modulus of continuity of order $r$ and $\beta>0$ is the smoothness parameter. 
Then, for $\ell=1,2$, and any positive numbers $\epsilon, p, \tau$ satisfying $d/(d+\tau) < p < 1$, we have the following excess risk bound of $\widehat{\LB}_{(-\ell)}$ with probability greater than $1-3e^{-\tau}-\Delta_\NC$:
\begin{align}		\label{eq-thm-ERB}
& 
\risk \big( \widehat{\LB}_{(-\ell)} \big) - \risk \big( \LB^* \big)
\\
&
\leq
c_1 \lambda_\NC\gamma_\NC^{-d} + c_2 \gamma_\NC^\beta 
+ c_3 \Big\{ \gamma_\NC^{(1-p)(1+\epsilon)d} \lambda_\NC^p \NC \Big\}^{-\frac{1}{2-p}} 
+ c_4 \NC^{-1/2}\tau^{1/2} + c_5\NC^{-1} \tau 
+ c_6  r_{\PS,\NC}r_{\mOR,\NC} \ . 
\nonumber			
\end{align}
Here, $c_1,\ldots,c_6$ are constants that are independent of $\NC$. 
\end{theorem}
\noindent To better interpret the result, suppose the bandwidth parameter $\gamma_{\NC}$ and the regularization parameter $\lambda_{\NC}$ are chosen with rates $\gamma_\NC \propto \NC^{ - 1/(2\beta + d) }$ and $\lambda_\NC \propto \NC^{ - (\beta+d)/(2\beta + d) }$. Under these choices, the bound in \eqref{eq-thm-ERB} reduces to $\risk \big( \widehat{\LB}_{(-\ell)} \big) - \risk \big( \LB^* \big)		=
O_P \big( \NC^{- \beta/(2\beta+d) }  \big)
+
O_P \big( r_{\PS,\NC}r_{\mOR,\NC}  \big) $. 
The first term, $\NC^{-\beta/(2\beta+d)}$, is related to the convergence rate of the SVM, which depends on the smoothness of the true {\OURR}\ $\LB^*$. If $\LB^*$ is very smooth (i.e., $\beta \to \infty$ with fixed $d$), the first term approximates the rate $O_P(\NC^{-1/2})$ from \citet{Chen2022}. The second term, $r_{\PS,\NC}r_{\mOR,\NC}$, represents the convergence rate of the estimated nuisance functions and is a second-order bias of the nuisance function estimators. It is important to note that the absence of the first-order bias is a consequence of employing the doubly robust loss function. If both nuisance functions are estimated at $o_P(\NC^{-1/4})$ rates, the second term attains $o_P(N^{-1/2})$ rate. Finally, we remark that the excess risk bound in \citet{Chen2022} depends on a parameter associated with a surrogate loss function, which, as discussed above, is used to address the lack of smoothness in their loss function. However, in our framework, we do not employ any surrogate loss function, and as a result, our excess risk bound \eqref{eq-thm-ERB} does not depend on rates stemming from the difference between the surrogate loss function and the true loss function.

\section{Simulation}									\label{sec:Simulation}

We conduct a simulation study to assess the finite-sample performance of the indirect and direct approaches of estimating the MRTP. We have $\NC=1000$ blocks where the block size $\NI_i$ is randomly drawn from the set $\{3,\ldots,22\}$ and the probability of drawing an element from this set is proportional to the empirical distribution of the size of the blocks in the training data for our case study in Section \ref{sec:SenegalData}. For unit $j$ in block $i$, we have four pre-treatment covariates $\bX_{ij} = (W_{ij1}, W_{ij2}, W_{ij3}, C_{i})$. The unit-level covariates $(W_{ij1}, W_{ij2}, W_{ij3})$ are generated from a 3-dimensional multivariate normal distribution with $\EXP \big( W_{ijk} \big) = 0$, $\VAR \big( W_{ijk} \big) = 1$, and $\text{Cov} (W_{ijk},W_{ijk'}) = 0.2$. The block-level covariate $C_i$ is generated from a standard normal distribution and is independent of $(W_{ij1}, W_{ij2}, W_{ij3})$. The treatment $A_{ij}$ is generated from the following model:
$A_{ij} \cond \bX_{ij} \sim {\rm Ber} \big( {\rm expit} \big\{
0.25 \big( -2 + W_{ij1} + W_{ij2} + W_{ij3} + 10 C_i \big)
\big\}\big)$. Similar to the block size, the treatment model is designed to mimic the training data for our data application where $\overline{A}_{i}$ between the simulated data and the training data are nearly identical. Lastly, the outcome $Y_{ij}$ is generated from a model that satisfies Assumptions \hyperlink{(A1)}{(A1)}-\hyperlink{(A5)}{(A5)}; see Section \ref{sec:supp-Simulation} of the Supplementary Material for details.

Since the estimation procedures for $\LB_\OV^*$ and $\LB_\SO^*$ are nearly identical, %except for the difference in the loss functions $\loss_\OV$ and $\loss_\SO$ associated with each {\OURR}\, 
we focus on $\LB_\OV^*$. For the direct approach, we unify the dimension of $\bX_i$ with $\mbX_i = \sum_{j} \bX_{ij} /\NI_i$, use the SVM described above, run 10-fold cross-validation for the SVM parameters, and do median-adjustment from 5 cross-fitted estimators, where each cross-fitted estimator is also median-adjusted from 3 undersampled estimators. For the indirect approach, we use two methods to estimate $\mu$, a linear regression model with $(\mA_i, \mbX_i)$ as regressors regularized by a Lasso penalty \citep{Tibshirani1996} and random forests \citep{Breiman2001}; these methods are referred to as Lasso and RF, respectively. For the target $\thr$, we vary between $0.67$ and $0.73$, which matches the observed range in the application in Section \ref{sec:SenegalData}. We evaluate the performance of each method on a test set consisting of 200 blocks. We repeat this entire process 50 times and, therefore, obtain 10000 {\OURR} estimates for each method and $\thr$. 

\begin{figure}[!htb]
\centering
\includegraphics[width=0.85\textwidth]{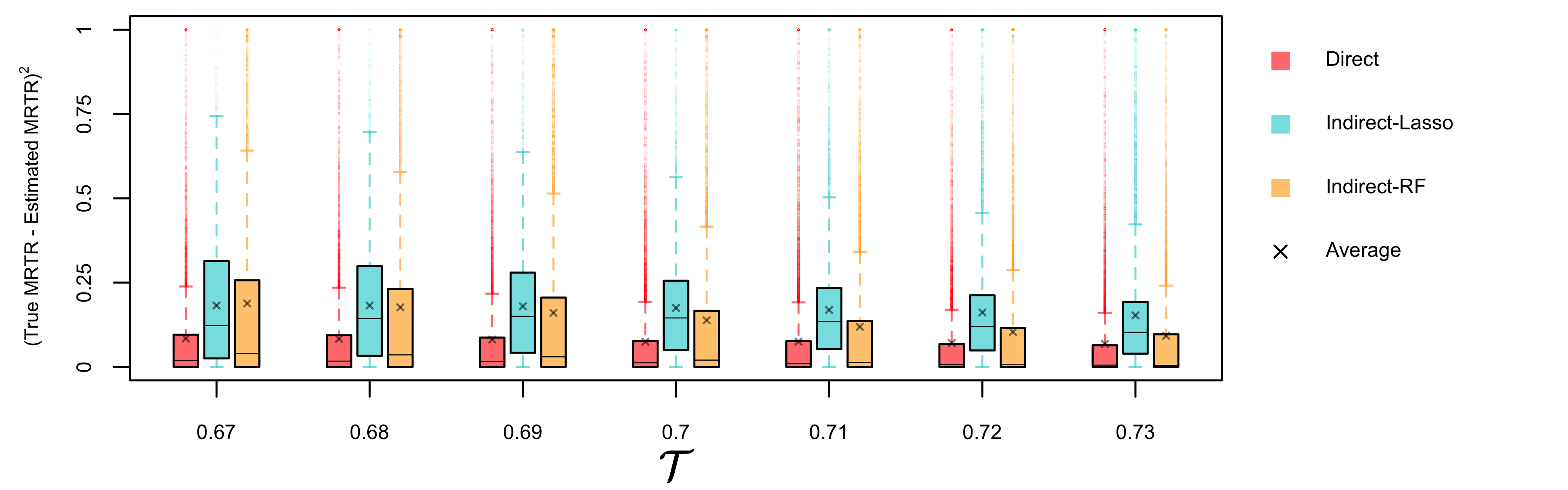}
\caption{\footnotesize Boxplots of the Squared Deviation Between the Predicted and True {\OURR}s. 
The  $x$ and $y$-axes show the threshold $\thr$ and the squared deviation between the predicted and true {\OURR}s, respectively. }	
\label{fig-1}
\vspace*{-0.3cm}
\end{figure}

Figure \ref{fig-1} shows the squared deviations between the estimated and the true {\OURR}s. The direct approach shows the best performance where its squared deviation between the predicted and the true {\OURR}s is generally the smallest across all threshold values $\thr$.  

\begin{figure}[!htb]
\centering
\includegraphics[width=0.85\textwidth]{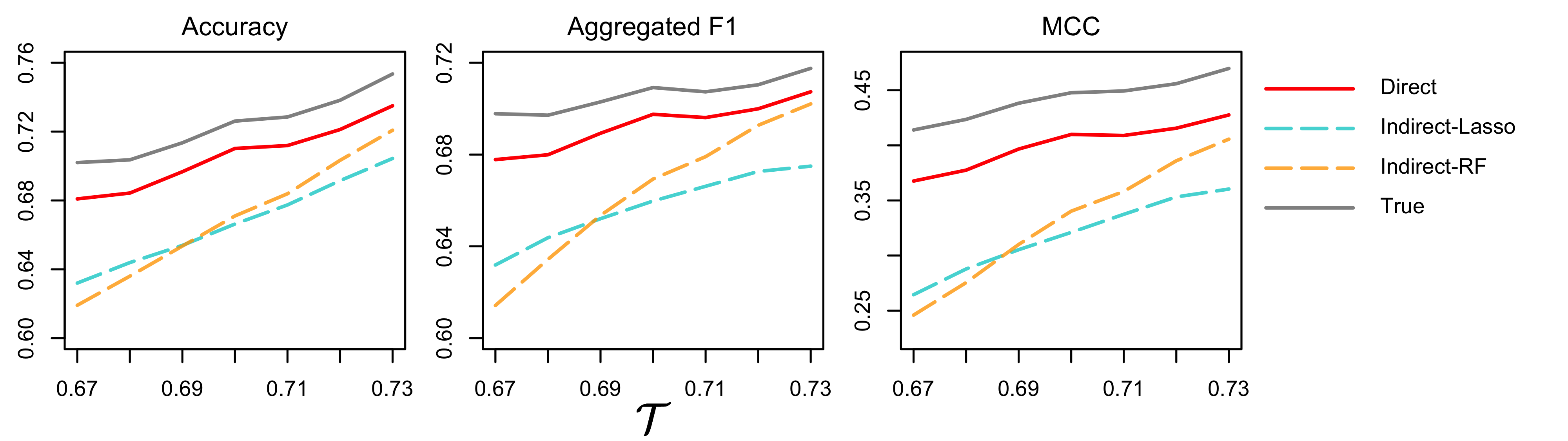}
\caption{\footnotesize Evaluation of {\OURR}s Using Different Measures of Classification Performance. The $x$-axis shows the threshold $\thr$. The $y$-axis shows the value of the measure that is specified at the top of each plot. A larger value in the $y$-axis implies better performance.}	
\label{fig-2}
\vspace*{-0.3cm}
\end{figure}

Next, we assess the methods using the following classification performance measures:  accuracy, two-sided F1 score, and the Matthews correlation coefficient (MCC) \citep{MCC}; see Section \ref{sec:supp-classification} of the Supplementary Material for details.  For every classification performance measure, a larger value means better performance. Figure \ref{fig-2} shows the result for each method. We also plot the true {\OURR}\ for comparison. The true {\OURR}\ performs the best across the different performance measures and all thresholds $\thr$, justifying that the measures are meaningful for evaluating the performance of the policies. Among the estimated policies,  
the direct approach generally performs well across most $\thr$ and the three performance measures. Combined with Figure \ref{fig-1}, we would generally recommend the direct approach in terms of bias and performance measures based on classification. 

\section{Application: WASH Policy in Senegal}											\label{sec:SenegalData}

\subsection{Background}							\label{sec:DataDesc}

We use the proposed methods to design a WASH policy for Senegal. Specifically, we use the 2014-2018 Senegal Demographic and Health Survey (DHS), which used a two-stage stratified survey design where households (i.e., the study units) are nested under census blocks. We restrict the sample to households with complete data on $\bO_{ij} = (Y_{ij}, A_{ij}, \bX_{ij})$. For household $j$ in block $i$, the outcome $Y_{ij}$ is binary where $Y_{ij}=1$ indicates that there is no child having diarrhea in the household and $Y_{ij} = 0$ indicates that there is at least one child having diarrhea in the household. The treatment $A_{ij}$ is binary where $A_{ij} =1$ indicates that the household has a WASH facility, defined as having a private water source or a private flush toilet, and $A_{ij} = 0$ indicates the opposite. The covariates $\bX_{ij}$ consist of the following nine household- and block-level characteristics: block size $(\NI_i)$, the indicator of whether census block $i$ is located in an urban area, number of household members, number of children in a household, the indicator of whether both parents in a household do not have jobs, whether parents ever attended schools, mother's age, and the average age of children. 

For training data, we use the 2014-2017 DHS, which contains $\NC = 1027$ census blocks with $\sum_{i} \NI_i = 13556$ households. For validation data, we use the 2018 DHS, which contains $213$ census blocks with $2859$ households. Also, for the validation data, we vary the diarrhea-free target threshold $\thr$ from $0.67$ to $0.73$, which is slightly above and below the average diarrhea-free incidence level of $0.689$ between 2014 and 2017. Then, for each $\thr$, we estimate the {\OURR}\ for the overall potential outcome. For both the direct and the indirect approaches, we use the same procedures as in the simulation study except that, for the direct approach, we use 100 cross-fitted estimators with median adjustment and 10 rounds of undersampling. In Section \ref{sec:BOAssessment} of the Supplementary Material, we discuss how the bounded cluster size assumption and Assumptions \hyperlink{(A1)}{(A1)}-\hyperlink{(A5)}{(A5)} are plausible in the Senegal DHS.

\subsection{Results} \label{sec:results} 

\subsubsection*{Finding 1: The direct approach is more accurate.}
\begin{figure}[!htb]
\centering
\includegraphics[width=0.85\textwidth]{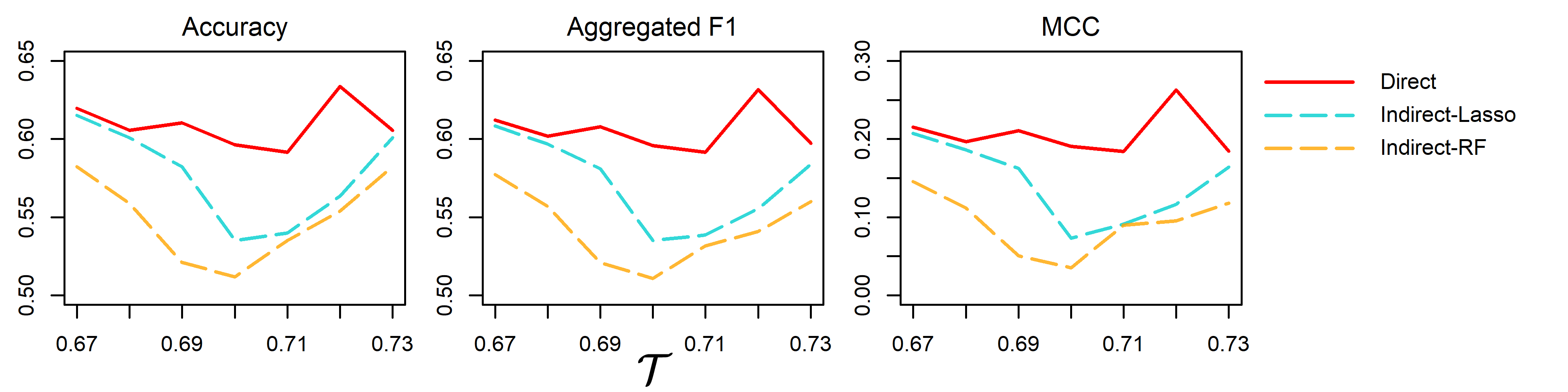}
\caption{\footnotesize Evaluation of the Estimated {\OURR}s in the 2018 Senegal DHS.	The $x$ and $y$-axes show the threshold $\thr$ and the magnitude of each classification performance measure, respectively. Higher $y$ values indicate better performance.}	
\label{fig-Data-1} 
\vspace*{-0.3cm}
\end{figure}

%For all of our empirical results below, we use the same estimators and classification performance measures used in the simulation study in Section \ref{sec:Simulation} except we cross-fit 100 times. 
Figure \ref{fig-Data-1} shows the classification performance measures. Across all performance measures, the direct approach performs better than the indirect approaches across all thresholds $\thr$. Combined with the results from the simulation study, the direct approach is more accurate (i.e., smaller deviance from the true {\OURR}) than the other two approaches.

\begin{figure}[!htb]
\centering
\includegraphics[width=0.85\textwidth]{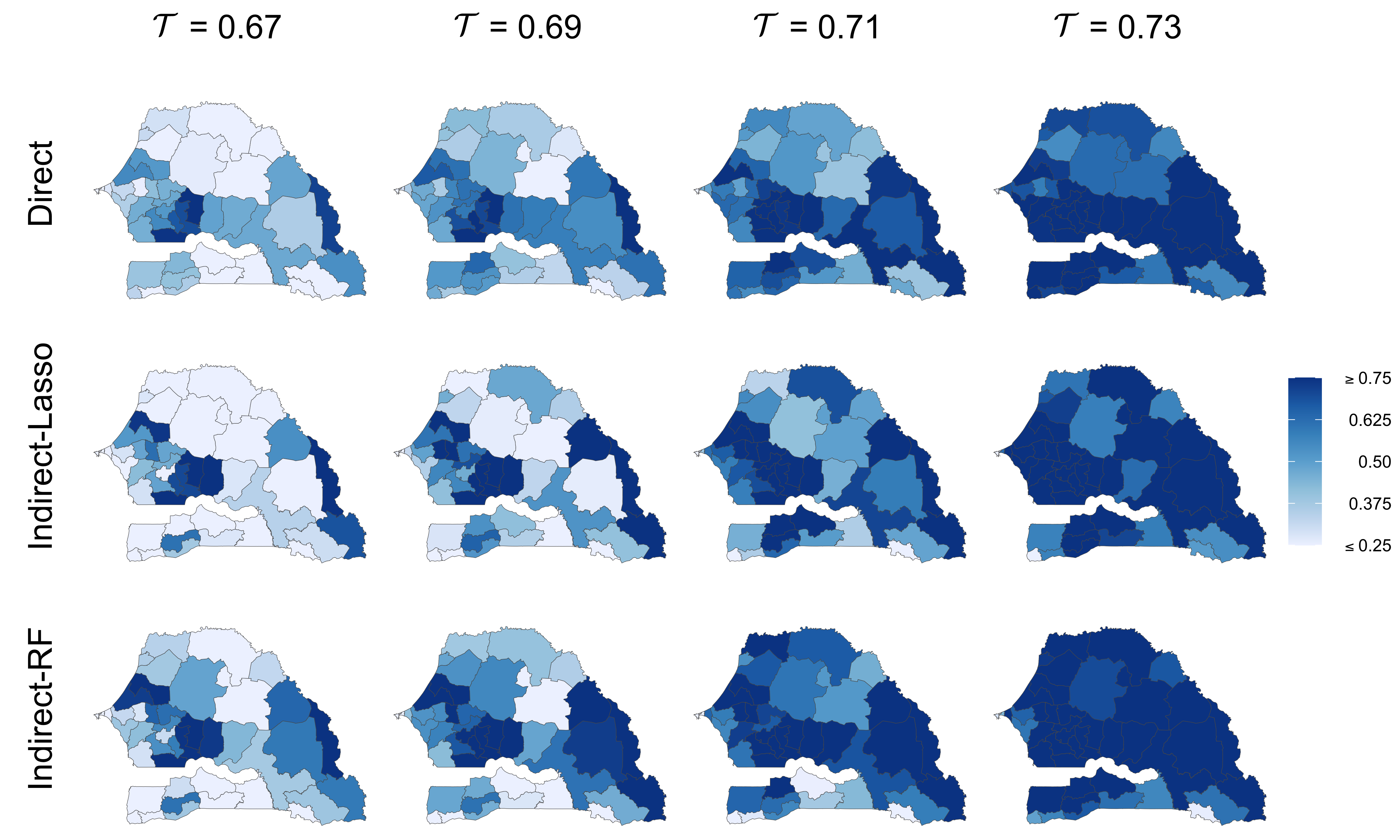}
\caption{\footnotesize Estimated {\OURR}s Across 45 Administrative Regions in the 2018 Senegal DHS. Each column corresponds to one of the four threshold values $\thr$. Each row corresponds to one of the allocation policies. Each heatmap shows the average of the estimated {\OURR}. Darker (lighter) blue color indicates that the estimated average {\OURR}\ in the administrative region is larger (smaller).}	
\label{fig-Data-2}
\vspace*{-0.3cm}
\end{figure}

Figure \ref{fig-Data-2} shows heatmaps of the estimated  {\OURR}s at four values of the thresholds. Specifically, the heatmap shows the weighted averages of the estimated {\OURR}s aggregated to 45 Senegalese administrative regions where the weight is based on the number of households in a census block (i.e., $\NI_i$); see Section \ref{sec:HeatMapDetail} of the Supplementary Material for more details behind the heatmaps. For each $\thr$, the estimated policy from the direct approach is generally more diffused than the indirect approach, implying that the direct approach generally prefers a more balanced, nationwide approach to allocating WASH facilities across Senegal compared to the indirect approach. Also, since the direct approach is more accurate than the indirect approach, the finding broadly suggests that an optimal WASH policy in Senegal is a diffuse approach where neighbors of an area with a high {\OURR}\ should also receive a relatively large amount of WASH facilities instead of a targeted, all-or-nothing approach where there are sharp discrepancies in the amount of WASH facilities allocated between two neighbors.

\subsubsection*{Finding 2: The direct approach is more resource-efficient.}
\begin{figure}[!htb]
\centering
\includegraphics[width=0.85\textwidth]{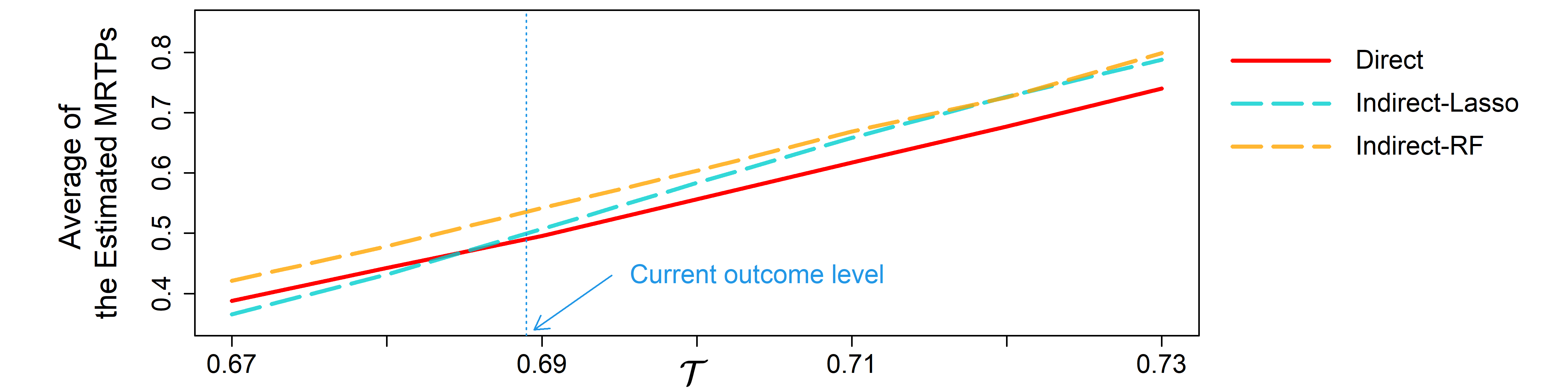}
\caption{\footnotesize Averages of Estimated {\OURR}s in 2018 Senegal DHS. The $x$ and $y$-axes show the threshold $\thr$ and the average of estimated {\OURR}s across census blocks. The dotted vertical line at $\thr=0.689$ shows the outcome level in the training data (i.e., between 2014 and 2017).}	
\label{fig-Data-3}
\vspace*{-0.3cm}
\end{figure}

Figure \ref{fig-Data-3} shows the weighted averages of estimated {\OURR}s needed to achieve the target outcome level $\thr$ where the weight is determined by the size of the census blocks. When $\thr$ is greater than 0.69, which is the outcome level in the training data, the direct approach uses fewer WASH facilities than the two indirect approaches. Combined with our first finding that the direct approach is more accurate, the direct approach is more resource-efficient than the indirect approaches in terms of allocating the right regions of Senegal to achieve higher accuracy. In other words, for thresholds greater than 0.69, the direct approach uses fewer resources and is more accurate than the other methods, suggesting that the indirect approaches are targeting the wrong regions of Senegal or using more resources than necessary to achieve the target $\thr$. Figure \ref{fig-Data-2} provides some evidence of this observation where the random forest is over-allocating WASH facilities across most of Senegal, even in areas that may not need additional WASH facilities to achieve $\thr$. In contrast, the direct approach generally allocates fewer resources throughout the nation while simultaneously achieving higher accuracy. Given that Senegalese policymakers would like to set $\thr$ higher than the current diarrhea-free incidence rate of 0.69, we would recommend the direct approach as it is not only accurate but also cost-efficient for allocating WASH facilities in Senegal.

\subsubsection*{Finding 3: Compared to current policy recommendations, the direct approach uses different characteristics to allocate WASH facilities.}

\begin{figure}[!htb]
\centering
\includegraphics[width=0.85\textwidth]{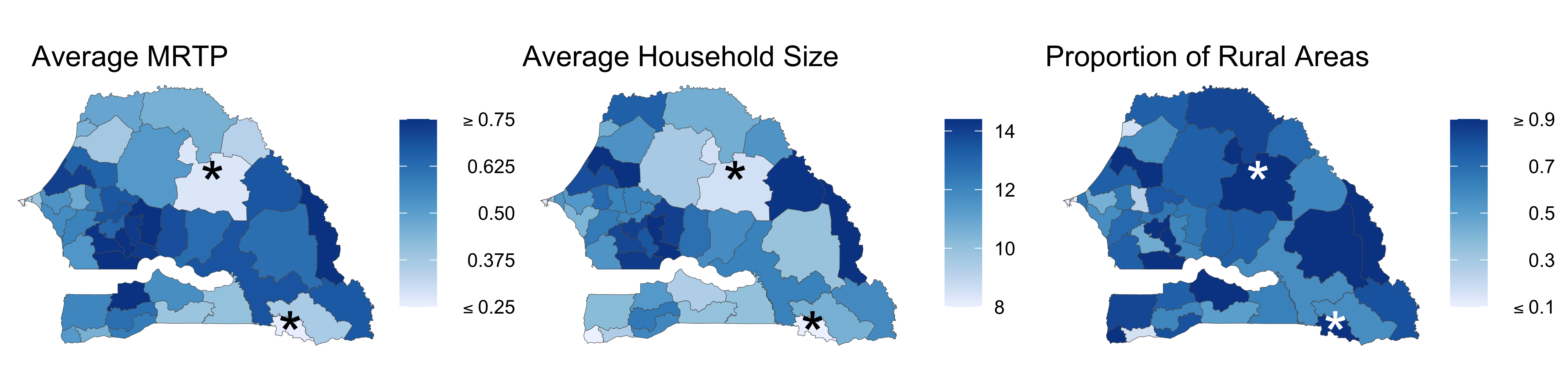}
\caption{\footnotesize Estimated {\OURR}s under $\thr=0.7$ (left), Household Size (middle), and Proportion of Rural Areas (right) in the 2018 Senegal DHS. Darker (lighter) blue color indicates a larger (smaller) value. The two regions with an asterisk $(*)$ are mostly rural, but are associated with low {\OURR}s and small household sizes.} 	
\label{fig-Data-4}
\vspace*{-0.3cm}
\end{figure}

In international development, one of the prevailing advice on the allocation of WASH facilities is to target rural areas \citep{RuralFirst1, RuralFirst2}. Specifically, these works suggest that nations use the proportion of rural areas to allocate WASH facilities, where a large proportion of rural areas would indicate that a large number of WASH facilities is needed. However, as depicted in Figure \ref{fig-Data-4}, the direct approach suggests that this should not be the main strategy for allocating WASH facilities. Instead, the allocation policy should focus on the average household size in a census block. For example, the regions in Figure \ref{fig-Data-4} marked with an asterisk $(*)$ are mostly rural areas but are associated with low {\OURR}\ estimates obtained from the direct approach. In contrast, the same regions have higher average household sizes and are associated with higher {\OURR}\ estimates. Overall, the {\OURR} estimate is more closely correlated with the average household size, where a census block with a large number of big households needs more WASH facilities. This positive correlation between the estimated {\OURR}\ and average household size also agrees with prior works in infectious diseases where the incidence of diarrhea is strongly correlated with household size  \citep{Diarrhea_HHsize1, Diarrhea_HHsize2, Diarrhea_HHsize3}. In short, if the target outcome is to reduce the incidence of diarrhea, the direct approach suggests that household size may be a better indicator to guide policy compared to the proportion of rural areas.

\subsubsection*{Finding 4: Comparing {\OURR}s under different outcomes can provide insights about the mechanism of the {\OURR} under the primary outcome.}

Using the methods developed in the paper, we can compare {\OURR}s from different potential outcomes by simply changing the loss functions. Importantly, these comparisons can reveal insights about the mechanism of the {\OURR} under the primary outcome. As an illustrative example, we compare the {\OURR} under the primary outcome that is discussed above (i.e., $\LB_{\OV}^*$) with the {\OURR}  under the spillover potential outcome (i.e., $\LB_\SO^*$).  

By Theorem 1 of  \citet{VWTT2011}, the difference $\LB_{D}^* = \LB_{\SO}^* - \LB_{\OV}^*$ serves as a proxy for the magnitude of the optimal policy's direct effect on the diarrhea incidence rate. A large $\LB_D^*$ would indicate that the optimal policy has a large direct effect on reducing diarrhea incidence whereas a small $\LB_D^*$ would indicate the opposite. Also, similar to Finding 3, by comparing $\LB_D^*$ as a function of an observed covariate, investigators can assess the heterogeneity of the optimal policy's direct effect.

\begin{figure}[!htb]
\centering
\includegraphics[width=0.85\textwidth]{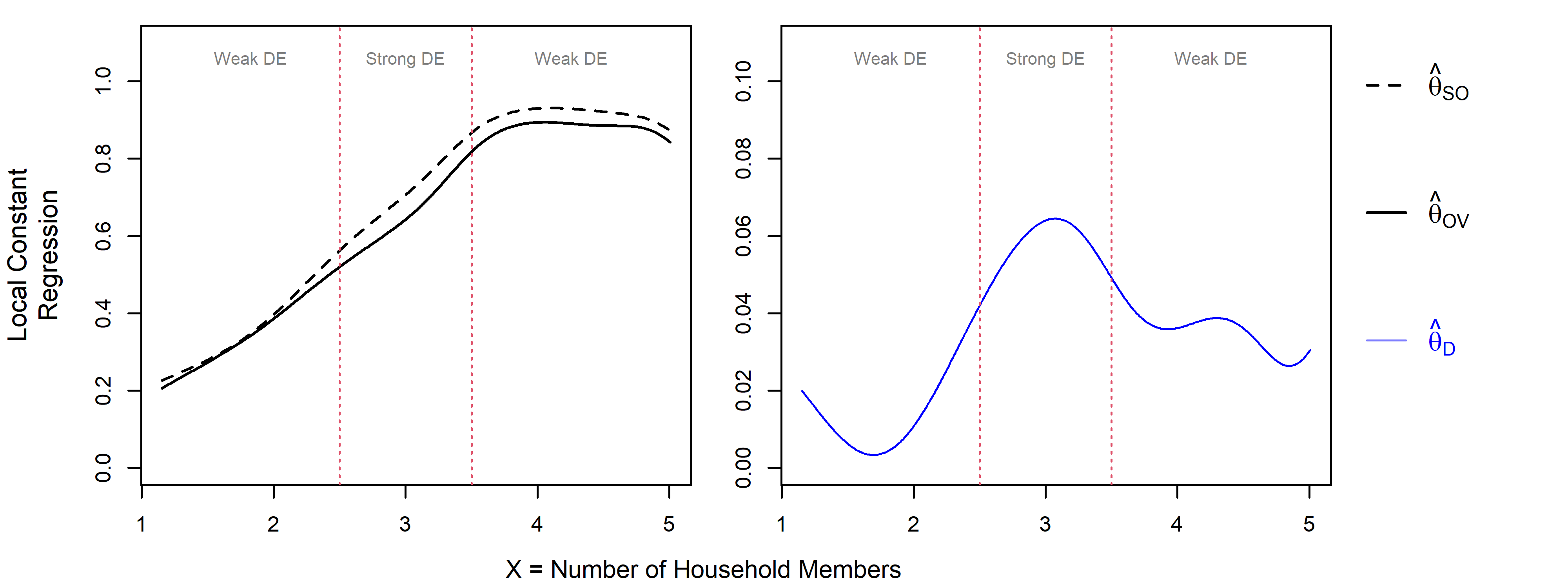}
\caption{\footnotesize Local Constant Regression Curves of the Estimated {\OURR}s $\widehat{\LB}_\SO$ (black solid line in the left panel), $\widehat{\LB}_\OV$ (black dashed line in the left panel), and the Difference of the Estimated {\OURR}s $\widehat{\LB}_D = \widehat{\LB}_\SO - \widehat{\LB}_\OV$ (blue solid line in the right panel). The $x$-axis is the average number of children per household. A larger $y$ value means that more WASH facilities are required in a census block to meet the threshold $\thr=0.7$. Note that the $y$-axes in the two plots are under two different scales for better visualization.} 	
\label{fig-Data-5}
\vspace*{-0.3cm}
\end{figure}

Figure \ref{fig-Data-5} shows one example of such heterogeneity with respect to the average number of children per household. The $y$-axis plots the estimated $\LB_D^*$ by plugging in estimates of $\LB_{\OV}^*$ and $\LB_{\SO}^*$ at $\thr=0.70$ and the $x$-axis plots the average number of children per household. The estimated $\LB_D^*$ shows a nonlinear relationship where $\LB_D^*$ is generally the largest when the average number of children per household is between 2.5 and 3.5. This suggests that the optimal policy's direct effect is the largest when the average number of children per household is of moderate size. We remark that similar nonlinear trends have been observed in prior works on international development, which examined the impact of household size on diarrhea incidence or a child's health outcome \citep{Svanes2002,Kandala2006, Kananura2022, Fenta2021}.

\section{Discussion}

In this paper, we study the identification and estimation of the {\OURR}, a policy that assigns the smallest level of treatment to achieve a target outcome level under partial interference. We propose two approaches for estimating {\OURR}, a simple indirect approach where a plug-in estimator of the outcome regression model is used and a direct approach where a novel, nonparametric, doubly robust loss function that incorporates partial interference is used inside of an empirical risk minimization problem. %We also show that a tempting, analysis strategy based on aggregated data at the cluster-level will be grossly misleading. 
We then use the proposed methods to estimate a WASH policy for Senegal by using the 2014-2018 Senegal DHS. We show that the direct approach is more accurate and resource-efficient than other methods. We also show a difference between our direct approach and what's currently recommended for allocating WASH facilities in international development.

We end by discussing a few limitations of our work. First, a limitation of the MRTP in \eqref{eq-Ours}, $\LB_{\OURR}^*$, is that it does not exactly tell which study units in a block should receive treatment. This is in contrast to the optimal allocation policy in \eqref{eq-OTA}, $\LB_{\OAP}^*$, that tells which study units should be treated. However, as we discussed in Section \ref{sec:intro}, $\LB_{\OURR}^*$ is still a valuable tool for practitioners as it provides an estimate of the minimum amount of treatment required to achieve the desired goal whereas $\LB_{\OAP}^*$ does not provide this information. It would be interesting future work to combine the strengths of both $\LB_{\OURR}^*$ and $\LB_{\OAP}^*$ to obtain a more comprehensive allocation policy for international development. 
Second, similar to any method in causal inference with observational data, our approach relies on some assumptions, including the usual identification Assumptions \hyperlink{(A1)}{(A1)}-\hyperlink{(A3)}{(A3)} and modeling Assumptions \hyperlink{(A4)}{(A4)}-\hyperlink{(A5)}{(A5)}. 
Policymakers should carefully evaluate the plausibility of these assumptions in the context of their research question.
For example, in Section \ref{sec:BOAssessment} of the Supplementary Material, we assess these assumptions in our data and found no significant issues.
If there is no strong evidence against these assumptions, the proposed policy can be a useful tool for policymakers.

%% file: Supp.tex
\appendix

\section*{Supplementary Material}

This document contains supplementary materials for ``Minimum Resource Threshold Policy Under Partial Interference.'' 
Section \ref{sec:A} presents additional results related to the main paper. 
Section \ref{sec:B} proves lemmas and theorems stated in the paper.

\section{Additional Details of the Main Paper}		\label{sec:A}

\subsection{Examples of Beneficial Intervention/Treatment for Most of the Population}		\label{sec:Beneficial Examples}

We provide plausible examples where the intervention/treatment seems beneficial for almost all members of the population.
\begin{itemize}

\item Prior works have shown that improving household access to improved water, sanitation, and hygiene (WASH) resources are critical to reduce rates of diarrhea-related diseases \citep{Diarrhea_New5, Diarrhea_New3}, especially among children \citep{Diarrhea_New6, McMichael2019}. Additionally, there is no biological rationale that well-managed WASH facilities cause diarrhea-related diseases. 

\item \citet{WaterPipe2012} studied the effect of getting easier access to piped water on various kinds of outcomes such as quality and quantity of water consumed, water-related time and financial costs incurred by the household. In particular, based on the results in Table 3 and related discussions, we can deduce that getting easier access to piped water results in a substantial increase in the quantity of water for most of the population. 

\item As discussed in page 10 of \citet{CohenDupas2010}, higher insecticide-treated bed nets (ITN) coverage rates would be beneficial for the population because the use of ITN in a household may have positive health externalities for neighboring households. 

\item \citet{COVID2022} conducted a meta-analysis of studying the effect of COVID-19 vaccines against SARS-CoV-2 infection. Their work and references therein suggest that COVID-19 vaccines are beneficial for most of the population. 

\item \citet{WORMS} studied the effect of school-level deworming projects on students' health status and academic achievements. They remarked that there were within- and across-school spillover effects, indicating that students who did not directly receive deworming treatment still benefited from those who did. Therefore, combined with the biological reasons, we can infer that deworming drugs are beneficial for most of a majority of  students.

\end{itemize}

\subsection{Examples of Real-world Applications Targeting a Certain Level of Outcomes}		\label{sec:Outcome Examples}

In this section, we present examples of real-world applications that target a specific level of outcomes rather than the maximum level of outcomes.
\begin{itemize}

\item High levels of protein in the urine, known as proteinuria, may be caused by diabetes, high blood pressure, autoimmune disorders, infections, and kidney diseases. Thus, for a normal adult, it is recommended to maintain the total urinary protein excretion less than 150 mg/day \citep{Carroll2000}.

\item HDL cholesterol is commonly referred to as ``good'' cholesterol because it plays a crucial role in removing harmful cholesterol from the bloodstream. As a result, maintaining high levels of HDL cholesterol is associated with a lower risk of cardiovascular disease. It is generally recommended to maintain HDL cholesterol levels above 60 mg/dL \citep{HDLC2002}.

\item Warfarin, a blood-thinning medication, is used to increase the international normalized ratio (INR), a measure of the time for the blood to clot, and it should be prescribed to keep patients' INR within the desired range, usually between 2 and 3,  according to the recommendations from American Heart Association \citep{Warfarin_AHA_2014}. 

\item Major medical associations recommend targeting proper ranges for chronic disease management measures such as hemoglobin level (male: 138-172 g/L; female: 121-151 g/L) \citep{TDI_Example_2019}. 

\item UN has established the Sustainable Development Goals in 2015 \citep{SDG1}, which consists of 17 specific goals. In particular, Goal 1 is to eradicate extreme poverty for all people everywhere by 2030 and Goal 3 is to ensure healthy lives and promoting well-being at all ages. Some specifics of these goals target certain levels of outcomes of interest as follows:
\begin{itemize}
\item 1.1: By 2030, eradicate extreme poverty for all people everywhere, currently measured as people living on less than \$1.25 a day
\item 1.2: By 2030, reduce at least by half the proportion of men, women and children of all ages living in poverty in all its dimensions according to national definitions

\item 3.1: By 2030, reduce the global maternal mortality ratio to less than 70 per 100,000 live births
\item 3.2: By 2030, end preventable deaths of newborns and children under 5 years of age, with 
all countries aiming to reduce neonatal mortality to at least as low as 12 per 1,000 live births 
and under-5 mortality to at least as low as 25 per 1,000 live births
\end{itemize}

\item The Global Technical Strategy for malaria 2016-2030 was adopted by the World Health Assembly in May 2015 \citep{Malaria2021}. It has set a target of reducing malaria incidence by 40, 75, and 90 percent by 2020, 2025, and 2030, respectively, compared with malaria incidence in 2015.  

\end{itemize}

\subsection{Application of Our Method to Other Real-World Examples}		\label{sec:supp:Extension}

Motivated from the examples in Sections \ref{sec:Beneficial Examples} and \ref{sec:Outcome Examples}, we lay out some concrete examples where our approach can be used.

\begin{itemize}

\item Motivated by \citet{CohenDupas2010} and \citet{Malaria2021}, we can study the minimum ITN coverage necessary to meet the thresholds set by the Global Technical Strategy for malaria control %2016-2030 by reducing malaria incidence below a village-specific threshold. 
Also, because an ITN is likely to reduce malaria incidence in both the household where it is installed and nearby (but not too far away) households, partial interference is a viable framework for modeling the effect of ITN installation on malaria incidence. Furthermore, there are biological reasons to believe that malaria incidence would exhibit a monotonic response to ITN coverage. Combined together, we can use our method to determine the Minimum Resource Threshold Policy (\OURR) of ITN coverage to achieve a desired malaria incidence level.

\item Motivated by \citet{WaterPipe2012}, we can study the minimum proportion of households with piped water that will meet or exceed the levels of existing hygiene and/or welfare indicators. For instance, as Tables 3 and 4 of \citet{WaterPipe2012} reported, one may use the numbers of baths and showers and the number of times a child fetched water in recent days as a basis for the hygiene and welfare indicators, respectively. Also, as suggested by \citet{WaterPipe2012}, these indicators are likely to show monotonic response to water pipe installation. Finally, partial interference is reasonable in this context because the hygiene and welfare indicators of a household are affected by piped water in nearby (but not too far away) households. Therefore, we can use our method to determine the smallest water pipe coverage necessary to achieve the desired hygiene and/or welfare levels. 

\item We can consider a policy for allocating water, sanitation, and hygiene (WASH) in developing countries to achieve the Sustainable Development Goals 3.2 by targeting under-5 mortality being lower than 25 per 1000 live births. The context is similar to the application of the main paper except that the outcome is under-5 mortality. As before, partial interference and monotonicity are reasonable assumptions for the context, and investigators may determine the \OURR of the amount of WASH facilities that achieves the desired under-5 mortality rate.

\end{itemize}

\subsection{A Graphical Illustration for the Setup}		\label{sec:illustration}

\begin{figure}[!htp]
\centering
\scalebox{0.6}{
\begin{tikzpicture}[
roundnode/.style={circle, draw=black, fill=white, very thick, minimum size=7mm},
squarednode/.style={rectangle, draw=white, fill=white, very thick, minimum size=7mm},
]
%Nodes

\draw[ultra thick, gray] (1,-2.5) ellipse (5.5cm and 4cm);

\node[squarednode] (C1) at (1,1.5) {{\Large  Cluster 1}};
\node[roundnode] (Cov1) at (-1,-0.25) {$X_{11}$};
\node[roundnode] (Cov2) at (3,-0.25) {$X_{12}$};
\node[roundnode] (Trt1) at (-3,-3) {$A_{11}$};
\node[roundnode] (Trt2) at (-1,-4-0.25) {$A_{12}$};
\node[roundnode] (Y1) at (5,-3) {$Y_{11}$};
\node[roundnode] (Y2) at (3,-4-0.25) {$Y_{12}$};

%Lines
\draw[-, thick] (Cov1) -- (Cov2);
\draw[-{Stealth[length=3mm]}, thick] (Cov1) -- (Trt1);
\draw[-{Stealth[length=3mm]}, thick] (Cov1) -- (Trt2);
\draw[-{Stealth[length=3mm]}, thick] (Cov2) -- (Trt1);
\draw[-{Stealth[length=3mm]}, thick] (Cov2) -- (Trt2);

\draw[-{Stealth[length=3mm]}, thick] (Cov1) -- (Y1);
\draw[-{Stealth[length=3mm]}, thick] (Cov1) -- (Y2);
\draw[-{Stealth[length=3mm]}, thick] (Cov2) -- (Y1);
\draw[-{Stealth[length=3mm]}, thick] (Cov2) -- (Y2);

\draw[-{Stealth[length=3mm]}, ultra thick, blue] (Trt1.east) to [out=0,in=180] (Y1.west);
\draw[-{Stealth[length=3mm]}, ultra thick, blue] (Trt2.east) to [out=0,in=180] (Y2.west);
\draw[-{Stealth[length=3mm]}, ultra thick, red] (Trt1.south) to [out=300,in=210] (Y2.south);
\draw[-{Stealth[length=3mm]}, ultra thick, red] (Trt2.south) to [out=280,in=260] (Y1.south);

\draw[ultra thick, gray] (1+13,-2.5) ellipse (5.5cm and 4cm);

\node[squarednode] (C2) at (1+13,1.5) {{\Large  Cluster 2}};
\node[roundnode] (Cov21) at (-1+13,-0.25) {$X_{21}$};
\node[roundnode] (Cov22) at (3+13,-0.25) {$X_{22}$};
\node[roundnode] (Trt21) at (-3+13,-3) {$A_{21}$};
\node[roundnode] (Trt22) at (-1+13,-4-0.25) {$A_{22}$};
\node[roundnode] (Y21) at (5+13,-3) {$Y_{21}$};
\node[roundnode] (Y22) at (3+13,-4-0.25) {$Y_{22}$};

%Lines
\draw[-, thick] (Cov21) -- (Cov22);
\draw[-{Stealth[length=3mm]}, thick] (Cov21) -- (Trt21);
\draw[-{Stealth[length=3mm]}, thick] (Cov21) -- (Trt22);
\draw[-{Stealth[length=3mm]}, thick] (Cov22) -- (Trt21);
\draw[-{Stealth[length=3mm]}, thick] (Cov22) -- (Trt22);

\draw[-{Stealth[length=3mm]}, thick] (Cov21) -- (Y21);
\draw[-{Stealth[length=3mm]}, thick] (Cov21) -- (Y22);
\draw[-{Stealth[length=3mm]}, thick] (Cov22) -- (Y21);
\draw[-{Stealth[length=3mm]}, thick] (Cov22) -- (Y22);

\draw[-{Stealth[length=3mm]}, ultra thick, blue] (Trt21.east) to [out=0,in=180] (Y21.west);
\draw[-{Stealth[length=3mm]}, ultra thick, blue] (Trt22.east) to [out=0,in=180] (Y22.west);
\draw[-{Stealth[length=3mm]}, ultra thick, red] (Trt21.south) to [out=300,in=210] (Y22.south);
\draw[-{Stealth[length=3mm]}, ultra thick, red] (Trt22.south) to [out=280,in=260] (Y21.south);

\end{tikzpicture}}
\caption{A Graphical Illustration for the Setup. The blue arrows from $A_{ij}$ to $Y_{ij}$ $(i,j=1,2)$ depict the direct effect of the treatment, and the red arrows from $A_{ij}$ to $Y_{ij'}$ $(i,j,j' = 1,2, j \neq j')$ depict the indirect effect of the treatment.  }
\label{fig-Illustration}
\end{figure}
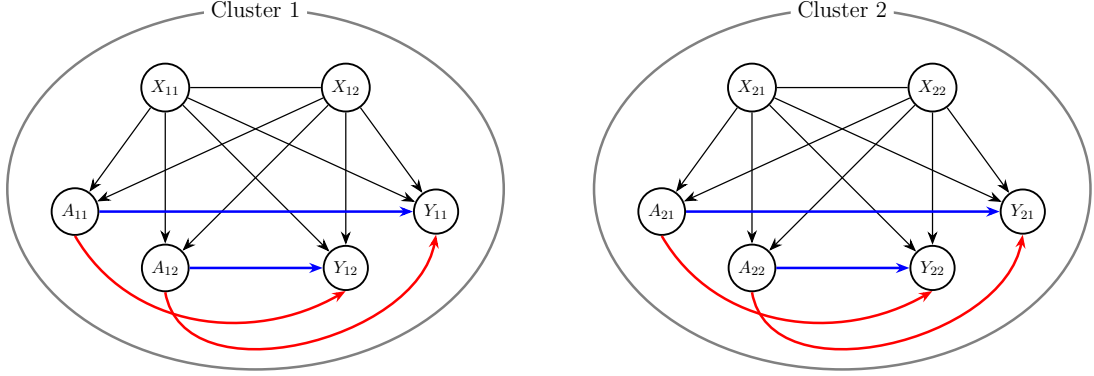

We provide a visual illustration for the setup in Figure \ref{fig-Illustration}. For simplicity, we consider $N=2$ clusters where each cluster has $\NI_i=2$ study units.  The black arrows from $A_{ij}$ to $Y_{ij}$ $(i,j=1,2)$ depict the direct effect of the treatment, and the red arrows from $A_{ij}$ to $Y_{ij'}$ $(i,j,j' = 1,2, j \neq j')$ depict the indirect effect of the treatment. No connection between two clusters illustrate the cluster-level independence.

\subsection{The (Mostly) Wrong Approach: Analysis With Aggregated, Cluster-Level Data}		\label{sec:aggregation}

We briefly discuss a tempting approach based on aggregating the data at the cluster-level. This aggregation approach has been discussed in the literature (e.g., Section 2.3 of \citet{Imbens2009} and \cite{Kilpatrick2021}) as a simple way to deal with interference. While this approach will clearly not work for estimating {\OURR}s like  $\LB_{\SO}^*(\bx_i)$ or other {\OURR}s targeting spillover-specific outcomes, from a practitioner's point of view, it is worth asking whether this approach can be used to estimate, or at least approximate, {\OURR}s like $\LB_{\OV}^* (\bx_i)$ which combine both the direct and spillover effects of treatment in a block. Unfortunately, as we illustrate below, this aggregation approach will lead to grossly misleading estimates of $\LB_{\OV}^* (\bx_i)$ except in very restrictive settings. 
% is only appropriate in very restrictive settings where there is no treatment effect heterogeneity and the treatment is completely randomized. In more realistic settings where there is treatment effect heterogeneity and/or the underlying interference model is complex, the approach will lead to grossly misleading estimates of $\LB_{\OV}^* (\bx_i)$.

Formally, following the above advice from the literature, suppose an investigator attempts to bypass the problem from interference at the unit/household-level by aggregating their data at the cluster/block-level. That is, for each cluster $i$, the investigator can consider $\mbO_i=(\mY_i,\mA_i,\mbX_i)$ to be the available data and  use existing techniques in the optimal treatment regime literature for a continuous treatment, such as \citet{Chen2016}, to obtain the minimum proportion of WASH facilities necessary to achieve a certain target $\mathcal{T}$. For example, given a cluster-level outcome model for the expected value of $\overline{Y}_i$ as a function of cluster-level variables $\overline{A}_i$ and $\overline{\mathbf{X}}_{i}$, the investigator can find the smallest $\overline{A}_i \in [0,1]$ where the expected outcome exceeds $\mathcal{T}$. %; see \citet{Chen2022} for another approach. 

Despite its simplicity, the above analysis is only appropriate in very restrictive settings, which we illustrate with an example. Suppose the treatment assignment depends on the measured covariates, and the outcome regression is given as $\EXP \big\{ \pot{Y_{ij}}{a_{ij},\ba_{\eij}} \cond \bX_{i} \big\}
=  \beta_1 a_{ij} + \beta_2 \ma_{\eij} +\bm{\beta}_3\T \bX_{ij} a_{ij}  + \bm{\beta}_4\T \bX_{ij} \ma_{\eij}$  
%, such as the true, unit-level outcome model is linear and satisfies
where $\beta_1, \ldots, \bm{\beta}_4$ are non-negative coefficients to guarantee Assumption \hyperlink{(A5)}{(A5)}. Some algebra reveals the average potential outcome at the cluster-level is
\begin{align}		\label{eq-simple}
&
\EXP \big\{ \pot{\mY_{i}}{\ba_i} \cond \bX_i \big\}
=  \bigg( \beta_1+\beta_2 + \frac{\NI_i  \bm{\beta}_4\T \mbX_{i}  }{\NI_i - 1} \bigg)  \ma_{i} + 
\bigg\{
\frac{(\NI_i-1) \bm{\beta}_3 + \bm{\beta}_4}{\NI_i-1}
\bigg\}\T
\bigg(
\frac{1}{\NI_i} \sum_{j=1}^{\NI_i} a_{ij} \bX_{ij} 
\bigg) \ .
\end{align}
If the investigator uses the aggregated, cluster-level data to estimate the {\OURR}, the resulting estimate will be biased because the cluster-level outcome model of $\overline{Y}_i$ given $\overline{A}_i$ and $\mbX_i$ is mis-specified. Or equivalently, there is an omitted variable bias because of the term $\sum_{j=1}^{\NI_i} a_{ij} \bX_{ij}$, which roughly measures the covariance between the unit-level treatment variable and the unit-level covariate. The magnitude and the direction of the bias will depend on (a) the magnitude of treatment effect heterogeneity, as measured by $\bm{\beta}_3$ and $\bm{\beta}_4$, and (b) the magnitude and the sign of the measured, unit-level confounding, as measured by the covariance of $A_{ij}$ and $\bX_{ij}$. 

More generally, if the outcome model is nonlinear, which is often the case in popular epidemiological models (e.g., \citet{Magal2014}), no amount of modeling with aggregated, cluster-level data $(\overline{Y}_i,  \overline{A}_i, \mbX_i)$ will completely remove this bias as the cluster-level data cannot capture both unit-level treatment heterogeneity and unit-level confounding. As a concrete example, suppose the treatment is completely randomized and there are no interactions between the covariates and the treatment, %, akin to the first example. 
but there exists non-linear relationship between the treatment on the outcome:  
\begin{align} 		\label{eq:wrong3}
\EXP \big\{ \pot{Y_{ij}}{a_{ij},\ba_{\eij}} \cond \bX_{i} \big\} 
=\beta_0 + \beta_1 a_{ij} + \beta_2 \big\{ (\ma_{\eij} - q_a)^p \big\}_+
\end{align}
where $\beta_1$ and $\beta_2$ are non-negative coefficients, $q_a \in [0,1]$, and $p$ is a positive integer.  Roughly speaking, the model states that the household's outcome can be affected by its peer households through a non-linear function $z \mapsto ( z -q_a)^p$  if at least $(100 \times q_a) \%$ of their peers are treated; see \citet{OR_Shape_NonLin1}, \citet{OR_Shape_NonLin2}, and \citet{OR_Shape_NonLin3} and references therein for other types of threshold models in networks.  As before, Assumptions \hyperlink{(A4)}{(A4)} and \hyperlink{(A5)}{(A5)} hold for this model and some algebra will reveal that the cluster-level outcome model will be mis-specified when using only aggregated, cluster-level data $(\mY_i, \mA_i, \mbX_i)$ due to the non-linearity of $\overline{a}_{\eij}$ in the household-level outcome model.  Consequently, the resulting {\OURR}\ with the cluster-level data will be biased.  

\begin{figure}[!htb]
\centering
\includegraphics[width=1\textwidth]{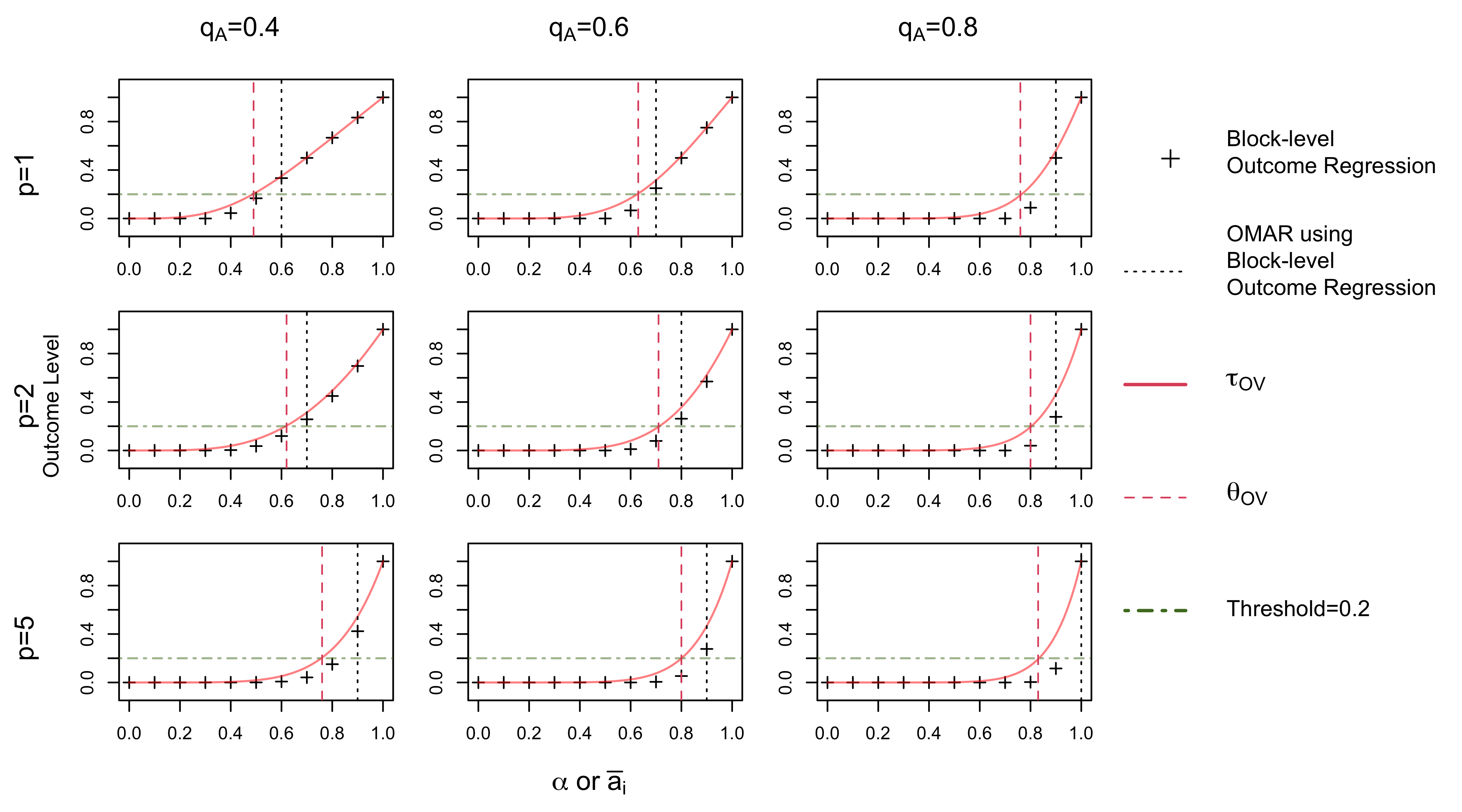}
\caption{Graphical Comparison between the {\OURR}s Based on the Aggregated Cluster-level Outcome Regression and $\tau_\OV(\alpha)$. 
Black dotted and red dashed lines indicate the {\OURR}\ based on the aggregated cluster-level outcome regression and that based on $\tau_\OV(\alpha)$ (i.e., $\LB_\OV$), respectively.}	
\label{fig-A-NaiveFail}
\end{figure}

We provide a graphical illustration of model \eqref{eq:wrong3}.  To demonstrate, we fix the cluster size $\NI_i=10$ and the coefficients $\beta_0=\beta_1=0$, and choose $\beta_2$ so that the range of the outcome regression becomes $[0,1]$. We consider three levels for $q_a \in \{0.4,0.6,0.8\}$ and $p \in \{1,2,5\}$, respectively, and we choose the threshold $\thr=0.2$. Figure \ref{fig-A-NaiveFail} visually presents the differences between the {\OURR}s based on $\tau_\OV$ and the aggregated cluster-level outcome regression. We find that the differences vary between 0.07 and 0.17. The toy example suggests that estimating the {\OURR}\ based on the aggregated outcome regression may yield significantly biased estimates of $\theta_\OV$ in \eqref{eq-def:lowerB}.

However, we mention that this simple aggregation approach may work under different assumptions. For instance, \citet{Kilpatrick2021} assumed that the cluster-level potential outcome only depends on the total number of treatment implemented in a cluster, i.e., the cluster-level stratified interference. This implies that the average potential outcome at the cluster-level has a form of $
\EXP \big\{ \overline{Y}_i^{(\ba_{ij})} \cond \mbX_i \big\}
=
\mu^{\dagger} \big( \ma_i, \mbX_i \big)$ for some function $\mu^{\dagger}$, which can be identified as (nonparametric) regression models of $\overline{Y}_i$ on $(\mA_i, \mbX_i)$. In turn, using their $g$-formula approach and/or the indirect approach in Section \ref{sec-Indirect}, we can get a valid estimate of $\LB_{\OV}^*$. We remark that the cluster-level stratified interference assumption lacks the necessary flexibility to define $\LB_{\SO}^*$. This is because it eliminates the possibility of having distinct cluster-level outcomes based on the treatment recipients, which is a critical aspect of interference.

Next, we compare the classification performance measures of the true policy of the main manuscript and the policy obtained from the aggregated cluster-level outcome regression. We consider the following simple data generating process:
\begin{align*}
& \NI_i = 10
\ ,	\quad
X_{ij} \sim \text{Ber}(0.5)
\ ,	\quad
A_{ij} \cond X_{ij} \sim \text{Ber}(0.5)
\\
&
Y_{ij} \cond ( A_{ij}, \bA_{\eij}, X_{ij} , \bX_{\eij} ) = 
A_{ij} + 0.5 \mA_{\eij} + 0.5 A_{ij} X_{ij} + \epsilon_{ij}
, \quad \epsilon_{ij} \sim N(0,1) \ .
\end{align*}
The aggregated cluster-level outcome regression is given as $\EXP \big( \mY_i \cond \mA_i, \overline{X}_i) = (1.5 + 0.5 \overline{X}_i) \mA_i $, and the simple policy is obtained as follows:
\begin{align*}
\theta_{\text{Simple}}^*(\overline{x}_i)
=
\min
\Big\{ a \in \{0,0.1,\ldots,1 \} \, \Big| \, 
\EXP \big( \mY_i \cond \mA_i=a, \overline{X}_i = \overline{x}_i ) \geq \theta
\Big\} \ .
\end{align*}	
The true policy $\theta_{\OV}^*(\alpha)$ is defined based on equation \eqref{eq-def:lowerB} of the main paper. 

\begin{figure}[!htb]
\centering
\includegraphics[width=1\textwidth]{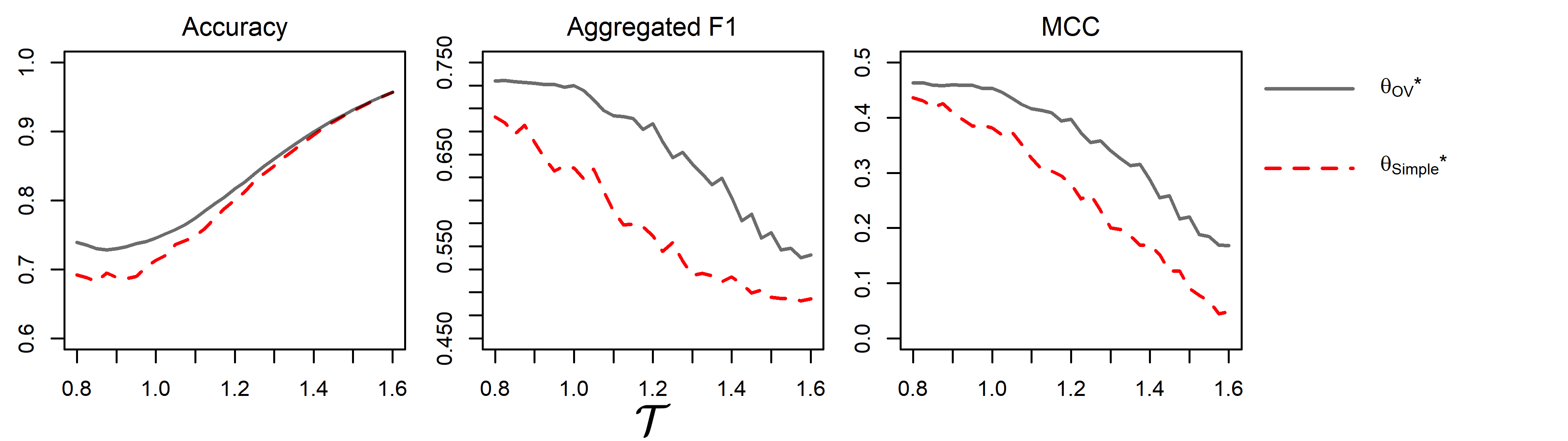}
\caption{Classification Performance Measures of $\theta_{\OV}^*$ and $\theta_{\text{Simple}}^*$.}
\label{fig-Simple-plot}
\end{figure}

We generate $N=10^5$ observations from the above data generating process, and compare the three classification performance measures of $\theta_{\text{Simple}}^*$ and $\theta_{\OV}^*$ across $\thr \in [0.8,1.6]$. We use the true outcome regressions $\mu^*$ and $\mu^\dagger$ to construct $\theta_{\OV}^*$ and $\theta_{\text{Simple}}^*$. As a result, the discrepancies in the performance measures can be attributed to the use of the aggregated approach instead of the approach proposed of the main paper. The range of $\thr$ has been chosen such that the lower bound of the interval is not significantly smaller than the average outcome of $\EXP \big( \mY_i \big) = 0.875$. Figure \ref{fig-Simple-plot} graphically summarizes the result. We find that $\theta_{\OV}^*$ uniformly yields better classification performance measures compared to $\theta_{\text{Simple}}^*$, suggesting that using the aggregated cluster-level outcome regression is suboptimal even in the simple model. 

In summary, an analysis based on aggregated, cluster-level data will often lead to biased estimates of {\OURR}. Also, if the investigator is interested in {\OURR}s based on spillover-specific outcomes such as $\LB_{\SO}^*(\bx_i)$, a cluster-level analysis is simply infeasible.

\subsection{Inverse Probability-Weighted and Outcome Regression-based Loss Functions}					\label{sec:otherloss}

We introduce the IPW and outcome regression-based loss functions. Specifically, we can replace $\psi_{\text{DR}}$ in \eqref{eq-L02} of the main paper with the following functions.
\begin{align*}
& 
\psi_{\text{IPW}} (a, s, \bO_{ij}, \bO_\eij )
=
\frac{Y_{ij} \ind ( A_{ij} = a,  S_\eij = s) }{\PS^* (a , s \cond \bX_i)}
\ , \quad
\psi_{\text{OR}} (a , s, \bO_{ij}, \bO_\eij)
=
\mu^* \bigg( a, \frac{s}{\NI_i-1}, \bX_{ij}, \bX_{\eij} \bigg)
\ .
\end{align*}

\subsection{Machine Learning Methods Used for the Outcome Regression Estimation}					\label{sec:supp-ORest}

As candidate machine learning methods, we include the following methods and R packages in our super learner library: linear regression (\texttt{glm}), Lasso/elastic net (\texttt{glmnet} \citep{glmnet}), spline (\texttt{earth} \citep{earth}, \texttt{polspline} \citep{polspline}), generalized additive model (\texttt{gam} \citep{gam}), boosting (\texttt{xgboost} \citep{xgboost}, \texttt{gbm} \citep{gbm}), random forest (\texttt{ranger} \citep{ranger}), and neural net (% \texttt{nnet} \citep{nnet}, and 
\texttt{RSNNS} \citep{RSNNS}). 
%We fit models from the methods in the library using the training data and we create an optimal weighted average of the methods using the test data performance. 

\subsection{Computation: Training the Support Vector Machine in \eqref{eq-SVM2} of the main Paper}						\label{sec:supp-SVMsolve}
This section presents the computational details on training the support vector machine in \eqref{eq-SVM2} of the main paper. The algorithm to train SVMs under this type of nonconvex function is already discussed in prior works \citep{DCalgorithm,Chen2016} and we present a summary of it for completeness. Also, to keep the notation clear, the discussion below assumes that the nuisance functions are known, but the identical computation algorithm is used to train the SVM when the nuisance functions are estimated.

To start off, we can decompose the loss function for the overall outcome case in equation  \eqref{eq-L} of the main paper into the difference of two convex function $\loss_+ (t, \bO_i)$ and $\loss_- (t, \bO_i)$, i.e., $ \loss(t, \bO_i) =  \loss_+(t,\bO_i) - \loss_-(t,\bO_i) $ for any $t$ and $\bO_i$ where
\begin{align*}
&
\loss_+ (t, \bO_i)
=
\begin{cases}
\losszo_+ (0, \bO_i) - 2 \delta t
& \text{if } -\infty < t < 0 
\\
\losszo_+ (t, \bO_i)
& \text{if }0 \leq t \leq 1 
\\
\losszo_+ (1, \bO_i) + \big( \overline{\delta} +  2 \delta \big) (t - 1) 
& \text{if }1 < t < \infty
\end{cases}
\\
&
\loss_- (t, \bO_i)
=
\begin{cases}
\losszo_- (0, \bO_i) - 2 \delta t - \delta + \delta e^t 
& \text{if }-\infty < t < 0 
\\
\losszo_- (t, \bO_i) 
& \text{if }0 \leq t \leq 1 
\\
\losszo_- (1, \bO_i) + \big( \overline{\delta} +  2 \delta \big) (t - 1) - \delta + \delta e^{1-t} 
& \text{if }1 < t < \infty
\end{cases} \\
&
\losszo_\pm (t, \bO_i)
= 
\frac{1}{\NI_i}
\sum_{j=1}^{\NI_i}
\sum_{a=0}^{1}
\sum_{s=0}^{\NI_i-1}
{\NI_i - 1 \choose s}
\psi_{\text{DR}}(a, s,\bO_{ij}, \bO_\eij)
\\
&
\hspace*{4cm}
\times 
\sum_{\ell=0}^{\NI_i - a - s} {\NI_i - a - s \choose \ell} 
\bigg\{ \frac{  (-1)^{\ell} }{\ell + a + s + 1} \bigg\}_\pm t^{\ell+a+s+1} + (\thr)_\pm t
\end{align*}
Here, $(a)_+ = \max(a,0)$, $(a)_-= - \min(a,0)$, and $\overline{\delta}$ is chosen as the maximum of the left derivatives of $ \losszo_+(t, \bO_i)$ and $ \losszo_-(t, \bO_i)$ at $t=1$, i.e., $ \overline{\delta} =
\max \big\{
\lim_{ \epsilon \downarrow 0} \nabla \losszo_+(1 - \epsilon, \bO_i)
,
\lim_{ \epsilon \downarrow 0} \nabla \losszo_-(1 - \epsilon, \bO_i)
\big\}$ and $\nabla \losszo_\pm(t, \bO_i) $ is the derivative of $\losszo_\pm(t, \bO_i)$ with respect to $t$. Critically, the two loss functions $ \loss_+$ and $ \loss_-$ are convex and non-decreasing in $t$. For the spillover outcome case, we use
\begin{align*}
& \losszo_{\SO,\pm} (t, \bO_i)
= 
\frac{1}{\NI_i}
\sum_{j=1}^{\NI_i}
\sum_{s=0}^{\NI_i-1}
{\NI_i - 1 \choose s}
\psi_{\text{DR}}(0, s,\bO_{ij}, \bO_\eij)
\sum_{\ell=0}^{\NI_i - s} {\NI_i - s \choose \ell} 
\bigg\{ \frac{  (-1)^{\ell} }{\ell + s + 1} \bigg\}_\pm t^{\ell+s+1} + (\thr)_\pm t
\end{align*}

Given the decomposition of the loss function into the difference of two convex functions, we use the DC algorithm \citep{DCalgorithm}, which is an iterative algorithm, to solve the original non-convex optimization problem; see Algorithm \ref{alg:DC} for details.
\begin{algorithm}[!htb]
\begin{algorithmic}[1]
\REQUIRE Initialize values $\Beta^{(0)} \in \R^\NC$, $b^{(0)} \in \R$. Set iteration number to zero, $j \leftarrow 0$.
\STATE Precompute the gradient $\nabla \loss_-(t,\bO_i)$ where
\begin{align*}
\nabla \loss_-(t,\bO_i)
=
\begin{cases}
\displaystyle{ \frac{\partial}{\partial t} \loss_- (t, \bO_i) } 
& t \neq 0 ,1
\\[0.25cm]
\displaystyle{ \frac{1}{2}\lim_{\epsilon \downarrow 0} \bigg\{  \frac{\partial}{\partial t} \loss_- (t+\epsilon, \bO_i) + \frac{\partial}{\partial t} \loss_- (t - \epsilon, \bO_i) \bigg\} }
& 	t = 0 ,1
\end{cases}
\end{align*}
\REPEAT
\STATE Let $\Beta^{(j+1)}$ and $b^{(j+1)}$ be the solution to the following convex optimization problem.
\begin{align*}
\hspace*{-0.5cm}
\begin{bmatrix}
\Beta^{(j+1)} \\ b^{(j+1)}
\end{bmatrix}
\in \argmin_{\Beta, b}
\bigg[
\frac{1}{\NC} \sum_{i=1}^\NC
\left\{  \begin{array}{l}
\loss_+ \big(  \kvec_i\T \Beta + b , \bO_i \big)  
\\
\quad
-
\nabla \loss_- \big( \kvec_i\T \Beta^{(j)} + b^{(j)}, \bO_i \big) \big( b + \kvec_i\T \Beta \big)
\end{array}
\right\}
+ \frac{\lambda_\NC}{2} \Beta\T \kmat \Beta
\bigg] \\[-1.25cm]
\end{align*}
\STATE $j \leftarrow j+1$
\UNTIL{convergence}
\RETURN $\big( \widehat{\Beta}, \widehat{b} \big)  \leftarrow \big( \Beta^{(j)},  b^{(j)} \big)$. 
\end{algorithmic}
\caption{DC Algorithm}
\label{alg:DC}
\end{algorithm}    

To initiate the DC algorithm, we choose the initial value as follows. First, for each $i$, let the solution be $r_i$, i.e., $r_i = \argmin_{t \in [0,1]} \loss ( t , \bO_i )$ which can be obtained from a grid-search. In words, $r_i$ is an approximate of $\widehat{\LB}(\bx_i)$ that are found by a grid-search. But, since $r_i$ is bounded in the unit interval, it may not be a suitable approximate of $\widetilde{\LB}(\bx_i)$, the SVM solution before the winsorization. As a consequence, directly using $r_i$ to construct initial points may lead to an estimate policy shrinking to a certain value, i.e., a policy does not reflect the heterogeneity induced by $\mbx_i$. To stretch $r_i$ outside of the unit interval, we consider the following steps.
\begin{itemize}
\item[(a)] Let $\phi$ and $\varphi$ be
\begin{align*}
& \phi (a, a', \bX_i) = \frac{1}{\NI_i} \sum_{j=1}^{\NI_i} \widehat{\mu} \Big( a, a', \bX_{ij}, \bX_\eij \Big)
\ , \
\varphi(\bX_i) = \frac{1}{\NI_i }\sum_{j=1}^{\NI_i} 
\frac{Y_{ij} - \widehat{\mu} (A_{ij}, \mA_\eij, \bX_{ij}, \bX_\eij)}{\widehat{e}(A_{ij}, S_\eij \cond \bX_i) } \ .
\end{align*}
\item[(b)] By only using the clusters with non-0 and non-1 $r_i$s, i.e., $r_i \in (0,1)$, we fit linear regression models where $\phi(a,a',\bX_i)$ and $\varphi(\bX_i)$ are regressed on $r_i$s. We choose $(a,a')$ from $\{0,1\} \otimes \{0,0.2,0.4,0.6,0.8,1\}= \{(0,0), (0,0.2),\ldots,(1,0.8),(1,1)\}$, i.e., 12 levels. Let $(\widehat{\beta}_{0,\text{model }k}, \widehat{\beta}_{1,\text{model }k})$ are the estimated regression coefficients from $k$th model.   

\item[(c)] Let $\widehat{r}_i$ be the adjusted initial points which are defined as follows. 
\begin{itemize}
\item[(c-1)] If $r_i \in (0,1)$, no adjustment is required, i.e., $\widehat{r}_i = r_i$.
\item[(c-2)]  For clusters having $r_i=1$, we use the largest prediction values obtained from the 13 regression models and 1, i.e., 
\begin{align*}
\widehat{r}_i
=
\max
\bigg\{
\frac{\phi(0,0, \bX_i) - \widehat{\beta}_{0,\text{model 1}}}{\widehat{\beta}_{1,\text{model 1}}},
\ldots,
\frac{\varphi(\bX_i) - \widehat{\beta}_{0,\text{model 13}}}{\widehat{\beta}_{1,\text{model 13}}},
1
\bigg\} \ .
\end{align*}
\item[(c-3)] Similarly, for clusters having $r_i=0$, we use the smallest prediction values obtained from the 13 regression models and 0, i.e., 
\begin{align*}
\widehat{r}_i
=
\min
\bigg\{
\frac{\phi(0,0, \bX_i) - \widehat{\beta}_{0,\text{model 1}}}{\widehat{\beta}_{1,\text{model 1}}},
\ldots,
\frac{\varphi(\bX_i) - \widehat{\beta}_{0,\text{model 13}}}{\widehat{\beta}_{1,\text{model 13}}},
0
\bigg\} \ .
\end{align*}
\end{itemize}

\end{itemize}

Second, we take $b^{(0)} = \sum_{i=1}^{\NC} \widehat{r}_i / \NC$ and $\Beta^{(0)}$ as a vector satisfying $ \widehat{r}_i = \kvec_i\T \Beta^{(0)} + b^{(0)}$ for all $i$; i.e., $\widehat{\bm{r}} = \kmat \Beta^{(0)} + b^{(0)} \bm{1} $ where $\widehat{\bm{r}} = [ \widehat{r}_1, \ldots, \widehat{r}_\NC ]\T \in \R^\NC$ and $\bm{1} = [1,\ldots,1]\T \in \R^\NC$. Even though the kernel matrix $\kmat$ is invertible due to the positive definiteness of the kernel function $\kernel$, the inverse of $\kmat$ cannot be obtained due to the numerical singularity. Under such case, we add a tiny value to diagonal of $K$ until its inverse can be obtained. In line 1, $\nabla \loss_-$ is a subgradient of $\loss_-$ that accounts for the non-differentiability of $\loss_-$ at $t=0$ and $t=1$.

The convex optimization in line 3 can be solved by using many standard algorithms and softwares. The iteration stops when $\big\| (\Beta^{(j+1)} , b^{(j+1)}) - (\Beta^{(j)} , b^{(j)} ) \big\|_2$ drops below some threshold value. We remark that because the objective function in \eqref{eq-SVM2} of the main paper is bounded below, the algorithm will always converge in finite steps \citep{DCalgorithm, Chen2016}.

\subsection{Details of Cross-validation}				\label{sec:supp-cv}

We present the details on how to choose the SVM parameters $\gamma$ and $\lambda$.  We consider a set of candidate values for $(\gamma_\ell,\lambda_\ell)$ where $\ell = 1,\ldots,K$. Without loss of generality, let the estimation data fold be $\mathcal{D}_1=\mathcal{D}_2^c$ and, as a consequence, observations in $\mathcal{D}_2$ is used to evaluate the estimated loss function $\widehat{\loss}_{(-1)}(t,\bO_i)$ for $i \in \mathcal{D}_2$. We further split $\mathcal{D}_2$ into training and tuning sets, denoted by $\mathcal{D}_{2,{\rm train}}$ and $\mathcal{D}_{2,{\rm tuning}}$, respectively, based on the number of cross-validation folds. For each candidate parameter $(\gamma_\ell,\lambda_\ell)$, we estimate the direct {\OURR}\ $\widehat{\LB}_\text{train} (\bX_i \con \ell)$ by only using the training set $\mathcal{D}_{2,{\rm train}}$ and obtain the empirical risk using the tuning set $\mathcal{D}_{2,{\rm tuning}}$. The optimal parameters $(\gamma^*, \lambda^*)$ are the minimizer of the average of the empirical risks across the tuning sets, i.e.
\begin{align*}
(\gamma^*, \lambda^*)
=
\argmin_{\ell=1,\ldots,K}
\frac{1}{\big| \mathcal{D}_{2,{\rm tuning}} \big| }
\sum_{i \in  \mathcal{D}_{2,{\rm tuning}} } 
\hspace*{-0.4cm}
\widehat{\loss}_{(-1)} \big( \widehat{\LB}_\text{train} (\bX_i \con \ell), \bO_i \big) \ .
\end{align*}

\subsection{Details of Undersampling and Cross-fitting Procedures}				\label{sec:supp-cf}

We discuss the details on how to negate the impact of a particular realization of undersampling procedure. We randomly choose a subset of observations so that the cluster sizes are (nearly) balanced, and we repeat the undersampling for $U$ times indexed by $u$. Let $\widehat{\mu}^{(u)}$ and $\widehat{e}^{(u)}$ be the estimated outcome regression and propensity score obtained from $u$th undersample. Then, we take the median-adjusted nuisance function across $U$ estimated functions as the final estimate of the nuisance function, i.e., $\widehat{\mu} := \median_{u=1,\ldots,U} \widehat{\mu}^{(u)} $ and $\widehat{e} := \median_{u=1,\ldots,U} \widehat{e}^{(u)}$.

Next, we discuss the median-adjustment of cross-fitting procedure. Once we split the data into two folds $\mathcal{D}_1$ and $\mathcal{D}_2$, we obtain two directly estimated policies $\widehat{\LB}_{(-\ell)}$ for $k=1,2$ where $\mathcal{D}_\ell^c$ is used as the estimation data fold and $\mathcal{D}_\ell$ is used as the evaluation data fold. Investigators may use either $\widehat{\LB}_{(-1)}$ or $\widehat{\LB}_{(-2)}$ as the final estimate of the {\OURR}, denoted by $\widehat{\LB}^{(F)}$. However, we recommend to use $\widehat{\LB}^{(F)} (\bx) = \winsor\big( \{ \widehat{\LB}_{(-1)} + \widehat{\LB}_{(-2)} \} /2 \big) (\bx)$, the winsorized policy of the average of two non-winsorized policies, for the new $\mbx$ as the estimate of the {\OURR}\ to fully use the data. If the evaluation point is one of the points in the data, i.e., $\bx = \bx_i$ for some $i \in \mathcal{D}_\ell$, we recommend using $\widehat{\LB}^{(F)} (\bx) = \widehat{\LB}_{(-\ell)}(\bx_i)$ because $\widehat{\LB}_{(-\ell)}$ does not depend on $i$ while $\widehat{\LB}_{(\ell)}$ depends on $i$ which may lead to an overfitted value. Second, %we observe that the estimated policy depends on the particular random split of the data which can severely affect the estimate in finite samples.
to construct a more robust estimate of the {\OURR}\ under cross fitting, we use the recommendation in \citet{Victor2018} to our setting by taking the mean or the median of multiple {\OURR}\ estimates. Specifically, we repeat the estimation of $\widehat{\LB}^{(F)}$ multiple times, say $T$ times, and obtain $\widehat{\LB}_t^{(F)}$ $(t=1,\ldots,T)$ where the sample partitions are randomly done across splits. We define the mean-{\OURR}\ estimate $\widehat{\LB}^{(F,{\rm mean})}(\bx) = \sum_{t=1}^T \widehat{\LB}_t^{(F)}(\bx)/T$ and the median-{\OURR}\ estimate $\widehat{\LB}^{(F,{\rm median})}(\bx) = \median_{t=1,\ldots,T} \widehat{\LB}_t^{(F)}(\bx)$. % to mitigate the impact from sample splitting.

\subsection{Details of the Data Generating Process of the Simulation}				\label{sec:supp-Simulation}

We provide details of the data generating process of the simulation in Section \ref{sec:Simulation}. First, we provide the distribution of the cluster size $\NI_i$, which is the same as the empirical distribution of $\NI_i$ in the dataset used in Section \ref{sec:Simulation}.
\begin{table}[!htp]
\renewcommand{\arraystretch}{1.3} \centering
\setlength{\tabcolsep}{4pt}
\footnotesize
\begin{tabular}{|c|c|c|c|c|c|c|c|c|c|c|}
\hline
$\NI_i$     & 3     & 4     & 5     & 6     & 7     & 8     & 9     & 10    & 11    & 12    \\ \hline
Frequency   & 4     & 4     & 11    & 22    & 30    & 55    & 61    & 76    & 80    & 88    \\ \hline
Probability & 0.004 & 0.004 & 0.011 & 0.021 & 0.029 & 0.054 & 0.059 & 0.074 & 0.078 & 0.086 \\ \hline \hline
$\NI_i$     & 13    & 14    & 15    & 16    & 17    & 18    & 19    & 20    & 21    & 22    \\ \hline
Frequency   & 92    & 90    & 113   & 86    & 72    & 71    & 39    & 22    & 6     & 5     \\ \hline
Probability & 0.090 & 0.088 & 0.110 & 0.084 & 0.070 & 0.069 & 0.038 & 0.021 & 0.006 & 0.005 \\ \hline
\end{tabular}
\end{table}

\begin{figure}[!htb]
\centering
\includegraphics[width=0.75\textwidth]{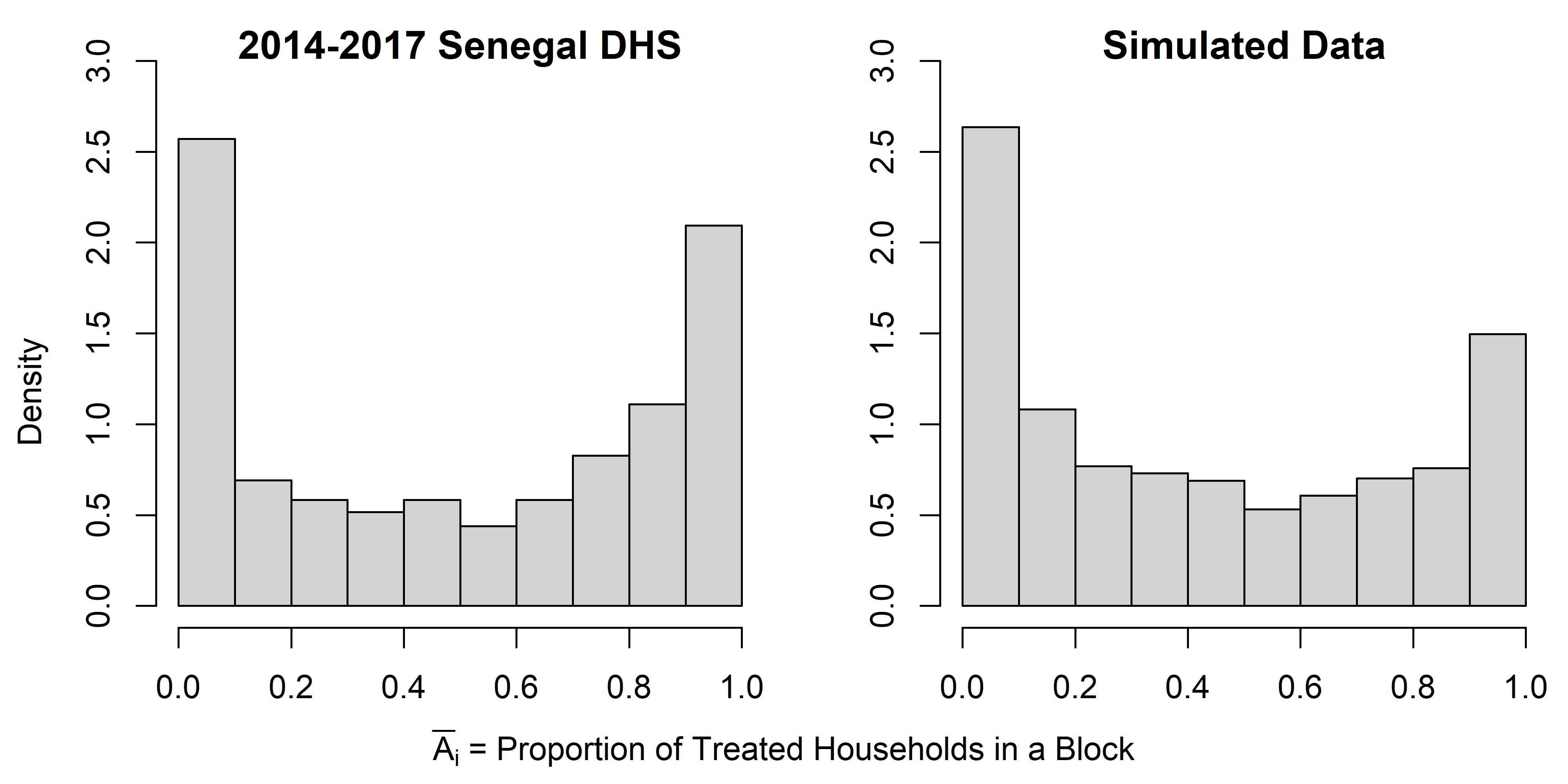}
\caption{Histograms of the treated households (i.e., $\overline{A}_i$) in the 2014-2017 Senegal DHS (left) and simulated datasets (right). The distribution of $\overline{A}_i$ in the simulation shows the distribution across 50 repetitions.}
\label{Fig-TreatmentProportion}
\end{figure}

Next, in Figure \ref{Fig-TreatmentProportion}, we provide graphical summaries of the distributions of $\overline{A}_i$ in the 2014-2017 Senegal DHS and the simulated datasets. The two distributions are similar to each other with the common support of $[0,1]$.

Lastly, we provide the details of the outcome regression model:
\begin{align*}
&
{Y}_{ij} \cond ( \bA_i, \bX_i )
\sim
{\rm Ber}
\left(
{\rm expit}
\left[
\begin{array}{l} 
\text{\footnotesize$
-0.35 + \big\{ 0.1 + 0.25 (C_i+W_{ij1})^2 \big\} A_{ij}
$
}
\\ 
\text{\footnotesize$
+
\big\{  0.05 + 0.15 ( \overline{W}_{\eij 2} + \overline{W}_{\eij 3} )^2 \big\} \overline{A}_\eij$}
\\ 
\text{\footnotesize$
+ 0.1 (C_i + \sum_{k=1}^3 W_{ijk} )
+ 0.25 (C_i^2 + \sum_{k=1}^3 W_{ijk}^2 )
+ 0.05 ( \sum_{k=1}^3 \overline{W}_{\eij k}) $}
\\ 
\end{array}\right]
\right) \ .
\end{align*}
Here, $\overline{W}_{\eij k} = \sum_{\ell \neq j} W_{ij k}/(\NI_i-1)$. We remark that the outcome model satisfies Assumptions \hyperlink{(A1)}{(A1)}-\hyperlink{(A5)}{(A5)}.

\subsection{Details of Classification Performance Measures}				\label{sec:supp-classification}

For an given {\OURR}\ $\LB$, we 
%refer to cluster $i$ as a true positive if $\mA_i > \LB(\mbX_i)$ and $\mY_i > \thr$, i.e., the observed treated proportion is greater than the true {\OURR}\ and the observed cluster-level response is also greater than the target response value. %In the context of a binary classification, two events $\{ \mA_i > \LB(\mbX_i) \}$ and $\{ \mY_i > \thr \}$ correspond to a predicted class and an observed class, respectively. 
%Similarly, we refer to cluster $i$ as a true negative if $\mA_i \leq \LB(\mbX_i)$ and $\mY_i \leq \thr$, a false positive if $\mA_i > \LB(\mbX_i)$ and $\mY_i \leq \thr$, or a false negative if $\mA_i \leq \LB(\mbX_i)$ and $\mY_i < \thr$. 
define the \textit{true positives} (TP), \textit{true negatives} (TN), \textit{false positives} (FP), and \textit{false negatives} (FN) as follows:
\begin{align}							\label{eq-TP}
&
{\rm TP} 
= 
\sum_{i \in \mathcal{D}_{\rm test}} \ind \big\{ \mY_i > \thr, \mA_i > \LB(\bX_i) \big\} 
\ , 
&&
{\rm TN} 
= 
\sum_{i \in \mathcal{D}_{\rm test}} \ind \big\{ \mY_i \leq \thr, \mA_i \leq \LB(\bX_i) \big\} 
\ , 
\\
&
{\rm FP} 
= 
\sum_{i \in \mathcal{D}_{\rm test}} \ind \big\{ \mY_i \leq \thr, \mA_i > \LB(\bX_i) \big\} 
\ , 
&&
{\rm FN} 
= 
\sum_{i \in \mathcal{D}_{\rm test}} \ind \big\{ \mY_i > \thr, \mA_i \leq \LB(\bX_i) \big\} \ .
\nonumber
\end{align}
Given these definitions, we use the following classification performance measures: accuracy, two-sided F1 score, and the Matthews correlation coefficient (MCC) \citep{MCC} which are defined as follows:
\begin{align*}
&
\text{Accuracy}
=
\frac{ {\rm TP} + {\rm TN} }{{\rm TP} + {\rm TN} + {\rm FP} + {\rm FN}}
\ ,
\
\text{F1}
=
\frac{ 2{\rm TP} }{2{\rm TP} + {\rm FP} + {\rm FN}}
+
\frac{ 2{\rm TN} }{2{\rm TN} + {\rm FP} + {\rm FN}}
\ ,
\\
&
\text{MCC}
=
\frac{{\rm TP}  \times {\rm TN}  
-
{\rm FP}  \times {\rm FN}  
}{
\{ (
{\rm TP}  + 
{\rm FP}  ) \times
(
{\rm TP}  + 
{\rm FN}  )
\times
(
{\rm TN}  + 
{\rm FP}  )
\times(
{\rm TN}  + 
{\rm FN}  ) \}^{1/2} } \ .
\end{align*}
The usual F1 score does not use true negatives in its score, i.e., 2TP/(2TP+FP+FN), and is sensitive to the definition of a positive label. For example, if we were to define the positive label as the opposite of the definition in equation \eqref{eq-TP}, i.e., positive label if $\overline{Y}_i \leq \thr$, the F1 score changes. To avoid this, we consider the two-sided F1 score, the average of the usual F1 score and the ``opposite'' F1 score, 2TN/(2TN+FP+FN).

\subsection{Assessment of Assumptions of the Main Paper} 	\label{sec:BOAssessment}

We take a moment to discuss the plausibility of the bounded cluster size $\NI_i$ assumption and Assumptions \hyperlink{(A1)}{(A1)}-\hyperlink{(A5)}{(A5)} in the Senegal DHS. 

\begin{itemize}
\item[(Bounded $\NI_i$)] The bounded block size assumption is plausible in the Senegal DHS  because the data was collected based on a stratified sampling design where a fixed number of households were sampled from each block \citepalias{SEN_DHS}. Also, the maximum number of households among $\NC=1027$ census blocks in the 2014-2017 Senegal DHS is $\UNI=22$, and the small value of $\UNI/\NC=0.021$ (i.e., an upper bound on $\NI_i/\NC$) suggests that the ``large $\NC$, small $\NI_i$'' asymptotic regime is a reasonable approximation for our data.

\item[(A1)] Assumption \hyperlink{(A1)}{(A1)} is plausible as long as households in different census blocks do not interact with each other. In the data,  99.15\% of the census blocks are geographically far apart from each other. The average and median distances among 22,578 pairs of census blocks in the 2018 Senegal DHS are 245.04km and 230.17km, respectively; only 192 (0.85\%) pairs of census blocks have distance smaller than 10km. Given this, we find that the partial interference assumption is plausible where interference likely occurs between households in the same census block and not across different census blocks. 

\item[(A2)] 
To check Assumptions \hyperlink{(A2)}{(A2)} and \hyperlink{(A3)}{(A3)}, we check covariate balance and overlap by using the binning approach in \citet{Hirano2004},  \citet{Overlap1} and \citet{Overlap2} for a continuous treatment variable. Algorithm \ref{alg:Balance} shows the details on the covariate balance assessment. 

\begin{algorithm}[!htb]
\begin{algorithmic}[1]
\STATE Divide $\sum_{i=1}^{\NC} \NI_i = 13556$ units into four groups:
\begin{align*}
& \mathcal{A}_k = \big\{ (i,j) \cond (A_{ij}, \mA_\eij) \in R_k \big\}
\ , \
R_k
=
\begin{cases}
\{ 0 \} \times [0, \alpha_0] \in \{0,1\} \times [0,1] & \text{ if } k=1
\\
\{ 0 \} \times (\alpha_0, 1] \in \{0,1\} \times [0,1] & \text{ if } k=2
\\
\{ 1 \} \times [0, \alpha_1] \in \{0,1\} \times [0,1] & \text{ if } k=3
\\
\{ 1 \} \times (\alpha_1, 1] \in \{0,1\} \times [0,1] & \text{ if } k=4
\end{cases}    		
\end{align*}
where $\alpha_0$ and $\alpha_1$ are chosen so that $\mathcal{A}_1, \ldots, \mathcal{A}_4$ have similar sizes. 

\FOR{$k=1,2,3,4$}

\STATE Obtain unadjusted $t$-statistics that compare the distribution of $\tbX_i$ between $\mathcal{A}_k$ and $\mathcal{A}_k^c$, i.e., 
\begin{align*}
\bigg\{ \widetilde{\bX}_{i,k} \, \bigg| \,
\widetilde{\bX}_{i,k}
=
\frac{ 
\sum_{j=1}^{\NI_i} \ind \{ (i,j) \in \mathcal{A}_k \} \bX_{ij}
}{ \sum_{j=1}^{\NI_i} \ind \{ (i,j) \in \mathcal{A}_k \} }
\bigg\}
\quad \text{v.} \quad
\bigg\{ \widetilde{\bX}_{i,k^c} \, \bigg| \,
\widetilde{\bX}_{i, k^c}
=
\frac{ 
\sum_{j=1}^{\NI_i} \ind \{ (i,j) \notin \mathcal{A}_k \} \bX_{ij}
}{ \sum_{j=1}^{\NI_i} \ind \{ (i,j) \notin \mathcal{A}_k \} }
\bigg\} 
\end{align*}

\STATE Calculate the estimated propensity score $
\widehat{e}_{ij,k} =
\widehat{P} \big\{ (A_{ij}, \mA_{\eij}) \in R_k \cond \bX_{ij}, \bX_{\eij}  \big\}$.

\STATE Let $-\infty = q_0 \leq q_1 \leq \ldots \leq q_9 \leq q_{10} = \infty$ be the deciles of  $\big\{ \widehat{e}_{ij,k} \cond (i,j) \in \mathcal{A}_{k} \big\}$.

\STATE Let $\mathcal{E}_{b,k}
=
\big\{
(i,j) \, \Big| \,
\widehat{e}_{ij,k} \in (q_{b-1}, q_b]
\big\}$ $(b=1,\ldots,10)$.

\STATE Obtain $t$-statistics that compare the distribution of $\tbX_i$ between $\mathcal{E}_{b,k} \cap \mathcal{A}_k$ and $\mathcal{E}_{b,k} \cap \mathcal{A}_k^c$, i.e., 
\begin{align*}
\bigg\{ \widetilde{\bX}_{i,k} \, \bigg| \,
\widetilde{\bX}_{i,k}
=
\frac{ 
\sum_{j=1}^{\NI_i} \ind \{ (i,j) \in \mathcal{E}_{b,k} \cap \mathcal{A}_k \} \bX_{ij}
}{ \sum_{j=1}^{\NI_i} \ind \{ (i,j) \in \mathcal{E}_{b,k} \cap \mathcal{A}_k \} }
\bigg\}
\quad \text{v.} \quad
\bigg\{ \widetilde{\bX}_{i,k^c} \, \bigg| \,
\widetilde{\bX}_{i, k^c}
=
\frac{ 
\sum_{j=1}^{\NI_i} \ind \{ (i,j) \in \mathcal{E}_{b,k} \cap \mathcal{A}_k^c \} \bX_{ij}
}{ \sum_{j=1}^{\NI_i} \ind \{ (i,j) \in \mathcal{E}_{b,k} \cap \mathcal{A}_k^c \} }
\bigg\} 
\end{align*}

\STATE Aggregate the $t$-statistics obtained in Step 7 with weights from the size of $\mathcal{E}_{1,k}$, $\ldots$, $\mathcal{E}_{10,k}$.

\STATE Obtain adjusted $t$-statistics by taking the median of $t$-statistics in Step 6 across multiple cross-fitting procedures.

\ENDFOR
\end{algorithmic}
\caption{Assessment of Covariate Balance}
\label{alg:Balance}
\end{algorithm}    

We use the median of the propensity score estimates from 100 cross-fitting procedures. As a consequence, we obtain the unadjusted/adjusted $t$-statistics in Figure \ref{fig:Balance}, which suggests covariate balance was satisfied for all cases.     

\begin{figure}[!htb]
\centering
\includegraphics[width=\textwidth]{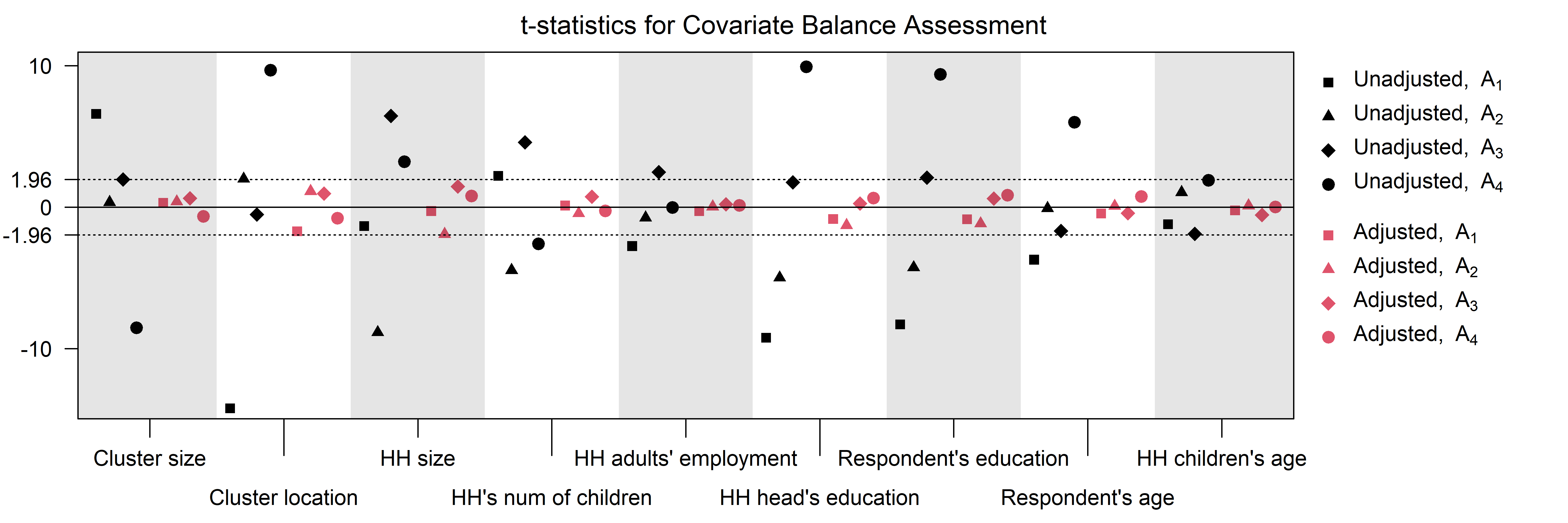}
\caption{Covariate Balance Assessment}	
\label{fig:Balance}
\end{figure}

\item[(A3)] Next, we assess the overlap assumption based on Algorithm \ref{alg:Overlap}.  Again, we use the median of the propensity score estimates from 100 cross-fitting procedures. 

\begin{algorithm}[!htb]
\begin{algorithmic}[1]
\STATE Divide $\sum_{i=1}^{\NC} \NI_i = 13556$ units into four groups:
\begin{align*}
& \mathcal{A}_k = \big\{ (i,j) \cond (A_{ij}, \mA_\eij) \in R_k \big\}
\ , \
R_k
=
\begin{cases}
\{ 0 \} \times [0, \alpha_0] \in \{0,1\} \times [0,1] & \text{ if } k=1
\\
\{ 0 \} \times (\alpha_0, 1] \in \{0,1\} \times [0,1] & \text{ if } k=2
\\
\{ 1 \} \times [0, \alpha_1] \in \{0,1\} \times [0,1] & \text{ if } k=3
\\
\{ 1 \} \times (\alpha_1, 1] \in \{0,1\} \times [0,1] & \text{ if } k=4
\end{cases}    		
\end{align*}
where $\alpha_0$ and $\alpha_1$ are chosen so that $\mathcal{A}_1, \ldots, \mathcal{A}_4$ have similar sizes.

\STATE Calculate the median of the estimated propensity scores obtained from multiple cross-fitting procedures, i.e.,
\begin{align*}
\widehat{e}_{ij,k}^{(\median)} =
\median_{s =1,\ldots,S}
\widehat{P}^{(s)} \big\{ (A_{ij}, \mA_{\eij}) \in R_k \cond \bX_{ij}, \bX_{\eij}  \big\}
\end{align*}
where the conditional probability $\widehat{P}^{(s)}$ is calculated from the estimated propensity score obtained from the $s$th cross-fitting procedure.

\STATE Compare histograms of $\widehat{e}_{ij,k}^{(\median)}(\bX_i)$ for $\mathcal{A}_k$ and $\mathcal{A}_k^c$ for each $k$.
\end{algorithmic}
\caption{Assessment of Overlap}
\label{alg:Overlap}
\end{algorithm}    

Figure \ref{fig-A-overlap} shows histograms that visually assess the overlap assumption. Based on the histograms, the overlap assumption seems to be satisfied or to be not severely violated. 

\begin{figure}[!htb]
\centering
\includegraphics[width=1\textwidth]{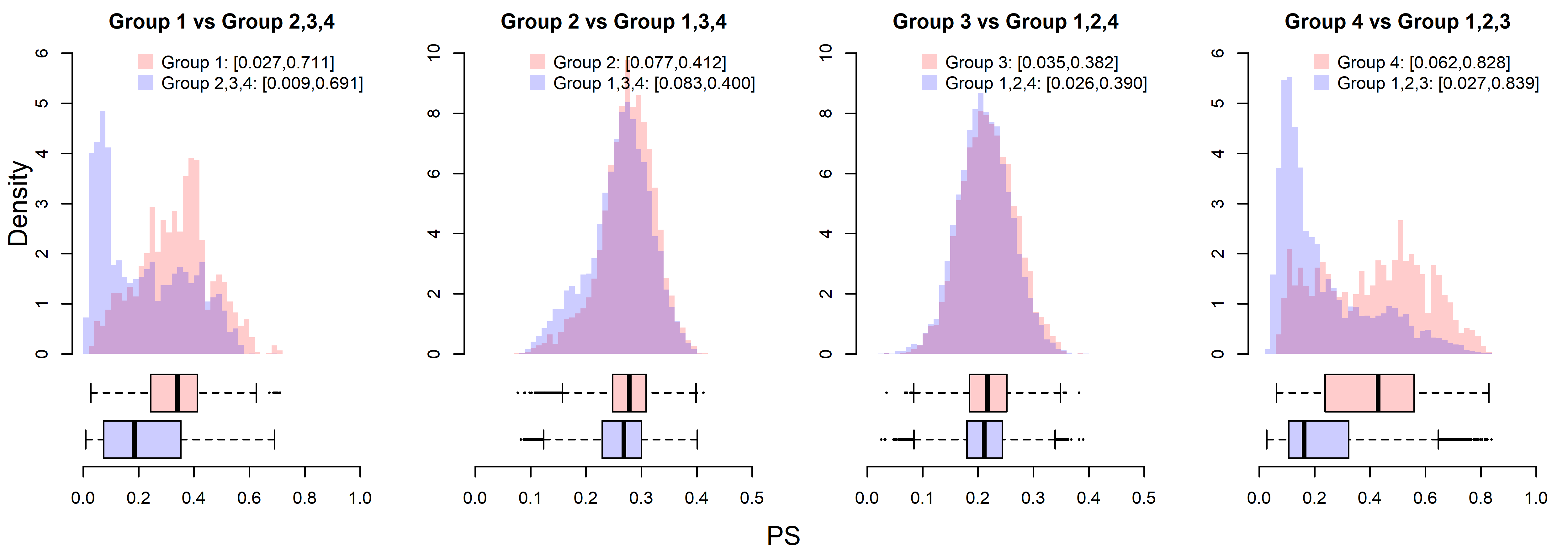}
\caption{Overlap Assessment. The numbers in brackets show the range of the estimated propensity scores for each group.}	
\label{fig-A-overlap}
\end{figure}		

Overall, all 9 observed covariates are balanced across different bins of treatment values and overlap is reasonable.

\item[(A4)] Assumption \hyperlink{(A4)}{(A4)} is plausible if the number of diarrhea-free children in a household can be reasonably approximated by a summary of peers' WASH status. However, the assumption may fail if a few households' presence (or absence) of WASH facilities is driving the incidence of diarrhea in the entire block, say if a few WASH-less households are located near communal water sources and they are primarily responsible for the diarrhea in the entire block. For example, if the census block has 20 households and the true response model for each household is $\EXP (Y_{ij} \cond \bA_i, \bX_i) = \beta_0 + \beta_1 A_{i1} + \beta_2 A_{ij} + \bm{\beta}_3\T \bX_{ij}$, i.e., every household $j$'s outcome depends on household $1$'s treatment status,  then $\EXP(\mY_i \cond \bA_i, \bX_i) = \beta_0 + \beta_1 A_{i1} + \beta_2 \mA_i + \bm{\beta}_3\T \mbX_i$ and Assumption \hyperlink{(A4)}{(A4)} is violated because the average response of block $i$ depends on the treatment status of household $1$.  Unfortunately, the data does not contain information about the location of households to test these hypothesized violations of Assumption \hyperlink{(A4)}{(A4)}. Instead, we visually diagnose the assumption by using a residual plot of the predicted values of the mean block-level response versus the observed block-level response. Specifically, let $\widehat{\epsilon}_{ij}^{(\text{median})} = Y_{ij} - \widehat{\mu}^{(\text{median})} (A_{ij}, \mA_\eij, \bX_{ij}, \bX_\eij)$ be the residuals where $\widehat{\mu}^{(\text{median})}$ is the median of the outcome regression from 100 cross-fitting procedures. We compare the residuals across the outcome regression estimate and the regressors $(A_{ij}, \mA_\eij, \bX_{ij}, \bX_\eij)$ and check whether the residuals deviate from zero in Figure \ref{fig-residual}. Since the dimension of $\bX_\eij$ varies, we use the average of $\bX_{\eij}$, i.e., $\overline{\bX}_\eij = \sum_{\ell \neq j } \bX_{i\ell}/(\NI_i-1)$. In general, the residuals are close to zero across the regressors, implying that the outcome regression under Assumption \hyperlink{(A4)}{(A4)} is not severely violated.  That is, while the diagnostic is not perfect, we find the predicted means do not show trends across the $x$-axis and Assumption \hyperlink{(A4)}{(A4)} could be plausible, subject to inherent limitations of the diagnostic plot.

\begin{figure}[!htb]
\centering
\includegraphics[width=\textwidth]{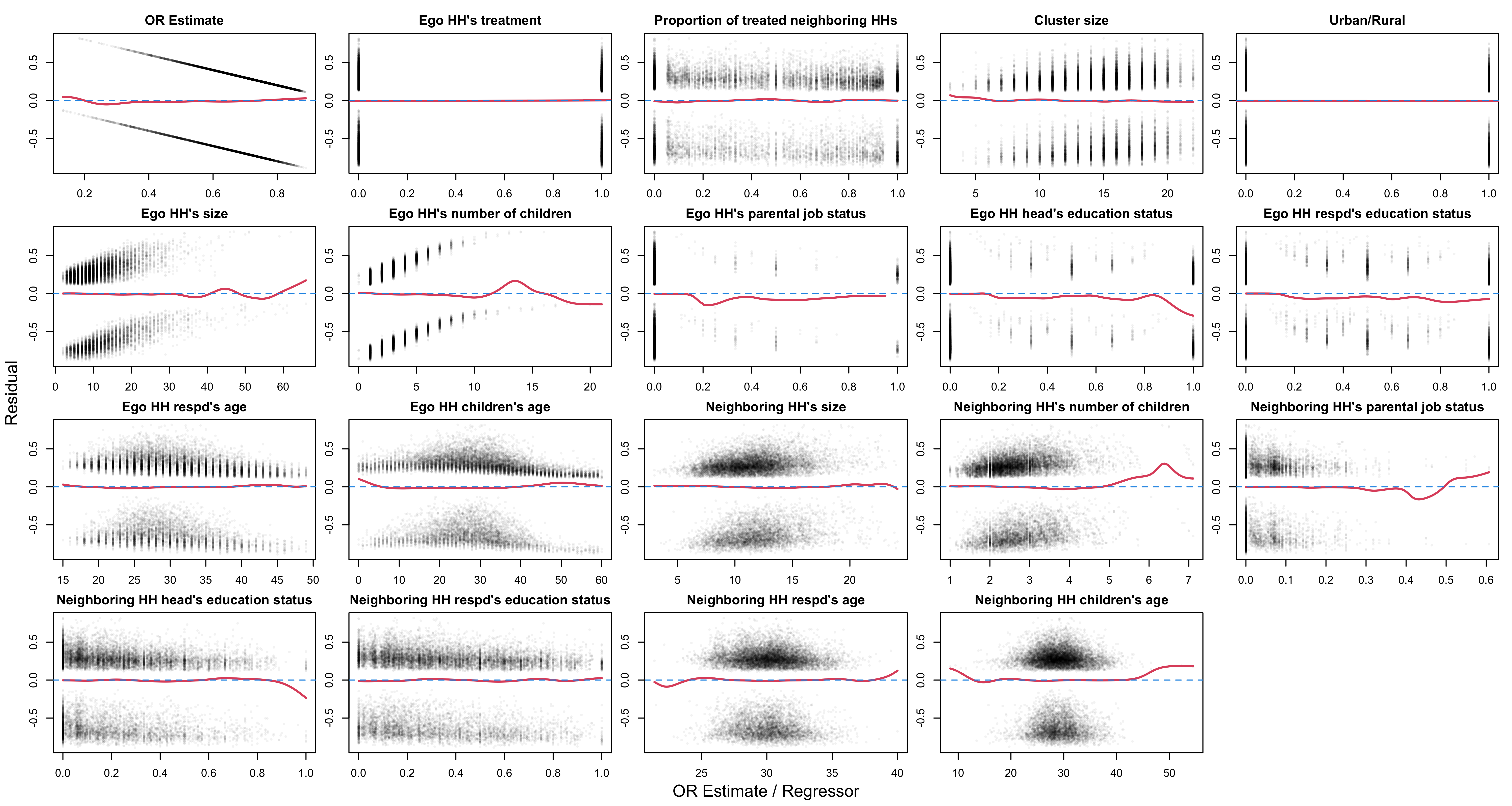}
\caption{Residual Plots. The $x$-axis shows the outcome regression estimate $\widehat{\mu}^{(\text{median})}$ (top left) and the regressors $(\mA_i, \mbX_i)$. The $y$-axis shows the residuals $\widehat{\epsilon}_i^{(\text{median})}$. The red curves are smoothing lines drawn for visual guidance. The blue dashed lines show the zero residual.}	
\label{fig-residual} 
\end{figure}

\item[(A5)] 
Finally, for Assumption \hyperlink{(A5)}{(A5)}, many prior works \citep{Diarrhea_New5, Diarrhea_New6, Diarrhea_New3, WASH_Effective2015, McMichael2019} suggest that installing WASH facilities will not have a negative impact on incidence of diarrhea; however, it may have a negative effect on other, non-health outcomes. Also, when we empirically assess Assumption \hyperlink{(A5)}{(A5)}, we find that the monotonicity assumption is rarely violated in the Senegal DHS and if violated, the deviation from monotonicity is small. Specifically, we first consider the difference between two cluster-level outcome regressions:
\begin{align*}
& 
V(a,a',s,s')
= 
\overline{\mu} \bigg(	 a' , \frac{s'}{\NI_i-1} , \bX_i \bigg) - \overline{\mu} \bigg( a , \frac{s}{\NI_i-1} , \bX_i \bigg)
\ , \
\overline{\mu} \big( a, a' , \bX_i \big)
=
\frac{1}{\NI_i}
\sum_{j=1}^{\NI_i}
\mu \big( a, a' , \bX_{ij}, \bX_\eij \big) \ .
\end{align*}
In particular, we focus on the variations contrasting two adjacent outcome regressions. As a consequence, there are $3\NI_i-2$ finest variations where $\NI_i-1$ variations have the form $V(0, 0, s, s+1)$, $\NI_i-1$ variations have the form $V(1, 1, s, s+1)$, and $\NI_i$ variations have the form $V(0,1,s,s)$; see the diagram below.
\begin{align*}
\begin{array}{ccccccc}
\overline{\mu} \big( 0, \frac{0}{\NI_i-1}, \bX_i \big) 
&
\stackrel{V_{i}(0, 0, 0,1)}{\rightarrow}
& \overline{\mu} \big( 0, \frac{1}{\NI_i-1}, \bX_i \big)
&
\stackrel{V_{i}(0 , 0 , 1, 2)}{\rightarrow}
&
\cdots
& 
\quad \quad \quad
\overline{\mu} \big( 0, \frac{\NI_i-1}{\NI_i-1}, \bX_i \big)
\\[0.2cm]
^{V_{i} (0,1,0, 0)} \downarrow
&
&
^{V_{i} (0,1,1, 1)} \downarrow 
& & &
^{V_{i} (0,1,\NI_i-1,\NI_i-1)} \downarrow 
\\
\overline{\mu}  \big( 1, \frac{0}{\NI_i-1}, \bX_i \big) 
&
\stackrel{V_{i}(1, 1, 0 , 1)}{\rightarrow}
& \overline{\mu} \big( 1, \frac{1}{\NI_i-1}, \bX_i \big) 
&
\stackrel{V_{i}(1 , 1 , 1, 2)}{\rightarrow}
&
\cdots
&
\quad \quad \quad
\overline{\mu} \big( 1, \frac{\NI_i-1}{\NI_i-1}, \bX_i \big) 
\end{array}
\end{align*} 	
In the Senegal DHS, we have $\sum_{i=1}^\NC (3\NI_i-2) = 38614$ variations in total. Let $\widehat{V}_{i}^{(t)} (a,a',s,s')$ be the estimated variation of cluster $i$ obtained from the $t$th cross-fitting procedure, and let $\widehat{V}_{i}^{(m)} (a,a',s,s')$ be the median of the variation, i.e.,
\begin{align*}
\widehat{V}_{i}^{(m)} (a,a',s,s')
=
\median
\Big\{ \widehat{V}_{i}^{(1)} (a,a',s,s') ,
\ldots, 
\widehat{V}_{i}^{(S)} (a,a',s,s') \Big\}
\end{align*}

Assumption \hyperlink{(A5)}{(A5)} can be empirically assessed by two means. First, out of 38614 variations, we count the number of times monotonicity is violated. Second, we measure the worst-case slope of the estimated $\mu$ as follows. Let $TV_i(a,a',s,s')$ be the absolute value of $V_i(a,a',s,s')$. Thus, the sum of 38614 $TV_i(a,a',s,s')$ is the total variation of the cluster-level outcome regression. We compute the relative magnitude of the slopes that are decreasing compared to the total variation, i.e., $\sum \ind( V < 0 )  TV  / \sum TV $. Overall, under the first assessment, we found that the monotonicity is violated $1.11\%$ of the time and under the second assessment, the relative magnitude of decreasing slopes is $6.01 \times 10^{-4}$. In short, the empirical validations show that the monotonicity assumption is rarely violated in the Senegal DHS and if violated, the deviation from monotonicity is small.

\end{itemize}

\subsection{Details of Figures \ref{fig-Data-2}-\ref{fig-Data-4} of the Main Paper}							\label{sec:HeatMapDetail}

We additionally describe how we draw Figures \ref{fig-Data-2}-\ref{fig-Data-4} of the main paper. The reported estimated {\OURR}s in Figures \ref{fig-Data-2} and \ref{fig-Data-4} are weighted average of the estimated {\OURR}s in each administrative region where weights are the number of households in a census block, i.e., census block size $\NI_i$. That is, the values represent $\overline{\LB}_g$s, which are defined as
\begin{align*}
\overline{\LB}_g
=
\frac{\sum_{i \in \mathcal{D}_{2018}} \ind \{ i \in \text{administrative area } g \} \cdot \NI_i \cdot \widehat{\LB}(\bx_i)}{\sum_{i \in \mathcal{D}_{2018}} \ind \{ i \in \text{administrative area } g \} \cdot \NI_i }
\end{align*}
where $\mathcal{D}_{2018}$ is the collection of census blocks in the 2018 Senegal DHS. In words, $\overline{\LB}_g$ is the proportion of households in administrative area $g$ that require WASH facilities. Similarly, the average household sizes in Figure \ref{fig-Data-4} represent 
\begin{align*}
\overline{\bar{x}}_{g, {\rm Household\, Size}}
&
=
\frac{\sum_{i \in \mathcal{D}_{2018}} \ind \{ i \in \text{administrative area } g \} \cdot \NI_i \cdot \bar{x}_{i,{\rm Household\, Size}}  }{\sum_{i \in \mathcal{D}_{2018}} \ind \{ i \in \text{administrative area } g \} \cdot \NI_i }
\\
&
=
\frac{\sum_{i \in \mathcal{D}_{2018}} \ind \{ i \in \text{administrative area } g \} \sum_{j=1}^{\NI_i} x_{ij,{\rm Household\, Size}}  }{\sum_{i \in \mathcal{D}_{2018}} \ind \{ i \in \text{administrative area } g \} \cdot \NI_i } \ .
\end{align*}
Again, $\overline{\bar{x}}_{g, {\rm Household\, Size}}$ is the average household wise in administrative area $g$. The proportions of rural area in Figure \ref{fig-Data-4} represent
\begin{align*}
\overline{ c }_{g,{\rm Rural}}
&
=
\frac{\sum_{i \in \mathcal{D}_{2018}} \ind \{ i \in \text{administrative area } g \} \cdot c_{i,{\rm Rural}}  }{\sum_{i \in \mathcal{D}_{2018}} \ind \{ i \in \text{administrative area } g \} } \ . 
\end{align*} 
Here $\overline{c}_{g, {\rm Rural}}$ is the proportion of the rural census blocks in administrative area $g$. Note that $\overline{\LB}_g$, $\overline{\bar{x}}_{g, {\rm Household\, Size}}$, and $\overline{ c }_{g,{\rm Rural}}$ do not address the geographical distance between census regions in different administrative areas. But, we believe that these statistics are geographically meaningful summaries to highlight the heterogeneity across administrative areas; see Figures 1 and 2 of \citet{Houngbonon2021} for similar summary statistics aggregated at Senegalese administrative areas.

Lastly, Figure \ref{fig-Data-3} shows the weighted average of the estimated {\OURR}s across all 45 administrative areas where weights are the number of households in a census block, i.e., census block size $\NI_i$. That is, the $y$-axis represents $\overline{\LB}$, which is defined as
\begin{align*}
\overline{\LB}
=
\frac{\sum_{i \in \mathcal{D}_{2018}} \NI_i \cdot \widehat{\LB}(\bx_i)}{\sum_{i \in \mathcal{D}_{2018}} \NI_i } \ .
\end{align*}
In words, $\overline{\LB}$ is the proportion of households in Senegal that require WASH facilities.

\section{Proof of Lemmas and Theorems}								\label{sec:B}

\subsection{Useful Lemmas}

\begin{lemma} 		\label{thm-supp-ER-1}
Suppose that $\tLB^*$ belongs to a Besov space on $\R^d$ with smoothness parameter $\beta>0$, i.e., $\mathcal{B}_{1,\infty}^\beta (\R^d) = \{ \tLB \in L_{\infty}(\R^d) \cond \sup_{t>0} t^{-\beta} \{ \omega_{r, L_1(\R^d) } (\tLB,t) \}	< \infty	,  r > \beta \}$ where $\omega_r$ is the modulus of continuity of order $r$. 
Then, for any positive $\epsilon, p, \tau$ satisfying $d/(d+\tau) < p < 1$, we have the following excess risk bound of $\widehat{\LB}$ with probability not less than $1-3e^{-\tau}$:
\begin{align*}
\risk \big( \widehat{\LB} \big) - \risk \big( \LB^* \big)
\leq
c_1 \lambda_\NC\gamma_\NC^{-d} + c_2 \gamma_\NC^\beta 
+ c_3 \Big\{ \gamma_\NC^{(1-p)(1+\epsilon)d} \lambda_\NC^p \NC \Big\}^{-\frac{1}{2-p}} 
+ c_4 \NC^{-1/2}\tau^{1/2} + c_5\NC^{-1} \tau 
\end{align*}
\end{lemma}
%		The first two terms in the excess risk bound from Theorem \ref{thm-supp-ER-1} correspond to the approximation error from using the RKHS $\RKHS_\kernel$ to obtain $\LB^*$. The last three terms correspond to the stochastic error from using finite amount of samples to train the optimal policy. If parameters $\gamma_\NC$ and $\lambda_\NC$ are carefully chosen, the upper bound of the excess risk converges to zero as $\NC$ grows. Specifically, if we take  to minimize the leading order of the excess risk bound, the excess risk bound becomes $\risk( \widehat{\LB} ) - \risk (\LB^*) = O_P ( \NC^{- \beta/(2\beta+d) } )$.

\begin{lemma}							\label{lem-EstimatedL}
Let $\widehat{\risk}_{(-\ell)}(\LB) = \EXP \big\{ \widehat{\loss}_{(-\ell)} \big( \LB(\bX_i), \bO_i \big) \cond \mathcal{D}_\ell^c \big\} $ be the estimated risk function where the expectation is taken with respect to $\bO_i$ while $\widehat{\loss}_{(-\ell)}$ is considered as a fixed function which is clarified by denoting $\mathcal{D}_\ell^c$ in the conditioning statement.  Let $\F^{[0,1]} = \big\{ f \cond f(\bx_i ) \in [0,1] \big\}$ be the collection of policies ranging over the unit interval. Under Assumption \hyperlink{(A1)}{(A1)}-\hyperlink{(A5)}{(A5)} and \hyperlink{(E1)}{(E1)}-\hyperlink{(E3)}{(E3)} of the main paper, we have $\big| \risk \big( \LB \big) - \widehat{\risk}_{(-\ell)} \big( \LB \big)  \big| \leq 0.5 c_6 r_\NC $ with probability greater than $1-\Delta_N$ for any $\LB \in \F^{[0,1]}$ where $c_6$ is a fixed constant, $r_\NC = r_{\PS,\NC}$ if the inverse probability-weighted loss function is used, $r_\NC = r_{\mOR,\NC}$ if the outcome regression loss function is used, and $r_\NC = r_{\PS,\NC}r_{\mOR,\NC}$ if the doubly robust loss function is used.
\end{lemma}

See Sections \ref{sec:proof:31} and \ref{proof-lem-EstimatedL} for the proof.

\subsection{Proof of Lemma \ref{lem-minimizer2}} \label{sec:supp:ProofLemma}

We only show the result about the overall outcome case because the result about the spillover outcome case is obtained from a similar manner.   Let $\F^{[0,1]} = \big\{ f \cond f(\bx_i ) \in [0,1] \big\}$ be the collection of policies ranging over the unit interval and $\winsor(\LB) \in \F^{[0,1]}$ be the winsorized function of $\LB \in \F$ over the unit interval, i.e.
\begin{align*}
\winsor(\LB) (\bx_i) 
= 
0 \cdot \ind \big\{ \LB (\bx_i) < 0 \big\}
+ \LB (\bx_i) \cdot \ind \big\{ 0 \leq \LB (\bx_i) \leq 1 \big\}
+ 1 \cdot \ind \big\{ 1 < \LB (\bx_i) \big\} \ .
\end{align*}
From the definition of $\loss$, we find $\loss(0, \bO_i) \leq \loss(t, \bO_i)$ for any $t \in (-\infty, 0)$ and $\loss(1, \bO_i) \leq \loss(t, \bO_i)$ for any $t \in (1,\infty)$. As a consequence, for any policy $\LB \in \F$ satisfies $\loss \big( \winsor(\LB)(\bX_i) , \bO_i \big) \leq \loss \big( \LB(\bX_i) , \bO_i \big)$ and $\risk (\winsor(\LB)) \leq \risk (\LB)$. This implies that $\LB'$, the minimizer of $\risk$, must belong to $\F^{[0,1]}$. 

For any function $\LB \in \F^{[0,1]}$, we find $\loss \big( \LB(\bX_i) , \bO_i \big) = \losszo \big( \LB(\bX_i) , \bO_i \big)$ and $\risk (\LB) = \risk^{[0,1]} (\LB) = \EXP \big\{ \losszo \big( \LB(\bX_i), \bO_i \big) \big\}$	 due to the constructions of $\loss$ and $\risk$. Combining the above results, we observe the following relationship.
\begin{align*}
\argmin_{\LB \in \F} R(\LB)
=
\argmin_{\LB \in \F^{[0,1]}} R(\LB)
=
\argmin_{\LB \in \F^{[0,1]}} R^{[0,1]} (\LB)
\end{align*}
Thus, it suffices to show that $\LB^*$ defined in \eqref{eq-def:lowerB} of the main paper minimizes $R^{[0,1]}$, which is represented as follows.	
{\footnotesize
\begin{align}
&
\risk^{[0,1]}(\LB)  - C_0
\nonumber
\\
&
= \EXP \big\{ \losszo \big( \LB(\bX_i), \bO_i \big) \big\}
- C_0
\nonumber
\\
&
=
\EXP \Bigg[
\frac{1}{\NI_i}
\sum_{j=1}^{\NI_i}
\sum_{a=0}^{1}
\sum_{s=0}^{\NI_i-1} {\NI_i - 1 \choose s} \psi_{\text{DR}} (a, s, \bO_{ij}, \bO_\eij) 
\sum_{\ell=0}^{\NI_i - a - s} {\NI_i - a - s \choose \ell} \frac{ (-1)^{\ell}  \big\{  \LB(\bX_i) \big\} ^{\ell + a + s + 1} }{\ell + a + s + 1} - \thr \big\{  \LB(\bX_i) \big\}
\Bigg]  
\nonumber
\\
&
=
\EXP \Bigg[
\int_{0}^{1} \bigg[
\frac{1}{\NI_i}
\sum_{j=1}^{\NI_i}
\sum_{a=0}^{1}
\sum_{s=0}^{\NI_i-1} {\NI_i - 1 \choose s}
\EXP \big\{ \psi_{\text{DR}} (a, s , \bO_{ij}, \bO_{\eij} ) \cond \bX_i \big\} \alpha^s (1-\alpha)^{\NI_i-a - s} - \thr
\bigg]
\ind \big\{ \alpha \leq \LB(\bX_i) \big\}
\, d \alpha
\Bigg]   
\nonumber
\\
&
=
\EXP \Bigg[
\int_{0}^{1} 
\bigg\{
\frac{1}{\NI_i}
\sum_{j=1}^{\NI_i}
\sum_{a=0}^{1}
\sum_{s=0}^{\NI_i-1} {\NI_i - 1 \choose s}
\mu^* \bigg( a,  \frac{s}{\NI_i-1}, \bX_{ij}, \bX_\eij \bigg) \alpha^s (1-\alpha)^{\NI_i-a - s}  - \thr
\bigg\}
\ind \big\{ \alpha \leq \LB(\bX_i) \big\}
\, d \alpha
\Bigg]  
\label{eq:EqLemma-1}
\ .
\end{align}}%
The first and second identities are trivial from the definition of $\risk^{[0,1]}$ and $\losszo$. The third identity is from the law of iterated expectation and the following algebra:
\begin{align*}
&
H(\bO_i)
\sum_{\ell=0}^{\NI_i - a - s} {\NI_i - a - s \choose \ell} \frac{ (-1)^{\ell}  t^{\ell + a + s + 1} }{\ell + a + s + 1} - \thr t
\\
& =
\int_{0}^{t} \Big\{
H(\bO_i)
\alpha^{s} (1-\alpha)^{\NI_i-a-s} - \thr
\Big\} \, d\alpha
\\
&
=
\int_{0}^{1} \Big\{
H(\bO_i)
\alpha^{s} (1-\alpha)^{\NI_i-a-s} - \thr
\Big\}
\ind\{ \alpha \leq t \}
\, d\alpha
\ , \ ^\forall H(\bO_i) \ , \ ^\forall t \in [0,1]\ . 
\end{align*}
The fourth identity is from $\EXP \big\{ \psi_{\text{DR}} (a , s , \bO_{ij}, \bO_\eij ) \cond \bX_i \big\} = \mu^*(a, \frac{s}{\NI_i-1}, \bX_{ij}, \bX_{\eij})$ for $s=0,1,\ldots,\NI_i-1$; we remark that any $\psi'$ satisfying  $\EXP \big\{ \psi' (a , s , \bO_{ij}, \bO_\eij ) \cond \bX_i \big\} = \mu^*(a, \frac{s}{\NI_i-1}, \bX_{ij}, \bX_{\eij})$ (e.g., inverse probability-weighted or outcome regression-based) can be used instead of $\psi_{\text{DR}}$. From the monotonicity condition \hyperlink{(A5)}{(A5)}, it is straightforward to check that the following sets are intervals if they are non-empty:
\begin{align*}
&
\mathcal{S}_- (\bX_i)
:=
\Bigg\{ \alpha \, \Bigg| \, 
\frac{1}{\NI_i}
\sum_{j=1}^{\NI_i}
\sum_{a=0}^{1}
\sum_{s=0}^{\NI_i-1} {\NI_i - 1 \choose s}
\EXP \big\{ \psi_{\text{DR}} (a, s , \bO_{ij}, \bO_{\eij} ) \cond \bX_i \big\} \alpha^s (1-\alpha)^{\NI_i -a - s}
< \thr \Bigg\}
\\
&
\mathcal{S}_+ (\bX_i)
:=
\Bigg\{ \alpha \, \Bigg| \, 
\frac{1}{\NI_i}
\sum_{j=1}^{\NI_i}
\sum_{a=0}^{1}
\sum_{s=0}^{\NI_i-1} {\NI_i - 1 \choose s}
\EXP \big\{ \psi_{\text{DR}} (a, s , \bO_{ij}, \bO_{\eij} ) \cond \bX_i \big\} \alpha^s (1-\alpha)^{\NI_i -a - s}
\geq \thr \Bigg\} \ .
\end{align*}
Since $\mathcal{S}_-(\bX_i)$ and $\mathcal{S}_+(\bX_i)$ are non-overlapping intervals, we establish that $\sup \mathcal{S}_-(\bX_i)$ and $\inf \mathcal{S}_+ (\bX_i)$ are equivalent. If $\mathcal{S}_-(\bX_i)$ and $\mathcal{S}_+(\bX_i)$ are empty, we define $\mathcal{S}_-(\bX_i)=\{0\}$ and $\mathcal{S}_+(\bX_i)=\{1\}$, respectively. In these cases, we also establish that $\sup \mathcal{S}_-(\bX_i)$ and $\inf \mathcal{S}_+ (\bX_i)$ are equivalent as $0$ or $1$, respectively. 	We remark that, without the monotonicity condition \hyperlink{(A5)}{(A5)}, these two sets may be disconnected sets, i.e., not intervals, and we cannot establish that $\sup \mathcal{S}_-(\bX_i)$ and $\inf \mathcal{S}_+ (\bX_i)$ are equivalent. 

The last representation \eqref{eq:EqLemma-1} suggests that
that $\risk^{[0,1]}(\LB)$ is minimized at $\LB'$ where
$\LB'(\bX_i) = \sup \mathcal{S}_-(\bX_i) = \inf \mathcal{S}_+ (\bX_i)$, i.e.,  
\begin{align*}
& 
\frac{1}{\NI_i}
\sum_{j=1}^{\NI_i}
\sum_{a=0}^{1}
\sum_{s=0}^{\NI_i-1} {\NI_i - 1 \choose s}
\EXP \big\{ \psi_{\text{DR}} (a, s , \bO_{ij}, \bO_{\eij} ) \cond \bX_i \big\} \alpha^s (1-\alpha)^{\NI_i -a - s}
\leq \thr \ \text{ for all }  \alpha \in \big[ 0, \LB'(\bX_i) \big] \ ,
\\
& 
\frac{1}{\NI_i}
\sum_{j=1}^{\NI_i}
\sum_{a=0}^{1}
\sum_{s=0}^{\NI_i-1} {\NI_i - 1 \choose s}
\EXP \big\{ \psi_{\text{DR}} (a, s , \bO_{ij}, \bO_{\eij} ) \cond \bX_i \big\} \alpha^s (1-\alpha)^{\NI_i -a - s}
\geq \thr  \ \text{ for all }  \alpha \in \big[ \LB'(\bX_i) , 1 \big] \ .
\end{align*}
As a consequence, $\LB'$ agrees with $\LB_{\OV}^*$ defined in \eqref{eq-def:lowerB} of the main paper. We can establish the result about $\LB_{\SO}^*$ by fixing $a=0$. This concludes the proof.

\subsection{Proof of Lemma \ref{thm-supp-ER-1}}							\label{sec:proof:31}

We only show the result about the overall outcome case because the result about the spillover outcome case is obtained from a similar manner.    The proof of \ref{thm-supp-ER-1} is similar to that of Theorem 2 of \citet{Chen2016} which use Theorem 7.23 of \citet{SVM} and Theorem 2.2 and 2.3 of \citet{SVM2013}. For completeness, we present a full exposition to our setting below. We first introduce Theorem 7.23 of \citet{SVM}. \\

\noindent \textbf{Theorem 7.23. }(\textit{Oracle Inequality for SVMs Using Benign Kernels}; \citet{SVM}) Let $\loss$ be a loss function having non-negative value. Also, let $\RKHS_\kernel$ be a separable RKHS of a measurable kernel $\kernel$ over $\X = \text{supp}(\tbX_i) \subset \R^d$. Let $P$ be a distribution on $\bO_i$. Furthermore, suppose the following conditions are satisfied.
\begin{itemize}
\item[\hypertarget{(C1)}{(C1)}] For all $(t,\bo_i)$, there exists a constant $B>0$ satisfying $\loss(t, \bo_i) \leq B$.
\item[\hypertarget{(C2)}{(C2)}] $\loss(t,\bo_i)$ is locally Lipschitz continuous with respect to $t$.
\item[\hypertarget{(C3)}{(C3)}] For all $(t,\bo_i)$, we have $\loss( \winsor_{c_0}(t),\bo_i) \leq \loss(t,\bo_i)$ where $ \winsor_{c_0}(t) =  t \cdot \ind \big\{ | t | \leq c_0 \big\} + \text{sign}(t) c_0 \cdot \ind \big\{ c_0 < | t | \big\}$.
\item[\hypertarget{(C4)}{(C4)}] $\EXP \big[ \big\{ \loss \big(  \winsor_{c_0}(\tLB) (\tbX_i), \bO_i \big) - \loss \big( \tLB^*(\tbX_i), \bO_i \big) \big\}^2 \big] \leq V \cdot \big[ \EXP \big\{  \loss \big(\winsor_{c_0}(\tLB)  (\tbX_i), \bO_i \big) - \loss \big( \tLB^*(\tbX_i), \bO_i \big) \big\} \big]^v$ is satisfied for constant $v \in [0,1]$, $V \geq B^{2-v}$, and for all $\tLB \in \RKHS_\kernel$. 
\item[\hypertarget{(C5)}{(C5)}] For fixed $\NC \geq 1$, there exists constants $p \in (0,1)$ and $a \geq B$ such that the dyadic entropy number $\EXP_{D_X \sim P_X^\NC} \big[ e_i \big( \text{identity map}: \RKHS_\kernel \rightarrow L_2(D_{\tbX}) \big) \big] \leq a \cdot i^{-\frac{1}{2p}}$ $(i \geq 1)$ where $e_i(A)$ is the entropy number of $A$.
\end{itemize}
We fix $\tLB_0 \in \RKHS_\kernel$ and a constant $B_0 \geq B$ such that $\loss\big( \tLB_0(\tbx_i), \bo_i \big) \leq B_0$ for any $\bo_i$. Then, for all fixed $\tau>0$ and $\lambda_\NC$, the SVM using $\RKHS_\kernel$ and $\loss$ satisfies
\begin{align}							\label{eq-SVM-proof1}
&
\lambda_\NC \big\| \LB \big\|_{\RKHS_\kernel}^2 + \trisk \big( \winsor_{c_0}( \LB ) \big) - \trisk \big( \tLB^* \big)
\\
&
\leq
9
\big\{ \lambda_\NC \big\| \tLB_0 \big\|_{\RKHS_\kernel}^2 + \trisk \big( \tLB_0 \big) - \trisk \big( \tLB^* \big) \big\}
+
K_0 \bigg( \frac{a^{2p}}{\lambda_\NC^p \NC} \bigg)^{\frac{1}{2-p-v+vp}}
+
3 \bigg( \frac{72 V \tau}{\NC} \bigg)^{\frac{1}{2-v}}
+
\frac{15 B_0 \tau}{\NC} \ .
\nonumber
\end{align}
with probability $P^\NC$ not less than $1-3e^{-\tau}$, where $K_0 \geq 1$ is a constant only depending on $p$, $c_0$, $B$, $v$, and $V$. \\

We verify Assumptions \hyperlink{(C1)}{(C1)}-\hyperlink{(C5)}{(C5)} as follows:\\	

\noindent \textbf{Verification of Assumption \hyperlink{(C1)}{(C1)}}: From Assumptions \hyperlink{(A1)}{(A1)}-\hyperlink{(A5)}{(A5)} and $\mY_i \in [0,1]$, we find that $\psi_\text{IPW}$, $\psi_\text{OR}$, and $\psi_\text{DR}$ are bounded and, as a consequence, $\losszo$ in \eqref{eq-L01} of the main paper is bounded. As a result, $\loss$ in \eqref{eq-L} of the main paper is bounded as well. \\

\noindent \textbf{Verification of Assumption \hyperlink{(C2)}{(C2)}}: We find the derivative of $\loss$ in \eqref{eq-L} of the main paper is
\begin{align*}
\nabla \loss(t, \bo_i)
=
\begin{cases}
\delta e^t 
&
t \in (-\infty,0)
\\
\sum_{s=0}^{\NI_i} \psi_\ell (s, \bo_i) t^s(1-t)^{\NI_i-s} - \thr
&
t \in (0,1)
\\
\delta e^{1-t}
&
t \in (0,\infty)
\end{cases}
\end{align*}
and we find that $\nabla \loss(t, \bO_i)$ is bounded for all $t$ except $t=0,1$. Moreover, $\loss(t, \bo_i)$ is continuous at $t=0$ and $t=1$. Thus, $\loss(t,\bo_i)$ is locally Lipschitz continuous with Lipschitz constant $B' = \sup_{(t,\bo_i)} \nabla \loss(t, \bo_i)$. \\

\noindent \textbf{Verification of Assumption \hyperlink{(C3)}{(C3)}}: We take $c_0=1$. It is trivial that $\loss ( \winsor_{c_0}(\tLB), \bo_i) \leq \loss(\tLB, \bo_i)$ from the form of $\loss$ in \eqref{eq-L} of the main paper. \\

\noindent \textbf{Verification of Assumption \hyperlink{(C4)}{(C4)}}: Note that $\loss(t, \bo_i) \leq B$. Thus, we find
\begin{align*}
\EXP \Big[ \big\{ \loss \big( \winsor_{c_0}(\tLB) (\tbX_i), \bO_i \big) - \loss \big( \tLB^*(\tbX_i), \bO_i \big) \big\}^2 \Big]
\leq
2 \EXP \Big[ \big\{ \loss \big( \winsor_{c_0}(\tLB) (\tbX_i), \bO_i \big) \big\}^2 + \big\{ \loss \big( \tLB^*(\tbX_i), \bO_i \big) \big\}^2 \Big] 
\leq 4B^2 \ .
\end{align*}
We take $v=0$ and $V=4B^2$ and the condition is satisfied. \\

\noindent \textbf{Verification of Assumption \hyperlink{(C5)}{(C5)}}: Since we use the Gaussian kernel, we can directly use Theorem 7.34 of \citet{SVM} which is given below. \\

\noindent \textbf{Theorem 7.34. }(\textit{Entropy Numbers for Gaussian Kernels}; \citet{SVM}) Let $\nu$ be a distribution on $\R^d$ having tail exponent $\tau \in (0,\infty]$. Then, for all $\epsilon > 0$ and $d/(d+\tau) < p < 1$, there exists a constant $c_{\epsilon,p} \geq 1$ such that
\begin{align*}
e_i \big( \text{identity map}: \RKHS_\kernel \rightarrow L_2(\nu) \big) 
\leq
c_{\epsilon,p} \gamma^{ - \frac{(1-p)(1+\epsilon)d}{2p} } i^{-\frac{1}{2p}}
\end{align*}
for all $i \geq 1$ and $\gamma \in (0, 1]$. \\

Therefore, Assumption \hyperlink{(C5)}{(C5)} holds with $a= c_{\epsilon,p} \gamma_\NC^{ - \frac{ (1-p)(1+\epsilon) d}{2p} }$.

As a consequence, the result in equation \eqref{eq-SVM-proof1} holds with $c_0=1$, $v=1$, $V=4B^2$, $B_0=B$, $a= c_{\epsilon,p} \gamma_\NC^{ - \frac{ (1-p)(1+\epsilon) d}{2p} }$. Moreover, we find $\loss(\winsor(\tLB)(\tbx_i), \bo_i) \leq \loss( \winsor_{c_0=1}(\tLB) (\tbx_i), \bo_i)$ since $\loss(0,\bo_i) \leq \loss(t,\bo_i)$ for $t \in [-1,0]$ and this leads $\trisk(\winsor(\tLB)) \leq \trisk ( \winsor_{c_0=1}(\tLB) )$. Thus, we find the following result holds with probability $P^\NC$ not less than $1-3e^{-\tau}$.
\begin{align}						\label{eq-SVM-proof2}
&
\trisk \big( \widehat{\tLB} \big) - \trisk \big( \tLB^* \big)
\\
\nonumber		
&
\leq
\trisk \big( \winsor_{c_0=1} ( \widetilde{\tLB} ) \big) - \trisk \big( \tLB^* \big)
\\
\nonumber
&
\leq
\lambda_\NC \big\| \widehat{\tLB} \big\|_{\RKHS_\kernel}^2 + \trisk \big( \winsor_{c_0=1} ( \widetilde{\tLB} ) \big) - \trisk \big( \tLB^* \big)
\\
\nonumber
& 
\leq
9
\big\{ \lambda_\NC \big\| \tLB_0 \big\|_{\RKHS_\kernel}^2 + \trisk \big( \tLB_0 \big) - \trisk \big( \tLB^* \big) \big\}
+
K_0 \bigg\{ \frac{ \gamma_\NC^{ - (1-p)(1+\epsilon) d} }{\lambda_\NC^p \NC} \bigg\}^{\frac{1}{2-p}}
+
36 \sqrt{2} B \sqrt{ \frac{\tau}{\NC} }
+
15 B \frac{ \tau}{\NC} \ .
\end{align}

The above result holds for any $\tLB_0 \in \RKHS_\kernel$, so we can further bound the approximation error $\lambda_\NC \big\| \tLB_0 \big\|_{\RKHS_\kernel}^2 + \trisk \big( \tLB_0 \big) -  \trisk \big( \tLB^* \big)$ by choosing $\tLB_0$ in a specific way which is presented in \citet{SVM2013}. We first define a function $Q_{r,\gamma} : \R^d \rightarrow \R$ as
\begin{align}								\label{eq-Qfunction}
Q_{r,\gamma} (\bm{z})
=
\sum_{j=1}^r
{r \choose j} (-1)^{1-j} \frac{1}{j^d} \bigg( \frac{2}{\gamma^2} \bigg)^{\frac{d}{2} } \kernel_{j\gamma/\sqrt{2} }(\bm{z})
\ , \
\kernel_\gamma(\bm{z}) 
=
\exp \big\{ - \gamma^2 \big\| \bm{z} \big\|_2^2 \big\}
\end{align}
for $r \in \{1,2,\ldots\}$ and $\gamma > 0$. Since the range of $\LB^*$ is bounded between $[0,1]$, we find $\tLB^* \in L_2(\R^d) \cap L_\infty(\R^d)$. Thus, we can define $\tLB_0$ by convolving $Q_{r,\gamma}$ with $\tLB^*$ as follows \citep{SVM2013}.
\begin{align*}
\tLB_0 (\tbx_i)
= \big( Q_{r,\gamma} * \tLB^* \big)(\tbx_i)
= \int_{\R^d} Q_{r,\gamma}(\tbx_i - \bm{z}) \tLB^*(\bm{z}) \, d\bm{z} \ .
\end{align*}
Next, we introduce theorem 2.2 and 2.3 of \citet{SVM2013}.\\

\noindent \textbf{Theorem 2.2. } Let us fix some $q \in [1,\infty)$. Furthermore, assume that $P_{\tbX}$ is a distribution on $\R^d$ that has a Lebesgue density $f_{\tbX} \in L_p(\R^d)$ for some $p \in [1,\infty]$. Let $\tLB: \R^d \rightarrow \R$ be such that $\tLB \in L_q(\R^d) \cap L_\infty(\R^d)$. Then, for $r \in \{1,2,\ldots\}$, $\gamma > 0$, and $s \geq 1$ with $1=s^{-1} + p^{-1}$, we have
\begin{align*}
\big\| Q_{r,\gamma} * \tLB - \tLB \big\|_{L_q(P_{\tbX})}^q 
\leq
C_{r,q} \cdot \big\| f_{\tbX} \big\|_{L_p(\R^d)} \cdot \omega_{r,L_{qs}(\R^d)}^q ( \tLB, \gamma/2)
\end{align*}
where $C_{r,q}$ is a constant only depending on $r$ and $q$. \\

\noindent  \textbf{Theorem 2.3. } Let $\tLB \in L_2(\R^d)$, $\RKHS_\kernel$ be the RKHS of the Gaussian kernel $\kernel$ with parameter $\gamma>0$, and $Q_{r,\gamma}$ be defined by \eqref{eq-Qfunction} for a fixed $r \in \{1,2,\ldots\}$. Then we have $Q_{r,\gamma} * \tLB \in \RKHS_\kernel$ with
\begin{align*}
\big\| Q_{r,\gamma} * \tLB \big\|_{\RKHS_\kernel} \leq \big( \gamma \sqrt{\pi} \big)^{ -\frac{d}{2} } \big( 2^r -1 \big) \big\| \tLB \big\| _{L_2(\R^d) } \ .
\end{align*}
Moreover, if $\tLB \in L_\infty(\R^d)$, we have $| Q_{r,\gamma} * \tLB | \leq (2^r - 1) \big\| \tLB \big\|_{L_\infty(\R^d) }$. \\

As a result, we obtain
\begin{align}						\label{eq-SVM-proof3}
&
\lambda_\NC \big\| \tLB_0 \big\|_{\RKHS_\kernel}^2 + \trisk \big( \tLB_0 \big) - \trisk \big( \tLB^* \big)
\\
\nonumber
&
=
\lambda_\NC \big\| Q_{r,\gamma_\NC} * \tLB^* \big\|_{\RKHS_\kernel}^2 + \trisk \big( Q_{r,\gamma_\NC} * \tLB^* \big) - \trisk \big( \tLB^* \big)
\\
\nonumber
& 
\leq
\lambda_\NC \big( \gamma_\NC \sqrt{\pi} \big)^{ - d } \big( 2^r -1 \big)^2 \big\| \tLB^* \big\| _{L_2(\R^d) }^2 + \trisk \big( Q_{r,\gamma_\NC} * \tLB^* \big) - \trisk \big( \tLB^* \big)
\\
\nonumber
& 
\leq
\lambda_\NC \big( \gamma_\NC \sqrt{\pi} \big)^{ - d } \big( 2^r -1 \big)^2 \big\| \tLB^* \big\| _{L_2(\R^d) }^2 + B' \cdot \big\| Q_{r,\gamma_\NC} * \tLB^* - \tLB^* \big\|_{L_1(P_{\tbX})}
\\
\nonumber
& 
\leq
\lambda_\NC \big( \gamma_\NC \sqrt{\pi} \big)^{ - d } \big( 2^r -1 \big)^2 \big\| \tLB^* \big\| _{L_2(\R^d) }^2 + B' \cdot C_{r,1} \cdot \big\| f_{\tbX} \big\|_{L_\infty(\R^d)} \cdot \omega_{r,L_1(\R^d)} ( \tLB, \gamma_\NC/2)
\\
\nonumber
&
\leq
\lambda_\NC \big( \gamma_\NC \sqrt{\pi} \big)^{ - d } \big( 2^r -1 \big)^2 \big\| \tLB^* \big\| _{L_2(\R^d) }^2 + B' c' \cdot C_{r,1} \cdot \big\| f_{\tbX} \big\|_{L_\infty(\R^d)} \gamma_\NC^{\beta}
\end{align}
The first equality is from the construction of $\tLB_0$. The first inequality is from Theorem 2.3 of \citet{SVM2013}. The second inequality is from Lipschitz continuity of $\loss$. The third inequality is from Theorem 2.2 of \citet{SVM2013} with $q=s=1$ and $p=\infty$. The last inequality holds for some constant $c'$ since $\tLB^* \in \mathcal{B}_{1,\infty}^\beta (\R^d)$ implies $\omega_{r,L_1(\R^d)}(\tLB^*, \gamma_\NC/2) \leq c' \gamma_\NC^{\beta}$ from the definition of a Besov space. Combining the results in \eqref{eq-SVM-proof2} and \eqref{eq-SVM-proof3}, we have the following result
\begin{align*}
\risk \big( \widehat{\LB} \big) - \risk \big( \LB^* \big)
&
\leq
9 \big\{ \lambda_\NC \big( \gamma_\NC \sqrt{\pi} \big)^{ - d } \big( 2^r -1 \big)^2 \big\| \tLB^* \big\| _{L_2(\R^d) }^2 + B' c' \cdot C_{r,1} \cdot \big\| f_{\tbX} \big\|_{L_\infty(\R^d)} \gamma_\NC^{\beta} \big\}
\\
& \hspace*{2cm}
+
K_0 \bigg\{ \frac{ \gamma_\NC^{ - (1-p)(1+\epsilon) d} }{\lambda_\NC^p \NC} \bigg\}^{\frac{1}{2-p}}
+
36 \sqrt{2} B \sqrt{ \frac{\tau}{\NC} }
+
15 B \frac{ \tau}{\NC}
\\
&
\leq
c_1 \lambda_\NC\gamma_\NC^{-d} + c_2 \gamma_\NC^\beta 
+ c_3 \Big\{ \gamma_\NC^{(1-p)(1+\epsilon)d} \lambda_\NC^p \NC \Big\}^{-\frac{1}{2-p}} 
+ c_4 \NC^{-1/2}\tau^{1/2} + c_5\NC^{-1} \tau \ .
\end{align*}

\subsection{Proof of Lemma \ref{lem-EstimatedL}}	\label{proof-lem-EstimatedL}

We only show the result about the overall outcome case because the result about the spillover outcome case is obtained from a similar manner.   We find the following result for $t \in [0,1]$.
\begin{align}				\label{eq-SVM2-proof1}
&
\Big|
\widehat{\loss}_{(-\ell)} (t, \bO_i) - \loss (t, \bO_i)
\Big|
\nonumber
\\
&
=
\Big|
\widehat{\losszo}_{(-\ell)} (t, \bO_i) - \losszo (t, \bO_i)
\Big|
\nonumber
\\
&
=
\Bigg|
\frac{1}{\NI_i}
\sum_{j=1}^{\NI_i}
\sum_{a=0}^{1}
\sum_{s=0}^{\NI_i-1} {\NI_i - 1 \choose s}
\Big\{ \widehat{\psi}_k (a , s, \bO_{ij}, \bO_{\eij} ) - \psi_k (a , s, \bO_{ij}, \bO_{\eij}) \Big\}
\sum_{\ell=0}^{\NI_i - a - s} {\NI_i - a - s \choose \ell} \frac{ (-1)^{\ell}  t^{\ell + a + s + 1} }{\ell + a + s + 1}
\Bigg|
\nonumber		
\\
&
\leq 
\frac{1}{\NI_i}
\sum_{j=1}^{\NI_i}
\sum_{a=0}^{1}
\sum_{s=0}^{\NI_i-1} {\NI_i - 1 \choose s}
\Big| \widehat{\psi}_k (a , s, \bO_{ij}, \bO_{\eij} ) - \psi_k (a , s, \bO_{ij}, \bO_{\eij}) \Big|
\sum_{\ell=0}^{\NI_i - a - s} {\NI_i - a - s \choose \ell} \frac{ t^{\ell + a + s + 1} }{\ell + a + s + 1}
\nonumber
\\
&
\leq 
C' \max_{(a,s) \in \{0,1\} \otimes \{ 0,\ldots, \UNI \} } \Big| \widehat{\psi}_k (a , s, \bO_{ij}, \bO_{\eij} ) - \psi_k (a , s, \bO_{ij}, \bO_{\eij}) \Big|  
\end{align} 
for some generic constant $C'$. The last inequality is from $t \in [0,1]$ and bounded $\NI_i$. Also, we find the following result for $t \in (-\infty, 0)$.
\begin{align*}
\Big|
\widehat{\loss}_{(-\ell)} (t, \bO_i) - \loss (t, \bO_i)
\Big|
=
&
\Big|
\widehat{\losszo}_{(-\ell)} (0, \bO_i) - \losszo (0, \bO_i)
\Big|
=
0 \ .
\end{align*}
Finally, we find the following result for $t \in (1,\infty)$ for all $s=0, 1, \ldots, \NI_i$.
\begin{align*}
\Big|
\widehat{\loss}_{(-\ell)} (t, \bO_i) - \loss (t, \bO_i)
\Big|
&
=		
\Big|
\widehat{\losszo}_{(-\ell)} (1, \bO_i) - \losszo (1, \bO_i)
\Big|
\\
&
\leq 
C' \max_{(a,s) \in \{0,1\} \otimes \{ 0,\ldots, \UNI \} } \Big| \widehat{\psi}_k (a , s, \bO_{ij}, \bO_{\eij} ) - \psi_k (a , s, \bO_{ij}, \bO_{\eij}) \Big|   
\end{align*}
where constant $C'$ is given in \eqref{eq-SVM2-proof1}. Therefore, for any $\LB$, we find
\begin{align*}
\Big| \risk \big( \LB \big) - \widehat{\risk}_{(-\ell)} \big( \LB \big)  \Big|
&
\leq
\EXP \bigg[
\Big| \loss \big( \LB (\bX_i) , \bO_i \big) - \widehat{\loss}_{(-\ell)} \big( \LB (\bX_i) , \bO_i \big)	\Big|
\, \bigg| \, \mathcal{D}_\ell^c
\bigg]
\\
&
\leq
\EXP \bigg[
\Big| \loss \big( \LB (\bX_i) , \bO_i \big) - \widehat{\loss}_{(-\ell)} \big( \LB (\bX_i) , \bO_i \big)	\Big|^2
\, \bigg| \, \mathcal{D}_\ell^c
\bigg]^{1/2}
\\
&
\leq
C'
\EXP \bigg[
\max_{(a,s) \in \{0,1\} \otimes \{ 0,\ldots, \UNI \} }
\Big| \widehat{\psi}_k (a , s, \bO_{ij}, \bO_\eij) - \psi_k (a , s, \bO_{ij}, \bO_\eij) \Big|  ^2
\, \bigg| \, \mathcal{D}_\ell^c
\bigg]^{1/2}
\\
&
\leq
C'
\max_{(a,s) \in \{0,1\} \otimes \{ 0,\ldots, \UNI \} }
\big\| \widehat{\psi}_k (a , s, \bO_{ij}, \bO_\eij) - \psi_k (a , s, \bO_{ij}, \bO_\eij) \big\|_{P,2}
\end{align*}
The first inequality is from the definition of $\risk$. The second inequality is from the Jensen's inequality. The third inequality is from the above results. The last inequality is from the definition of $\| \cdot \|_{P,2}$. Therefore, it suffices to bound $\big\| \widehat{\psi}_k (a , s, \bO_{ij}, \bO_\eij) - \psi_k (a , s, \bO_{ij}, \bO_\eij) \big\|_{P,2}$ for all three types of $\psi_k $ where $k \in \{\text{IPW},\text{OR},\text{DR}\}$.

First, the difference between $\widehat{\psi}_{\text{IPW}}$ and $\psi_{\text{IPW}}(a, s, \bO_{ij}, \bO_\eij)$ is
\begin{align*}
&
\Big| \widehat{\psi}_{\text{IPW}}(a , s, \bO_{ij}, \bO_\eij) - \psi_{\text{IPW}}(a , s, \bO_{ij}, \bO_\eij) \Big|
\\
&
=
\Bigg|
\frac{Y_{ij} \ind(A_{ij} = a , S_\eij = s) }{ \widehat{\PS}_{(-\ell)} \big( a, s \cond \bX_i \big) } 
-
\frac{Y_{ij} \ind(A_{ij} = a , S_\eij = s) }{ \PS^* \big( a, s \cond \bX_i \big) } 
\Bigg|
\\
&
\leq
\Big| Y_{ij} \ind(A_{ij} = a , S_\eij = s) \Big| \frac{ \big|\PS^* \big( a , s \cond \bX_i \big) - \widehat{\PS}_{(-\ell)} \big( a , s \cond \bX_i \big) \big| }{\widehat{\PS}_{(-\ell)} \big( a , s \cond \bX_i \big) \PS^* \big( a , s \cond \bX_i \big)}
\\
& \leq \frac{\big|\PS^* \big( a , s \cond \bX_i \big) - \widehat{\PS}_{(-\ell)} \big( a , s \cond \bX_i \big) \big|}{c c'} \ .
\end{align*}
The upper bound is from the bounded outcome and Assumptions \hyperlink{(A3)}{(A3)} and  \hyperlink{(E1)}{(E1)}. Thus, we find the following result with probability greater than $1-\Delta_\NC$:
\begin{align*} 
    \big\| \widehat{\psi}_{\text{IPW}}( a , s, \bO_{ij} , \bO_\eij) - \psi_{\text{IPW}}( a , s, \bO_{ij} , \bO_\eij) \big\|_{P,2} 
  &  \leq 
    \frac{1}{cc'}
    \big\| 
    \PS^*( a , s \cond \bX_i) - \widehat{\PS}_{(-\ell)}( a , s  \cond \bX_i) \big\|_{P,2} 
    \\
   & \leq 
    C_{\text{IPW}} \cdot r_{\PS,\NC} (a,s)
\end{align*}
where $C_{\text{IPW}}=1/(cc')$.

Second, we study the difference between $\widehat{\psi}_\text{OR}$ and $\psi_\text{OR}$ which is 
\begin{align*}
& \big| \widehat{\psi}_\text{OR} ( a , s, \bO_{ij} , \bO_\eij) - \psi_\text{OR} ( a , s, \bO_{ij} , \bO_\eij) \big| \\
& 
= \big| \widehat{\mOR}_{(-\ell)}(a , s / ( \NI_i - 1), \bX_{ij}, \bX_{\eij}) - \mOR^* (a , s / (\NI_i-1), \bX_{ij}, \bX_\eij) \big| \ .
\end{align*}
Therefore, we find the following result with probability greater than $1-\Delta_\NC$.
\begin{align*}
&
\big\| \widehat{\psi}_\text{OR} ( a , s, \bO_{ij} , \bO_\eij) - \psi_\text{OR} ( a , s, \bO_{ij} , \bO_\eij) \big\|_{P,2} \\
&
= \bigg\| \widehat{\mOR}_{(-\ell)} \bigg( a , \frac{s}{\NI_i-1}, \bX_{ij}, \bX_\eij \bigg) - \mOR^* \bigg( a , \frac{s}{\NI_i-1}, \bX_{ij}, \bX_\eij \bigg) \bigg\|_{P,2}
= C_{\text{OR}} \cdot r_{\mOR,\NC}
\end{align*}
where $C_{\text{OR}}=1$.

Lastly, we prove the result when $\psi_\text{DR}$ is chosen. From \eqref{eq-SVM2-proof1}, we have
\begin{align*}
&
- C' \Big[ \widehat{\psi}_\text{DR}(  a , s, \bO_{ij} , \bO_\eij ) 
- \psi_\text{DR}( a , s, \bO_{ij} , \bO_\eij ) \Big]
\\
&
\leq
\widehat{\loss}_{(-\ell)} (t, \bO_i) - \loss (t, \bO_i)
\leq
C' \Big[ \widehat{\psi}_\text{DR}( a , s, \bO_{ij} , \bO_\eij) - \psi_\text{DR}( a , s, \bO_{ij} , \bO_\eij) \Big]
\end{align*}
where the sign of $C'$ is chosen to satisfy the inequality above. The expectation of $\widehat{\psi}_\text{DR} - \psi_\text{DR}$ is
\begin{align*}
&
\EXP \Big\{ \widehat{\psi}_\text{DR}(  a , s, \bO_{ij} , \bO_\eij ) 
- \psi_\text{DR}(  a , s, \bO_{ij} , \bO_\eij ) \, \Big| \, \mathcal{D}_\ell^c \Big\}
\\
& = 
\EXP \Bigg[
\Bigg\{ 
\frac{  Y_{ij} - \widehat{\mOR}_{(-\ell)}( a , \frac{s}{\NI_i-1} , \bX_{ij}, \bX_\eij) }{\widehat{\PS}_{(-\ell)}( a , s \cond \bX_i)}
-
\frac{ Y_{ij} - \mOR(a , \frac{s}{\NI_i-1} ,\bX_{ij}, \bX_\eij)  }{\PS^*(a , s \cond \bX_i)}
\Bigg\}  \ind ( A_{ij} = a ,  S_\eij = s)
\\
& \hspace*{2cm}
+
\widehat{\mOR}_{(-\ell)} \Big( a , \frac{s}{\NI_i-1} ,\bX_{ij}, \bX_\eij \Big) - \mOR^* \Big( a , \frac{s}{\NI_i-1} ,\bX_{ij}, \bX_\eij \Big) 
\, \Bigg| \, \mathcal{D}_\ell^c
\Bigg]
\\
& = 
\EXP \Bigg[
\frac{ \big\{ \mOR^* ( a , \frac{s}{\NI_i-1} , \bX_{ij}, \bX_\eij ) - \widehat{\mOR}_{(-\ell)}(a , \frac{s}{\NI_i-1} ,\bX_{ij}, \bX_\eij )\big\} \PS^*(a , s \cond \bX_i) }{\widehat{\PS}_{(-\ell)}( a , s \cond \bX_i)}
\\
& \hspace*{2cm}
+
\widehat{\mOR}_{(-\ell)} \Big( a , \frac{s}{\NI_i-1} ,\bX_{ij}, \bX_\eij \Big) - \mOR^*  \Big( a , \frac{s}{\NI_i-1},\bX_{ij}, \bX_\eij \Big)
\, \Bigg| \, \mathcal{D}_\ell^c
\Bigg]
\\
& = 
\EXP \Bigg[
\frac{ \big\{ \mOR^* (a , \frac{s}{\NI_i-1}, \bX_{ij}, \bX_\eij ) - \widehat{\mOR}_{(-\ell)}(a , \frac{s}{\NI_i-1}, \bX_{ij}, \bX_\eij  )\big\} \big\{ \PS^*(a , s \cond \bX_i) - \widehat{\PS}_{(-\ell)}(a , s \cond \bX_i) \big\} }{\widehat{\PS}_{(-\ell)}( a , s \cond \bX_i)}
\, \Bigg| \, \mathcal{D}_\ell^c
\Bigg] \ .
\end{align*}
The equalities are straightforward from the definition of $\psi_\text{DR}$ and the law of total expectation. Since $c' \leq \widehat{\PS}_{(-\ell)}$, we find
\begin{align*}
&
\EXP \Bigg[
\frac{ \big\{ \mOR^* (a , \frac{s}{\NI_i-1}, \bX_{ij}, \bX_\eij ) - \widehat{\mOR}_{(-\ell)}(a , \frac{s}{\NI_i-1}, \bX_{ij}, \bX_\eij ) \big\} \big\{ \PS^*(a , s \cond \bX_i) - \widehat{\PS}_{(-\ell)}(a , s \cond \bX_i) \big\} }{\widehat{\PS}_{(-\ell)}( a , s \cond \bX_i)}
\, \Bigg| \, \mathcal{D}_\ell^c
\Bigg]
\\
&
\leq
\frac{1}{c'}
\EXP \bigg[
\bigg| \mOR\bigg( a , \frac{s}{\NI_i-1}, \bX_{ij}, \bX_\eij \bigg) 
- \widehat{\mOR}_{(-\ell)} \bigg( a , \frac{s}{\NI_i-1}, \bX_{ij}, \bX_\eij \bigg) \bigg| \bigg| \PS(a, s \cond \bX_i) - \widehat{\PS}_{(-\ell)}(a,s \cond \bX_i) \bigg|
\, \bigg| \, \mathcal{D}_\ell^c
\bigg]
\\
&
\leq
\frac{1}{c'}
\Bigg\| \widehat{\mOR}_{(-\ell)} \bigg( a , \frac{s}{\NI_i-1}, \bX_{ij}, \bX_\eij \bigg)
 - \mOR^* \bigg( a , \frac{s}{\NI_i-1}, \bX_{ij}, \bX_\eij \bigg) \Bigg\|_{P,2}
 \Big\| \PS^*( a , s \cond \bX_i) - \widehat{\PS}_{(-\ell)}( a , s  \cond \bX_i) \Big\|_{P,2}  \ .
\end{align*}
The first inequality is straightforward. The second inquality is from the H\"older's inequality. From the last line, we find $\EXP \big\{ \widehat{\psi}_\text{DR}( a , s, \bO_{ij} , \bO_\eij) - \psi_\text{DR}( a , s, \bO_{ij} , \bO_\eij) \, \big| \, \mathcal{D}_\ell^c \big\} = O_P(r_{\PS,\NC} r_{\mOR,\NC})$. As a result, we have the following result with probability greater than $1-\Delta_\NC$.
\begin{align*}
&
\Big| \risk \big( \LB \big) - \widehat{\risk}_{(-\ell)} \big( \LB \big)  \Big|
\\
&
=
\bigg| \EXP \Big\{ \loss \big( \LB (\bX_i) , \bO_i \big) - \widehat{\loss}_{(-\ell)} \big( \LB (\bX_i) , \bO_i \big)	
\, \Big| \, \mathcal{D}_\ell^c
\Big\}
\bigg|
\\
&
\leq C'
\max_{(a,s) \in \{0,1\} \otimes \{ 0,\ldots, \UNI \} }
\bigg|
\EXP \Big\{ \widehat{\psi}_\text{DR}( a , s, \bO_{ij} , \bO_\eij) - \psi_\text{DR}( a , s, \bO_{ij} , \bO_\eij) \, \Big| \, \mathcal{D}_\ell^c \Big\}  \bigg|
\\
&
\leq
\frac{C}{c'}
\max_{(a,s) \in \{0,1\} \otimes \{ 0,\ldots, \UNI \} }
\left[
\begin{array}{l}
\Big\| \mOR^*(a , \frac{s}{\NI_i-1}, \bX_{ij}, \bX_\eij ) - \widehat{\mOR}_{(-\ell)}(a , \frac{s}{\NI_i-1}, \bX_{ij}, \bX_\eij )\Big\|_{P,2} 
\\
\quad \times 
\Big\| \PS^* (a , s \cond \bX_i) - \widehat{\PS}_{(-\ell)}(a , s \cond \bX_i) \Big\|_{P,2}
\end{array} 
\right]
\\
& 
\leq 
C_{\text{DR}} \cdot r_{\PS,\NC} r_{\mOR,\NC} 
\end{align*}	
where $C_{\text{DR}}=C/c'$.

Combining the established results, we have the following results  with probability greater than $1-\Delta_\NC$.
\begin{align*}
\Big| \risk \big( \LB \big) - \widehat{\risk}_{(-\ell)} \big( \LB \big)  \Big| 
\leq
\begin{cases}
C_{\text{IPW}} \cdot r_{\PS,\NC} & \text{ if } \psi_\text{IPW} \text{ is chosen}
\\
C_{\text{OR}} \cdot r_{\mOR,\NC} & \text{ if } \psi_\text{OR} \text{ is chosen}
\\
C_{\text{DR}} \cdot r_{\PS,\NC} r_{\mOR,\NC} & \text{ if } \psi_\text{DR} \text{ is chosen}
\end{cases}
\ .
\end{align*}
This implies $\big| \risk \big( \LB \big) - \widehat{\risk}_{(-\ell)} \big( \LB \big)  \big| \leq 0.5 c_6 r_\NC$ with probability greater than $1-\Delta_N$ where $c_6=2 \cdot \max\{ C_{\text{IPW}}, C_{\text{OR}}, C_{\text{DR}}\}$, $r_\NC = r_{\PS,\NC}$ if the inverse probability-weighted loss function is used, $r_\NC = r_{\mOR,\NC}$ if the outcome regression loss function is used, and $r_\NC = r_{\PS,\NC}r_{\mOR,\NC}$ if the doubly robust loss function is used.

\subsection{Proof of Theorem \ref{thm-ER-3}}

We only show the result related to the overall outcome case because the result related to the spillover outcome case is obtained in a similar manner.  We start with defining the risk function and the {\OURR}\ associated with the estimated loss function. Let $\widehat{\risk}_{(-\ell)}(\LB) = \EXP \big\{ \widehat{\loss}_{(-\ell)} \big( \LB(\bX_i), \bO_i \big) \cond \mathcal{D}_\ell^c \big\} $ be the estimated risk function where the expectation is taken with respect to $\bO_i$ while $\widehat{\loss}_{(-\ell)}$ is considered as a fixed function which is clarified by denoting $\mathcal{D}_\ell^c$ in the conditioning statement. Accordingly, let $\LB_{(-\ell)}^*$ be the approximated {\OURR}\ which is the minimizer of $\widehat{\risk}_{(-\ell)}(\LB)$, i.e., $\widehat{\risk}_{(-\ell)}(\LB_{(-\ell)}^*) \leq \widehat{\risk}_{(-\ell)}(\LB) $ for all $\LB \in \F$. We remark that $\LB_{(-\ell)}^* \in \F^{[0,1]}= \big\{ f \cond f(\bx_i ) \in [0,1] \big\}$, the collection of policies ranging over the unit interval.  Using $\LB_{(-\ell)}^*$ as the intermediate quantities, we can establish the excess risk of $\widehat{\LB}_{(-\ell)}$.

We decompose the excess risk as follows.
\begin{align*}
&
\Big| \risk \big( \widehat{\LB}_{(-\ell)} \big) - \risk \big( \LB^* \big) \Big|
\\
&
=
\underbrace{ \Big| \risk \big( \widehat{\LB}_{(-\ell)} \big) - \widehat{\risk}_{(-\ell)} \big( \widehat{\LB}_{(-\ell)} \big) \Big| }_{(A)}
+
\underbrace{ \Big| \widehat{\risk}_{(-\ell)} \big( \widehat{\LB}_{(-\ell)} \big) - \widehat{\risk}_{(-\ell)} \big( \LB_{(-\ell)}^* \big) \Big| }_{(B)}
+
\underbrace{ \Big| \widehat{\risk}_{(-\ell)} \big( \LB_{(-\ell)}^* \big) - \risk \big( \LB^* \big) \Big| }_{(C)}
\ .
\end{align*}
In the rest of the proof, we bound terms $(A)$, $(B)$, and $(C)$. 

From Lemma \ref{lem-EstimatedL} of the Supplementary Material, we find the upper bound of $(A)$ with probability greater than $1-\Delta_N$:
\begin{align*}
\Big| \risk \big( \widehat{\LB}_{(-\ell)} \big) - \widehat{\risk}_{(-\ell)} \big( \widehat{\LB}_{(-\ell)} \big) \Big| 
\leq 0.5 c_6  r_\NC
\ .
\end{align*}

Next, we bound $(B)$. Since $\widehat{\loss}_{(-\ell)}$ satisfies Assumptions \hyperlink{(C1)}{(C1)}-\hyperlink{(C5)}{(C5)} and $\LB_{(-\ell)}^*$ belongs to a Besov space $\mathcal{B}_{1,\infty}^\beta (\R^d)$, Theorem \ref{thm-supp-ER-1} can be applied. Hence, the following result is satisfied with probability greater than $1-3 e^{-\tau}$:
\begin{align*}
&
\Big| \widehat{\risk}_{(-\ell)} \big( \widehat{\LB}_{(-\ell)} \big) - \widehat{\risk}_{(-\ell)} \big( \LB_{(-\ell)}^* \big) \Big|
\\
& 
\leq
c_1 \lambda_\NC\gamma_\NC^{-d} + c_2 \gamma_\NC^\beta 
+ c_3 \Big\{ \gamma_\NC^{(1-p)(1+\epsilon)d} \lambda_\NC^p \NC \Big\}^{-\frac{1}{2-p}} 
+ c_4 \NC^{-1/2}\tau^{1/2} + c_5\NC^{-1} \tau 
 \ .
\end{align*}

Lastly, we bound $(C)$. Since $\LB^*$ is the minimizer of $\risk(\LB)$,  we find
\begin{align*}
\widehat{\risk}_{(-\ell)} \big( \LB_{(-\ell)}^* \big)
=
\risk  \big( \LB_{(-\ell)}^* \big)
+ 
\widehat{\risk}_{(-\ell)} \big( \LB_{(-\ell)}^* \big)
- \risk  \big( \LB_{(-\ell)}^* \big)
\geq
\risk  \big( \LB^* \big)
+ 
\widehat{\risk}_{(-\ell)} \big( \LB_{(-\ell)}^* \big)
- \risk  \big( \LB_{(-\ell)}^* \big) \ ,
\end{align*}
and this implies $\widehat{\risk}_{(-\ell)} \big( \LB_{(-\ell)}^* \big)
- \risk  \big( \LB_{(-\ell)}^* \big) \leq \widehat{\risk}_{(-\ell)} \big( \LB_{(-\ell)}^* \big) - \risk  \big( \LB^* \big) $. Similarly, since $\LB_{(-\ell)}^*$ is the minimizer of $\widehat{\risk}_{(-\ell)}(\LB)$,  we find
\begin{align*}
\risk \big( \LB^* \big)
=
\widehat{\risk}_{(-\ell)} \big( \LB^* \big) 
+
\risk \big( \LB^* \big)
-
\widehat{\risk}_{(-\ell)} \big( \LB^* \big) 
\geq
\widehat{\risk}_{(-\ell)} \big( \LB_{(-\ell)}^* \big) 
+
\risk \big( \LB^* \big)
-
\widehat{\risk}_{(-\ell)} \big( \LB^* \big) \ ,
\end{align*}
and this implies $\risk \big( \LB^* \big)
-
\widehat{\risk}_{(-\ell)} \big( \LB^* \big) \leq \risk \big( \LB^* \big) - \widehat{\risk}_{(-\ell)} \big( \LB_{(-\ell)}^* \big)$, i.e., $ \widehat{\risk}_{(-\ell)} \big( \LB_{(-\ell)}^* \big) - \risk \big( \LB^* \big)  \leq 
\widehat{\risk}_{(-\ell)} \big( \LB^* \big) - \risk \big( \LB^* \big)$. Combining two results, we have
\begin{align*}
\Big| \widehat{\risk}_{(-\ell)} \big( \LB_{(-\ell)}^* \big) - \risk \big( \LB^* \big) \Big|
\leq \max \Big\{
\Big| \widehat{\risk}_{(-\ell)} \big( \LB_{(-\ell)}^* \big) - \risk  \big( \LB_{(-\ell)}^* \big) \Big|
,
\Big| \widehat{\risk}_{(-\ell)} \big( \LB^* \big) - \risk \big( \LB^* \big) \Big|
\Big\} \ .
\end{align*}
From Lemma \ref{lem-EstimatedL} of the Supplementary Material, the right hand side of the above term is upper bounded by $0.5 c_6  r_\NC$ with probability greater than $1-\Delta_\NC$ because $\LB_{(-\ell)}^*, \LB^* \in \F^{[0,1]}$. As a result, we find an upper bound of $(C)$, which is $	\big| \widehat{\risk}_{(-\ell)} \big( \LB_{(-\ell)}^* \big) - \risk \big( \LB^* \big) \big| = O_P(r_\NC)$. We remark that $(A)$ and $(C)$ are both $O_P(r_\NC)$ with probability greater than $1-\Delta_\NC$. This concludes the desired result.